\documentclass[useAMS,usenatbib]{mn2e}

\usepackage{psfig, epsf, epsfig}

\title[ ]{A mechanism of bar formation
in disk galaxies: synchronization of apsidal precession}
\author[K. Bekki]
{Kenji Bekki${}^1$\thanks{E-mail:
kenji.bekki@uwa.edu.au} \\
${}^1$ICRAR M468
The University of Western Australia
35 Stirling Hwy, Crawley
Western Australia 6009, Australia}

\begin{document}

\date{Accepted, Received 2005 February 20; in original form }

\pagerange{\pageref{firstpage}--\pageref{lastpage}} \pubyear{2005}

\maketitle

\label{firstpage}

\begin{abstract}

We discuss the mechanism(s) of bar formation in isolated and tidally
interacting disk galaxies using
the results of idealized collisionless Nbody simulations of the galaxies.
In order to better understand the mechanism, we investigate
orbital eccentricities ($e$),
epochs of apocenter passages ($t_{\rm a}$),
azimuthal angles at $t_{\rm a}$ ($\varphi_{\rm a}$),
precession rates ($\Omega_{\rm pre}$),  
for individual stars, 
as well as bar strengths represented by
relative $m=2$ Fourier amplitude ($A_2$) and 
bar pattern speeds ($\Omega_{\rm bar}$).
The main results are as follows.
A significant fraction of stars with initially
different $\varphi_{\rm a}$ and $\Omega_{\rm pre}$  in an isolated disk galaxy 
can  have similar values
within several dynamical timescales. 
This synchronization of $\varphi_{\rm a}$ and $\Omega_{\rm pre}$,
which is referred to as apsidal precession synchronization  (``APS'')
in the present study,
is caused by the enhanced strength of the tangential component of gravitational force.
A weak seed bar ($A_2<0.1$) is first formed through APS in local regions of a disk,
then the bar grows due to APS.
In the bar growth phase ($0.1<A_2<0.4$),
APS can proceed more efficiently due to stronger tangential
force from the bar  so that it can  enhance the bar strength further.
This positive feedback loop in APS 
is the key physical mechanism of bar growth in isolated
stellar disks.
Bar formation can be severely suppressed in disks with lower disk mass
fractions and/or higher $Q$ parameters due to much less efficient APS.
APS proceeds more rapidly and more efficiently due to  strong
tidal perturbation 
in the formation of tidal bars compared to spontaneous bar formation.
\end{abstract}


\section{Introduction}

Bars are the fundamental galactic structure that influence galaxy
evolution in various ways (e.g., Sellwood 2014, S14).
One of major aims in observational and theoretical studies
of bars has been to reveal their various roles in galaxy 
formation and evolution.
For example, previous numerical and theoretical
studies of bars  revealed that
bars can induce the rapid inward transfer of cold gas to
the inner regions of galaxies and consequently
trigger gas fueling to the central starbursts and AGN
(e.g., Noguchi 1989; Shlosman et al. 1989;
Fukunaga \& Tosa 1991; Wada \& Habe 1992;  Heller \& Shlosman 1994;
Berentzen et al. 1998; Athanassoula et al. 2013; Spinoso et al. 2017).
Observational studies tried to find
a physical link  between recent
enhanced star formation in central regions of disk galaxies and 
presence (or absence) of bars (e.g., Hawarden et al. 1986;
Pompea \& Rieke 1990;  Aguerri 1999;
Coelho \& Gadotti 2011;
Ellison e tal. 2011; Perez \& Sanchez-Blazquez 2011;
Fraser-McKelvie et al. 2020).
Radial metallicity gradients of stars have been suggested to be
flattened due to mixing of stellar populations by dynamical
action of bars on stellar disks (e.g., Friedli et al. 1994).

Long-term dynamical action of bars on disk field stars 
can change mass, angular momentum, and energy of stars
and gas: bars can be one of main drivers for secular evolution
of disk galaxies (e.g., Kormendy 2013; Laurikainen et al. 2013).
Resonant dynamical interaction between bars and disk field stars
can be a e major mechanisms of outer rings  observed in a significant
fraction ($\approx 20$\%) of 
disk galaxies
(e.g., Buta \& Combes 1996). 
Boxy and peanut-shaped bulges can be the edge-on views of three-dimensional
thick stellar bars within disk galaxies (e.g., Combes \& Sanders 1981).
Thus understanding the physics of bar formation and evolution
can lead us further
to our deeper understanding of the essential mechanisms for 
structures of galaxies, origin of starbursts, and bulge formation.

It has long been a key theoretical question in bar formation in what physical
conditions bars are formed in galaxies
(e.g., Ostriker \& Peebles 1970; Athanassoula \& Sellwood 1986, AS86;
see a classic review by 
Sellwood  \& Wilkinson 1993).
Numerical simulations have so far revealed the physical parameters 
of galaxies required for
bar formation in disk galaxies, such as the ratio of total disk mass
to the total system mass (e.g., Hohl 1976; Carlberg \& Freeman 1985),
degrees of random motion (e.g., AS86),
initial rotation curves (e.g., Sellwood 1981; Combes \& Elmegreen 1993),
gas mass fractions (e.g., Shlosman \& Noguchi 1993; Bournaud et  al. 2005),
accretion rates of cold gaseous components (Sellwood \& Carlberg 1984),
and the compactness of galaxies (e.g., Efstathiou et al. 1982).
However, it is not so clear  
why these parameters can determine whether stellar bars
can be formed or not.
Also, although a number of the bar formation paths have been proposed,
such as global bar instability (e.g., Hohl 1976)
tidal interaction (e.g., Noguchi 1987; \L okas 2018, 2020), and  galaxy merging
(e.g., Cavanagh \& Bekki 2020),
it has not been clearly understood (i)  what mechanisms can trigger
bar formation in these formation paths
and (ii) why bar properties (e.g., bar pattern speed)
are different in these paths.
For example, it has not yet been clarified why strong tidal force in interacting
galaxies can trigger the
formation of stellar bars in disks that cannot form bars in isolation
(Noguchi 1987).

\begin{table}
\centering
\begin{minipage}{80mm}
\caption{Brief description of physical meanings for symbols 
used in  this study.}
\begin{tabular}{ll}
{Symbol}
& {Physical meaning}\\
$\Omega_{\rm bar}$  & bar pattern speed  \\
$\Omega_{\rm pre}$  & (apsidal) precession frequency  \\
$\Omega_{\rm pre,p}$  & peak $\Omega_{\rm pre}$ in N$(\Omega_{\rm pre})$
distribution function \\
$\Omega$  & angular frequency  \\
$\kappa$  & epicyclic frequency  \\
$T_{\rm r}$  & radial period  \\
$R_{\rm a}$   &  apocenter distance \\
$R_{\rm p}$   &  pericenter distance \\
$ \varphi_{\rm a}$ & azimuthal angle at $R=R_{\rm a}$ \\
$ t_{\rm a}$ & time at $R=R_{\rm a}$ \\
$e$ & orbital eccentricity \\
$v_{\varphi}$ & azimuthal component of velocity  \\
$v_{\rm c}$ & circular velocity   \\
$f_{\rm v}$ & ratio of $v_{\varphi}$ to $v_{\rm c}$   \\
$N(\Omega_{\rm pre})$  & number distribution of $\Omega_{\rm pre}$  \\
$N(\varphi_{\rm a})$  & number distribution of $\varphi_{\rm a}$  \\
$N_{\rm m}(\varphi_{\rm a})$  & mean of $N(\varphi_{\rm a})$ \\
$N_{\rm max}(\varphi_{\rm a})$  & maximum of $N(\varphi_{\rm a})$ \\
$N(e)$  & number distribution of $e$  \\
$\Sigma(t_{\rm a},\varphi_{\rm a})$ &  phase space density  at
($t_{\rm a},\varphi_{\rm a}$) \\
$\Sigma(\Omega_{\rm pre},e)$ &  phase space density  at
($\Omega_{\rm pre},e$) \\
$A_{\rm m}$  & m-th relative Fourier mode strength   \\
$A_{\rm 2}$  & m=2 relative Fourier mode strength (bar strength)  \\
$F_{\rm t}$  & normalized strength of tangential force ($f_{\rm t}/f$)  \\
$P_{\rm syn}$  &  degree of (apsidal) precession synchronization  \\
$S_{\rm aps}$  &  power of (apsidal) precession synchronization  \\
$F_{\rm a}$  &  fraction of particles with similar $t_{\rm a}$   \\
$f_{\rm d}$  & disk mass fraction  \\
$f_{\rm b}$  & bulge mass fraction  \\
$R_{\rm d}$  & disk size  \\
$a_{\rm s}$  & scale length for a stellar disk  \\
$t_{\rm dyn}$  & dynamical time scale at $R=R_{\rm d}$  \\
\end{tabular}
\end{minipage}
\end{table}

Contopoulos (1980) and Contopoulos \& Papayannopoulos (1980)
investigated the orbital properties of stars in barred potentials for
different bar strengths and pattern speeds  and different 
mass distributions of the systems
in order to understand the major orbital populations of bars
and the physical origin of bar lengths.
Although they revealed the basic orbital families of
barred systems, such as ``${\rm x}_1$'' orbits along bars and the roles
of inner and outer Lindblad resonances in determining
the locations and lengths of bars within disks,
they  did not discuss how the assumed initial bars are formed
in disk galaxies.
Lynden-Bell (1979, L79) proposed that initially weak stellar bars
can gravitationally capture 
the resonant orbits of stars ($\Omega_{\rm bar}=\Omega - 0.5 \kappa$) to fully
develop: see more discussion on six bar formation mechanisms including
this orbit trapping  in Lynden-Bell (1996).
However, the formation mechanism for such  weak bars
and the preferential capture of resonant
orbits by existing bars have not been investigated at all by
recent numerical simulations of bar formation.

AS86 investigated whether or not random motion of stars can suppress
global bar instabilities in disk galaxies  using numerical
simulations of four different disk models with different $Q$ parameters
and halo mass fractions. They also tried to find a correlation
between the growth rate of the simulated bars
and the net growth factor (NGF) predicted from the swing amplification
theory proposed by Toomre (1981, T81).
Although they found a positive correlation between the two, which
is consistent with the predictions by T81,
the simulated growth rates show a large dispersion for a given NFS factor,
in particular, for $\ln {\rm NSF}>0.8$ (see their Fig. 5).
Furthermore, the simulated growth rates are by a factor of $\sim 2$ smaller
than the local estimates based on the swing amplification theory  (their Fig. 6).
This inconsistency could be due largely to the assumptions adopted
in estimating the growth factors,
however, the origin of the inconsistency needs to be clarified.
It would be possible   that the positive correlation
is explained
by other theories of bar formation.
Thus, it would be fair to say that the physical mechanism of bar 
formation is yet to be fully
understood.

Each star in a disk galaxy can have a Rosetta-shaped orbit
in which the line connecting between apocenter and pericenter
(i.e., line of apsides)  slowly
rotates over many dynamical timescales:
this apsidal precession is simply referred to as precession
in the present study, as in Binney \& Tremaine (1987).
 The rate of this precession and 
and the direction of the line of apsides can be quite different between
different stars in disk galaxies.
Accordingly,
barred disk galaxies can be described as self-gravitating systems in which
(i) the  directions of the lines of apsides 
are aligned to a large extent
and
(ii) precession rates of these stars are very similar (S14):
disk galaxies with no bars,
on the other hand, have a much smaller fraction of  such synchronized orbits.
In short, bar formation can be described as (apsidal) precession synchronization,
which is referred to as ``APS'' from now on in the present study.
Since  APS process was not extensively investigated
at all  in previous numerical simulations of bar formation,
it is not so clear how and why APS can occur during bar 
formation in disk galaxies.

The purpose of this paper is thus to investigate (i) how precession of stars
can be synchronized during bar formation in disk galaxies
and (ii) what causes such synchronization
using idealized collisionless Nbody simulations of the galaxies.
In order to do so, we investigate the orbital properties of all stars in
a simulation, such as orbital eccentricities  ($e$),
apocenter distances ($R_{\rm a}$), epochs of apocenter passages
($t_{\rm a}$),
azimuthal angles at 
apocenter passages ($\varphi_{\rm a}$),
precession rates ($\Omega_{\rm pre}$), angular velocities ($\Omega$), 
radial period ($T_{\rm r}$).
Although there are a number of bar formation paths,
we focus exclusively on the bar formation through (i) global bar instability in
isolated disk galaxies and (ii) tidal perturbation during galaxy interaction with
companion galaxies. 
We mainly investigate the time evolution of $\varphi_{\rm a}$,
$\Omega_{\rm pre}$, and $e$ of stars
in disk galaxies with forming bars
for the two formation paths.
Thus, the present study is complementary
to previous theoretical studies of bar formation
(e.g., L79 and T81), which did not investigate how the alignment/synchronization
of $\varphi_{\rm a}$ and $\Omega_{\rm pre}$
occurs during
bar formation.

We consider that the tangential (azimuthal) component of gravitational force
is crucial for APS in bar formation in the present study. We therefore introduce the
following parameter:
\begin{equation}
F_{\rm t}= \frac{f_{\rm t}}{f},
\end{equation}
where $f$ is the total gravitational force and $f_{\rm t}$ 
is the tangential (azimuthal) 
component of the force.
This normalized tangential force  ($F_{\rm t}$)
is rather small ($\approx 0.01$) in the initial disk models with almost
spherical dark matter halos and very smooth stellar disks (i.e., no/little
local density inhomogeneity).
However, 
as shown later in this paper,
$F_{\rm t}$ can change much more dramatically
than the radial force so that it can change
$\varphi_{\rm a}$ and $\Omega_{\rm pre}$ of stars.
We thus  demonstrate how the time evolution of $F_{\rm t}$ can cause
significant changes in 
$\varphi_{\rm a}$ and $\Omega_{\rm pre}$ in order to understand
the physical origin of APS (thus bar formation).

The plan of the paper is as follows.
We describe the models for disk galaxies
and the methods to quantify the orbital properties of stars and 
the physical properties of the simulated bars
in \S 2.
We present the results of numerical simulations of
bar formation and reveal the major mechanism of bar formation both for
isolated and tidal models
in \S 3.
Based on these results,
we discuss what determines
the  lengths ($R_{\rm bar}$) and the  pattern speeds ($\Omega_{\rm bar}$)
of stellar bars in disk galaxies
in \S 4.
We summarize the essential mechanisms of bar formation revealed from the
present simulations
in \S 5.
In this paper, we do not discuss other key issues related to bar
formation and evolution, such as dynamical interaction between
bars and dark matter halos (e.g., Weinberg 1985; Debattista \& Sellwood 2000).
and redshift evolution of bars
(e.g.,  Abraham et al. 1999; Jogee et al. 2004; Sheth et al. 2008; Cavanagh et al. 2022).

\begin{table}
\centering
\begin{minipage}{90mm}
\caption{A summary for model parameters 
in isolated (``I'')  and interacting (``T'') disk galaxies.}
\begin{tabular}{llllll}
Model & $f_{\rm d}$ & $f_{\rm b}$ & $Q$ & $R_{\rm vir}/R_{\rm d}$ & $c$  \\ 
IA1 & 0.35 & 0.0 & 1.5 & 14.0 & 10 \\
IA2 & 0.04 & 0.0 & 1.5 & 14.0 & 10 \\
IA3 & 0.11 & 0.0 & 1.5  & 14.0 & 10 \\
IA4 & 0.14 & 0.0 & 1.5  & 14.0 & 10\\
IA5 & 0.17 & 0.0 & 1.5  & 14.0 & 10 \\
IA6 & 0.25 & 0.0 & 1.5  & 14.0 & 10\\
IA7 & 0.28 & 0.0 & 1.5  & 14.0 & 10\\
IA8 & 0.42 & 0.0 & 1.5  & 14.0 & 10\\
IA9 & 0.53 & 0.0 & 1.5  & 14.0 & 10\\
IA10 & 0.70 & 0.0 & 1.5  & 14.0 & 10\\
IA11 & 0.35 & 0.17 & 1.5  & 14.0 & 10\\
IA12 & 0.35 & 0.3 & 1.5  & 14.0 & 10\\
IA13 & 0.35 & 0.5 & 1.5  & 14.0 & 10\\
IA14 & 0.35 & 1.0 & 1.5  & 14.0 & 10\\
IA15 & 0.35 & 0.0 & 1.0  & 14.0 & 10\\
IA16 & 0.35 & 0.0 & 2.0  & 14.0 & 10\\
IA17 & 0.35 & 0.0 & 3.0  & 14.0 & 10\\
IB1 & 0.11 & 0.0 & 1.5  & 19.8 & 10\\
IB2 & 0.41 & 0.0 & 1.5 & 19.8 & 10\\
IB3 & 0.41 & 1.0 & 1.5 & 19.8 & 10 \\
IB4 & 0.41 & 2.0 & 1.5 & 19.8 & 10 \\
IB5 & 0.33 & 0.0 & 1.5 & 19.8 & 10 \\
IC1 & 0.09 & 0.0 & 1.5  & 14.0 & 16\\
IC2 & 0.14 & 0.0 & 1.5 & 14.0 & 16 \\
IC3 & 0.28 & 0.0 & 1.5 & 14.0 & 16 \\
TA1 & 0.35 & 0.0 & 1.5 & 14.0 & 10 \\
TA2 & 0.11 & 0.0 & 1.5 & 14.0 & 10 \\
TA3 & 0.11 & 0.17 & 1.5 & 14.0 & 10 \\
TA4 & 0.14 & 0.0 & 1.5& 14.0 & 10 \\
TA5 & 0.17 & 0.0 & 1.5 & 14.0 & 10\\
TB1 & 0.41 & 0.3 & 1.5 & 19.8 & 10\\
TB2 & 0.41 & 0.5 & 1.5 & 19.8 & 10\\
TB3 & 0.41 & 2.0 & 1.5 & 19.8 & 10\\
TC1 & 0.09 & 0.0 & 1.5 & 14.0 & 16\\
TC2 & 0.14 & 0.0 & 1.5 & 14.0 & 16\\
\end{tabular}
\end{minipage}
\end{table}

\begin{figure*}
\psfig{file=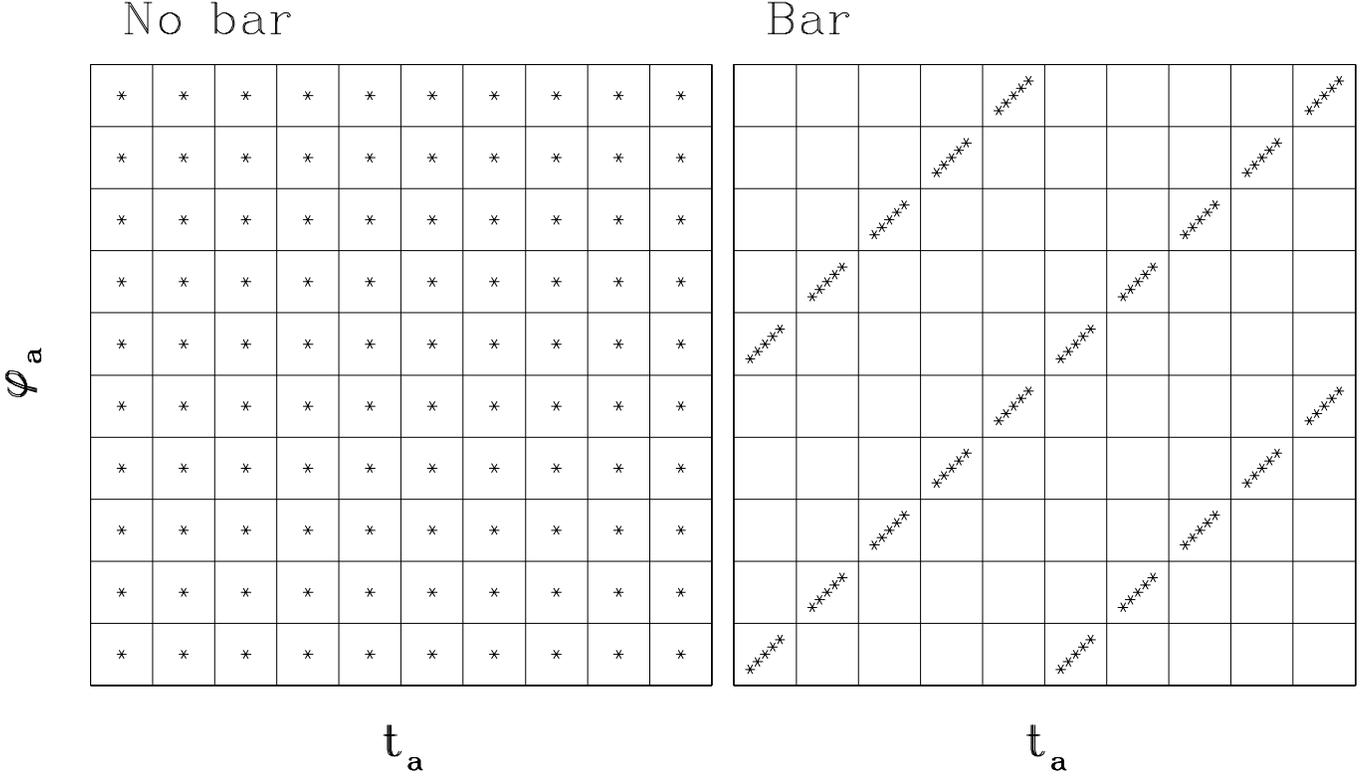,width=18.0cm}
\caption{
An illustration of bar formation in the $(t_{\rm a},\varphi_{\rm a})$ phase space,
where $t_{\rm a}$ and $\varphi_{\rm a}$ are the epoch of the apocenter passage
of a star and the azimuthal angle ($\varphi$)  of the star at $t_{\rm a}$, respectively.
 Just for convenience,
100 stellar particles are plotted by asterisk marks in the $10 \times 10$ grid points.
In stellar disks without bars, stellar particles can have a wide range of 
$\varphi_{\rm a}$ for a given $t_{\rm a}$. 
The particles in stellar disks with bars, on the other hand, show two distinct 
grid points in $(t_{\rm a},\varphi_{\rm a})$ (for a given $t_{\rm a}$)
where most of the particles reside
in due to synchronization of apsidal precession rates of
the particles during bar formation. 
Concentration of stellar particles in a limited number of grid points of
this  $(t_{\rm a},\varphi_{\rm a})$ phase space
is one of the two  manifestations of apsidal precession synchronization (APS)
that leads to bar formation. Other manifestation is the strong concentration
of particles in the  $(\Omega_{\rm pre},e)$ phase space,
which is discussed in the main text.
One of main purposes in this study is to understand how APS
can occur during dynamical evolution of disk galaxies.
We use this $(t_{\rm a},\varphi_{\rm a})$ phase space rather 
than $(t_{\rm p},\varphi_{\rm p})$ one, where $t_{\rm p}$ and $\varphi_{\rm p}$
are the epoch of pericenter passage and the azimuthal angle at $t_{\rm p}$,
respectively, mainly because the outer appearance of a bar depends
on the distribution of $\varphi_{\rm a}$.
}
\label{Figure. 1}
\end{figure*}

\begin{figure*}
\psfig{file=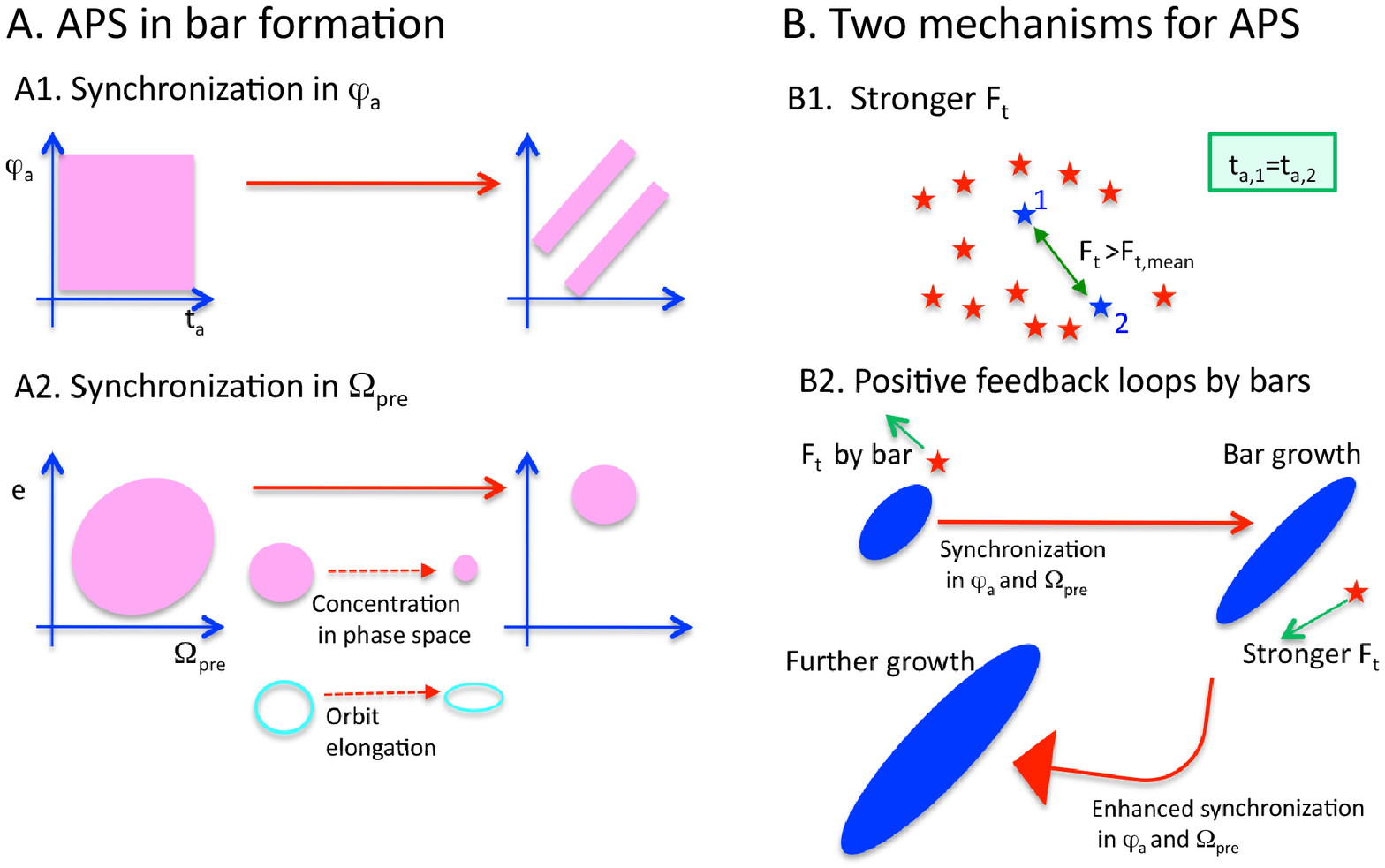,width=18.0cm,height=9cm}
\caption{
An illustration of (A) two key physical processes in APS leading to bar
formation, i.e., synchronization in $\varphi_{\rm a}$ and $\Omega_{\rm pre}$
(left half) and (B) two physical mechanisms that can cause APS
(right half). In A1 and A2, $e$ and $\Omega_{\rm pre}$ represent
orbital eccentricities and precession rates of stars, respectively:
see Fig. 1 for the explanations of $t_{\rm a}$ and $\varphi_{\rm a}$.
In B1 and B2,
$F_{\rm t}$ and $F_{\rm t, mean}$ are the tangential (azimuthal) component
of gravitational force between two nearby particles with
similar $t_{\rm a}$ ($t_{\rm a, 1} \approx t_{\rm a, 2}$) and the  mean 
of $F_{\rm t}$ in 
a local region of a disk, respectively.
Bar formation in disk galaxies requires synchronization
of $\varphi_{\rm a}$ (A1) and $\Omega_{\rm pre}$  (A2)
and elongation of orbits (A2) in their stars.
Also, strengthened 
$F_{\rm t}$ 
is considered to cause APS in the present study. There are two ways to
strengthen $F_{\rm t}$ in stellar disks.
One is mutual gravitational interaction between  particles 
that are close each other and have similar $t_{\rm a}$ (B1).
The other is dynamical action of existing
stellar bars on disk field stars (B2).
Stronger $F_{\rm t}$ by existing bars can cause more efficient
synchronization of 
$\varphi_{\rm a}$  and $\Omega_{\rm pre}$ so that the bars can have more 
stars with similar $\varphi_{\rm a}$ and $\Omega_{\rm pre}$ 
($\approx \Omega_{\rm bar}$).
As a result of this, bars can grow stronger  to have stronger  $F_{\rm t}$ on 
stars. The strengthened $F_{\rm t}$ can cause 
even more efficient synchronization of $\varphi_{\rm a}$  and $\Omega_{\rm pre}$,
which allow the bars to grow further. Thus, this positive feedback loop 
is the key mechanism of bar growth in disk galaxies.
}
\label{Figure. 2}
\end{figure*}

\begin{figure*}
\psfig{file=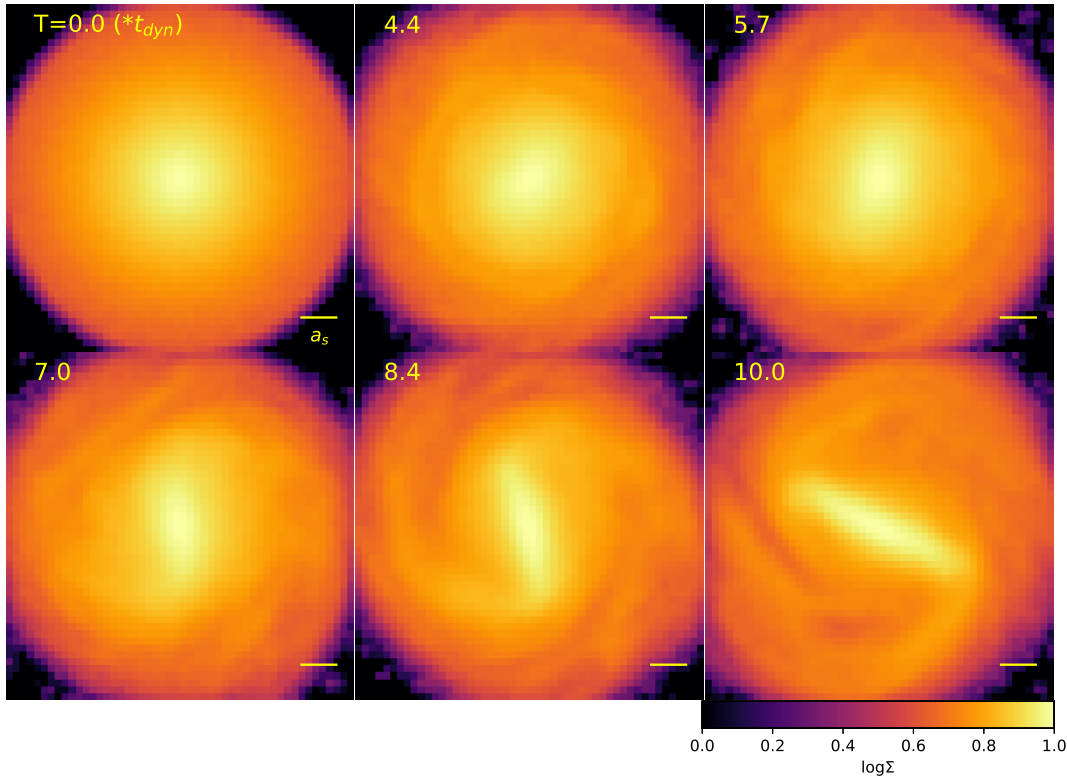,width=18.0cm}
\caption{
Time evolution of the surface mass density of stellar particles
($\Sigma$ in logarithmic scale)  
projected onto the $x$-$z$ plane in the fiducial model
(IA1). Time in units of $t_{\rm dyn}$
is given in the upper left corner of each panel. The scale bar shown in the lower right
corner of each panel indicate $a_{\rm s}$, which is the scale radius of the initial
exponential stellar disk.The mass density is normalized to range from 0 to 1 
in each panel so that
the density evolution can be clearly seen.
}
\label{Figure. 3}
\end{figure*}

\begin{figure*}
\psfig{file=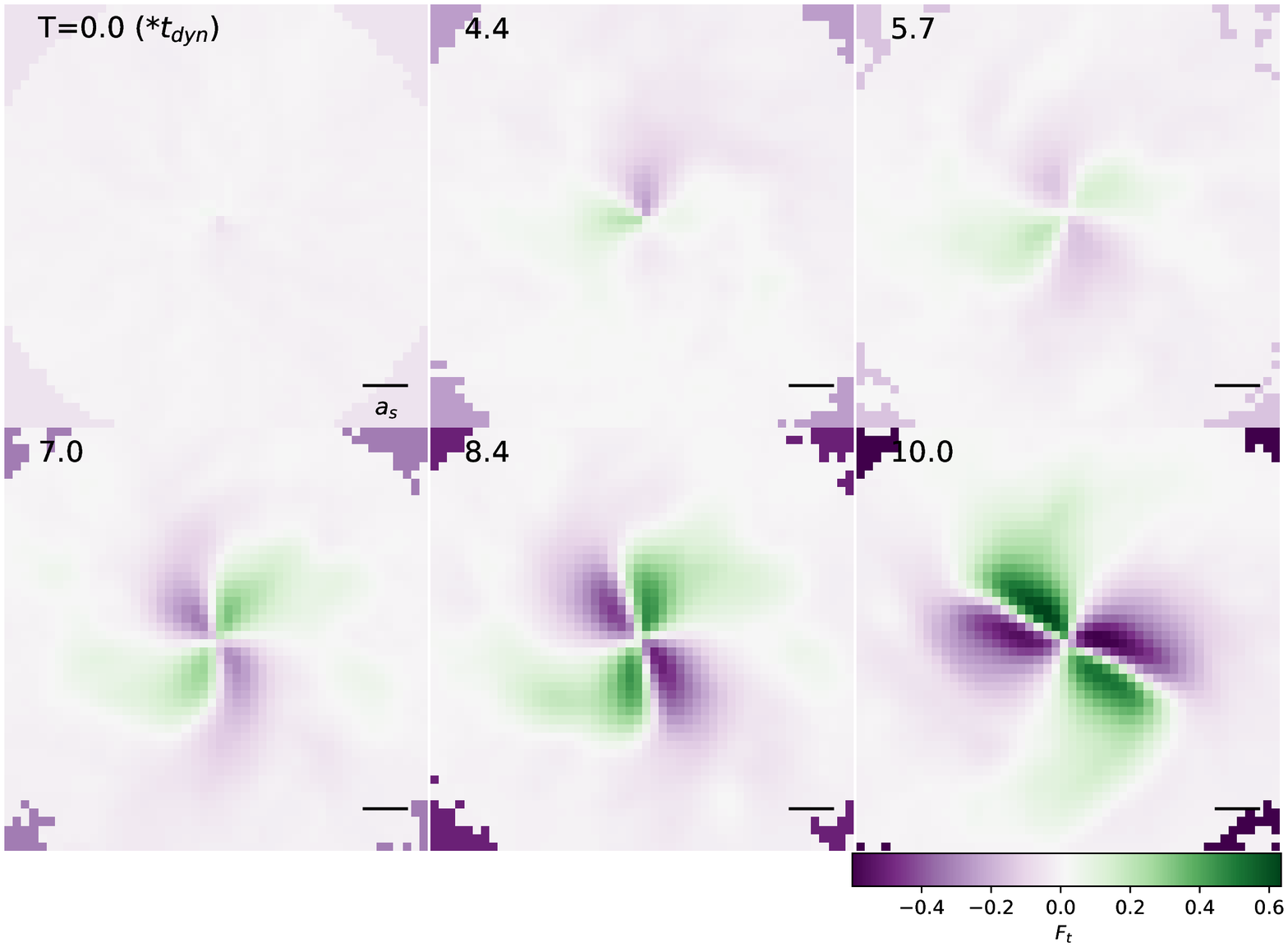,width=18.0cm}
\caption{
The same as Fig. 1 but for the 2D maps for the relative strengths of the 
tangential
components of gravitational force ($F_{\rm t}$).
}
\label{Figure. 4}
\end{figure*}

\section{Models}

\subsection{Isolated and interacting galaxy models}
We investigate the details of bar formation processes in both
(i)  isolated disk galaxy model
and (ii) tidally interaction one, because the physical mechanism(s) of bar
formation can be quite different between the two different models.
Since we have used and described the details of these models in our
previous papers (e.g., Bekki 2015), we give them in Appendix A.
We use our original code that can be run on GPU
(Graphics Processing Unit)
clusters (Bekki 2013, 2015) 
to perform purely collisionless Nbody simulations of bar formation:
the details of the code are given  in Appendix A.
In order to understand the physical mechanisms of bar formation more clearly,
we also run  comparative models with fixed gravitational potentials in addition to the
above-mentioned full Nbody simulations: the details are given also in Appendix A.

\begin{figure*}
\psfig{file=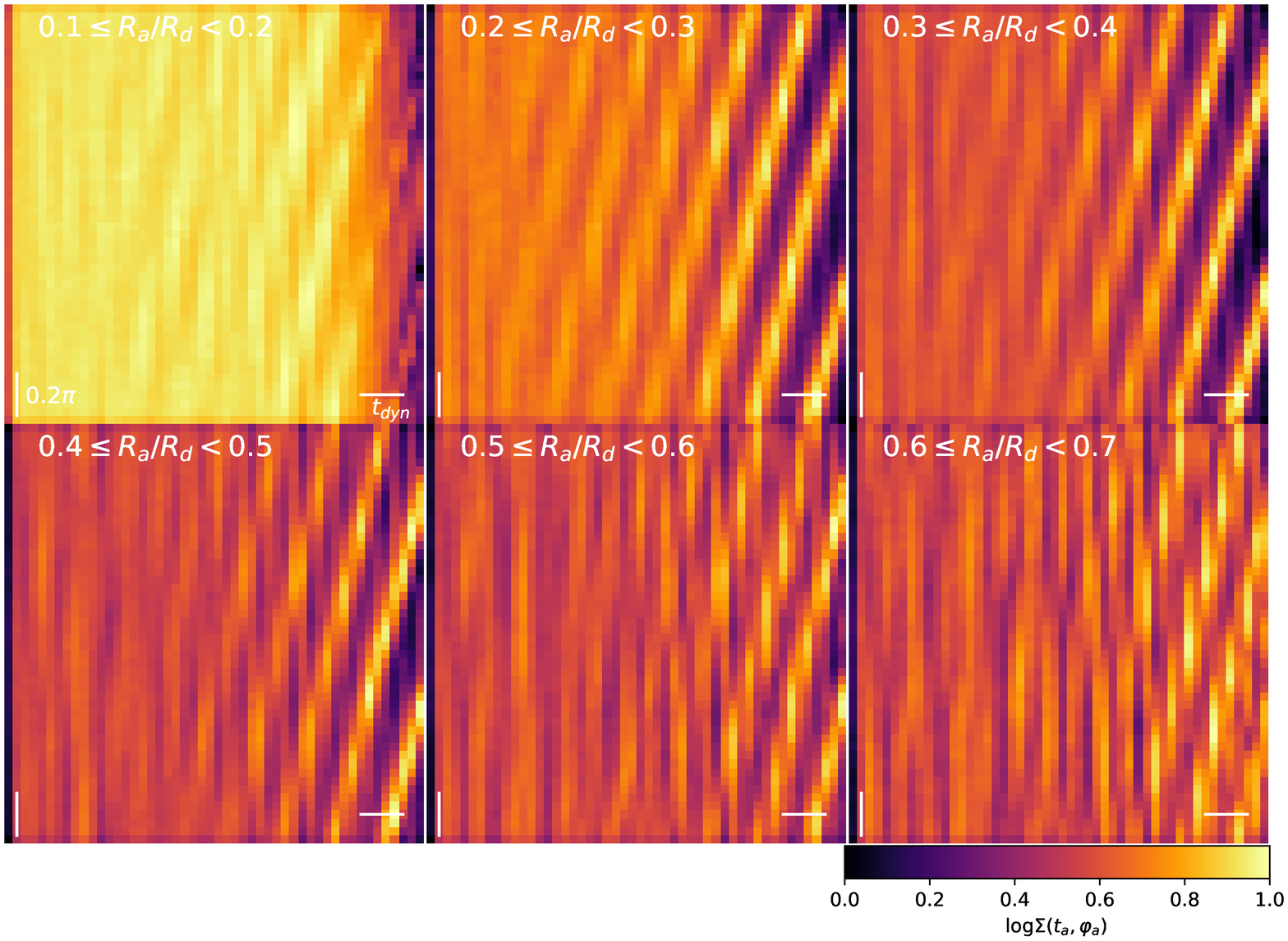,width=18cm}
\caption{
Surface number densities of stellar particles on the $t_{\rm a}-\varphi_{\rm a}$ plane
($\Sigma(t_{\rm a},\varphi_{\rm a})$ on logarithmic scale) for six radial bins
in the fiducial
model.
The horizontal and vertical axes represent $t_{\rm a}$ and $\varphi_{\rm a}$,
respectively, and the horizontal and vertical scale bars indicates
$t_{\rm dyn}$ and $0.2 \pi$, respectively.
The radial range for each frame is given in the upper left corner for each panel,
and $\Sigma(t_{\rm a},\varphi_{\rm a})$  is normalized
to range from 0 to 1 for each radial bin. Therefore, a density contrast
in each frame can be clearly seen (and overemphasized in some bins,
e.g., $t_{\rm a}/t_{\rm dyn}>8$ for $0.1 \le R_{\rm a}/R_{\rm d} < 0.2$).
Clearly, two distinct peaks for a given $t_{\rm a}$ can be clearly seen
in the later dynamical evolution of the disk ($>6t_{\rm dyn}$)
for $0.2 \le R_{\rm a}/R_{\rm d} < 0.4$,
which is a manifestation of bar formation in this model.
Such two distinct peaks become less significant in
the outer parts of the stellar disk ($0.5 \le R_{\rm a}/R_{\rm d}$).
}
\label{Figure. 5}
\end{figure*}

\begin{figure}
\psfig{file=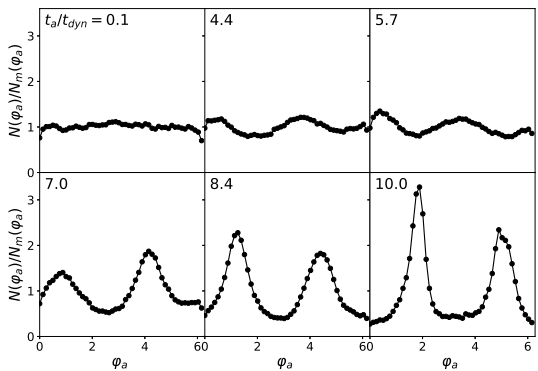,width=8.5cm}
\caption{
Normalized $\varphi_{\rm a}$  distributions
($N(\varphi_{\rm a})$) of stars at $0.2R_{\rm d}\le R \le 0.3R_{\rm d}$
for six different $t_{\rm a}$ (shown in the upper left corner)
in the fiducial model.
Since  $N(\varphi_{\rm a})$ is normalized by the mean $N(\varphi_{\rm a}$)
at the adopted $R$ range ($N_{\rm m}(\varphi_{\rm a})$),
this plot can indicate the degree of $\varphi_{\rm a}$ synchronization
at each time step. 
Clearly, the profiles show
two distinct peaks separated by $\approx \pi$ (radian),
which are characteristic of a stellar bar formed through APS,  at later epochs
(e.g, $t_{\rm a}/t_{\rm dyn}=7.0$, 8.4, and 10.0).
}
\label{Figure. 6}
\end{figure}

\begin{figure}
\psfig{file=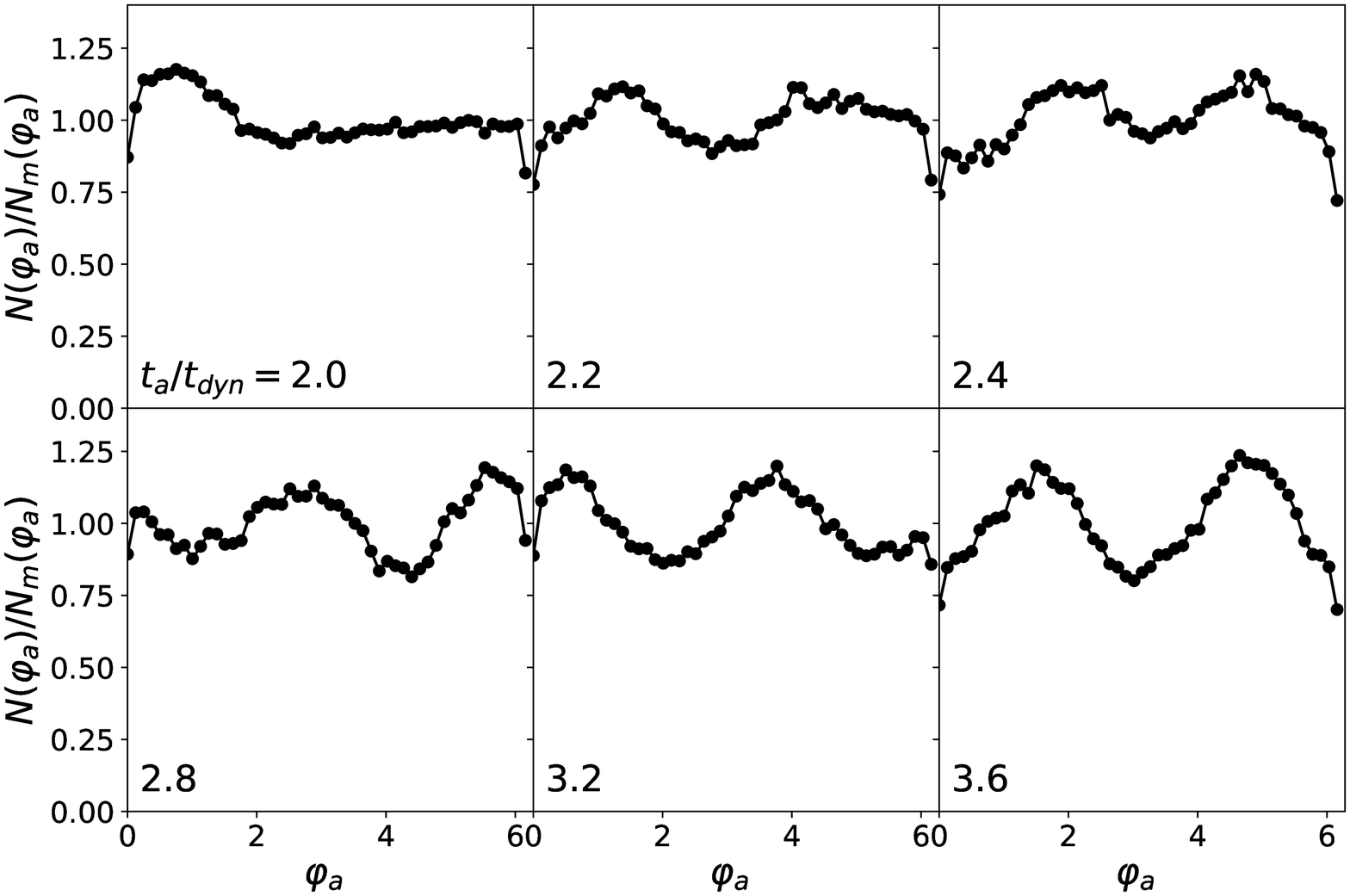,width=8.5cm}
\caption{
The same as Fig. 6 but for a different set of $t_{\rm a}$.
Two main peaks separated by roughly $\pi$ appear at $t_{\rm a}/t_{\rm dyn}=2.2$
and 2.4
whereas only one single broad peak can be seen at $t_{\rm a}/t_{\rm dyn}=2.0$.
We therefore consider that
$t_{\rm a}/t_{\rm dyn} \approx 2.2$ can
correspond to the formation phase of a seed bar
owing to the weakly synchronized $\varphi_{\rm a}$
in the fiducial model.
}
\label{Figure. 7}
\end{figure}

\subsection{Orbital properties of stellar particles}

In order to better understand the bar formation processes,
we investigate the orbital properties of individual stars,
such as  apocenter and pericenter distances
of orbits ($R_{\rm a}$
and $R_{\rm p}$, respectively), 
azimuthal angles at $R_{\rm a}$ and $R_{\rm p}$
($\varphi_{\rm a}$ and $\varphi_{\rm p}$, respectively),
 orbital eccentricities
($e$), radial and angular frequencies ($T_{\rm r}$ and $\Omega$,
respectively), and precession frequencies ($\Omega_{\rm pre}$) 
in a simulation.
The method to estimate the bar strengths and patter speeds and the 
evolution of the strengths and the speeds are given in
Appendix B.
Each star can have about ten
apocenter (pericenter) passages at least in a simulation,
these orbital properties are estimated at each apocenter passage.
For example, $e$ of a stellar particle 
after its $k$-th apocenter and pericenter passages is defined
as follows:
\begin{equation}
e_{k}=\frac{ R_{\rm a, \it k} - R_{\rm p, \it k} }
{ R_{\rm a, \it k} + R_{\rm p, \it k} } .
\end{equation}
The precession rate of a star  at its $k$-th apocenter passage
is  estimated as follows:
\begin{equation}
\Omega_{\rm pre, \it k}=\frac{ \varphi_{\rm a, \it k} - \varphi_{\rm a, \it k-1} }
{ T_{\rm r, \it k} },
\end{equation}
where $\varphi_{\rm a, \it, k}$ and $\varphi_{\rm a, \it, k-1}$
are $\varphi_{\rm a}$ at $k$-th and $k-1$ apocenter passages,
and $T_{\rm r, \it k}$ is the radial period at the $k$-th apocenter passage
(i.e., the time interval between the two apocenter passages),
which is simply defined as
\begin{equation}
T_{\rm r, \it k}= t_{\rm a, \it k}-t_{\rm a, \it k-1}.
\end{equation}

\subsection{Relative strengths of the radial and  azimuthal components of gravitational force}

As mentioned in \S 1,
we investigate the relative strength of tangential component
of gravitational force ($F_{\rm t}$)
for each individual stellar particle at
every $0.1 t_{\rm dyn}$ in
a simulation in order to discuss how orbital evolution
of stars  can be influenced by evolving $F_{\rm t}$.
The relative strength of the radial component is defined as follows:
\begin{equation}
F_{\rm r}= \frac{f_{\rm r}}{f},
\end{equation}
where $f_{\rm r}$ is the radial component of gravitational force.
This $F_{\rm r}$ does not change much (almost $\sim 1$) in
the early disk evolution, however, $F_{\rm t}$ can change dramatically
from initially very small values ($<0.01$) to $\approx 0.5$ in the final growth
phases of bars.
Time evolution of $F_{\rm t}$ could be also quite different
in spontaneous and tidal formation of stellar bars,
which could end up with different bar properties in the two bar formation
modes. In the present study, we use our original GPU-based direct Nbody simulation
code that does not use the tree method adopted in our other simulations
(e.g., Bekki et al. 2005). Therefore, we can more accurately estimate $F_{\rm t}$ for
all particles, which is an advantage of the present study.

\subsection{Quantifying the degrees of precession synchronization}
We consider that
the synchronization of apsidal precession among stars in disk galaxies is
one of physical processes that are  crucial for bar formation.
We try to quantify the degree of this synchronization by investigating
the phase space density of stars ($\Sigma(t_{\rm a}, \varphi_{\rm a}$)) in 
the $t_{\rm a}-\varphi_{\rm a}$ plane
($0 \le t_{\rm a} \le 10t_{\rm dyn}$).
We first count the number of stellar particles for each of the 
$50 \times 50$ grid points
in the $t_{\rm a}-\varphi_{\rm a}$ plane ($N(\varphi_{\rm a})$ and thereby
derive the phase space density  as follows:
\begin{equation}
\Sigma(t_{\rm a}, \varphi_{\rm a})
=\frac{ N(t_{\rm a}, \varphi_{\rm a}) }{ 0.04t_{\rm dyn}\pi }.
\end{equation}
We here use the grid sizes of $0.2 t_{\rm dyn}$ and $0.04 \pi$ for
$t_{\rm a}$ and  $\varphi_{\rm a}$, respectively,
mainly because the sizes are small enough to resolve the 
time evolution of characteristic patterns of
$\Sigma(t_{\rm a}, \varphi_{\rm a}$).

Although there could be a number of ways to quantity the degree
 of APS
($P_{\rm syn}$) in the simulated stellar disks, 
we adopt the following method. We first count the number
of stellar particles in each $\varphi$ grid point ($N(\varphi)$) at
each $t_{\rm a}$ grid  and estimate
the mean ($N_{\rm m}(\varphi$)) and maximum ($N_{\rm max}(\varphi)$).
The degree of APS is accordingly as follows:
\begin{equation}
P_{\rm syn}
=\frac{ N_{\rm max}(\varphi_{\rm a}) }
{ N_{\rm m}(\varphi_{\rm a}) }.
\end{equation}
Accordingly, stellar disk with a larger number of stars
with similar $\varphi_{\rm a}$
can have  a  higher $P_{\rm syn}$ value.
A barred galaxy has two distinct peaks in the 
$N_{\rm m}(\varphi_{\rm a})$ distribution and thus shows
a  high $P_{\rm syn}$ value.
It should be stressed that both (i)
 disks with bi-symmetric spiral arms yet without bars
and (ii)  prolate stellar systems without figure rotation
are unable to have high $P_{\rm syn}$, though they can show
high $A_2$. This means that we need to investigate $P_{\rm syn}$ 
(not just $A_2$) in order
to confirm bar formation through APS.

The time evolution rate of   $P_{\rm syn}$ in a disk galaxy
with a forming bar corresponds to
the bar growth rate. Accordingly, the time derivation of $P_{\rm syn}$
can be used to discuss what controls the bar growth rates
in disk galaxies. We here introduce the following ``$S$'' parameter,
which describes the power of a disk galaxy to grow its stellar bar through
physical processes (e.g., global instability or tidal interaction):
\begin{equation}
S_{\rm aps}=
\frac{ dP_{\rm syn}  }{ dt_0 }
=\frac{ P_{\rm syn}(t+dt_0) - P_{\rm syn}(t) }
{ dt_0 },
\end{equation}
where $dt_0$ is not the time step width of a simulation but the
time difference between two time steps at which $P_{\rm syn}$
are estimated. Since stellar bars grows slowly in most of the present models
(except for the tidal models),
we adopt a large $dt_0$ of $2t_{\rm dyn}$.
This $S$ parameter can depend on a number of disk properties such as 
disk mass fractions, Toomre's $Q$ parameters, and strengths of tidal perturbation.
We consider that bar growth rates can also  depend on the strengths of bars
(quantified by $A_2$) as follows:
\begin{equation}
S_{\rm aps}=F_{\rm aps}(A_2),
\end{equation}
where we derive the functional form of $F_{\rm aps}$ from the results
of the present simulations.
We cross-correlate between $A_2$ and $S_{\rm aps}$ in a simulated barred galaxy
in order to discuss whether and how the bar growth rate is determined by
the bar strength itself.
This investigation is still useful in understanding bar formation,
though many previous works already investigated the evolution of m=2 amplitudes
in disk galaxies using simulations
with different spatial resolutions (e.g., Dubinski et al. 2009).

\subsection{Parameter study for understanding bar formation}

Although we investigate the physical processes of bar formation  in 35 models
with different model parameters, we mainly describe the results of 
the fiducial model (IA1). The models are divided into the two main categories,
``I'' and ``T'' depending on whether they are for isolated or tidally
interacting models, and each of the two are divided into sub-categories,
``A'', ``B'', and ``C'' depending of the details of dark matter properties.
The models IA (TA) have $R_{\rm vir}/R_{\rm d}=14$ and $c=10$, which
are regarded as the standard model for dark matter halos of Milky Way
type disk galaxies  in the present study.
The models IB (TB) have $R_{\rm vir}/R_{\rm d}=19.8$ and $c=10$, which means
that the dark matter halos are more diffuse compared to
other models (i.e., $f_{\rm d}$ is 
higher for a given mass of dark matter). 
More compact dark matter models (IC and TC) with $R_{\rm vir}/R_{\rm d}=14$ 
and $c=16$ are investigated to discuss bar formation in
low-mass disk galaxies.
Three key parameters in the present study are $f_{\rm d}$, $f_{\rm b}$,
and $Q$ for the above 6 types of models, and the values are given
for each model in Table 2.

We mainly investigate the distributions of stellar particles in the two
phase spaces,   $(t_{\rm a},\varphi_{\rm a})$ and 
 $(\Omega_{\rm pre},e)$,  because bar formation can be clearly
manifested in the strong concentration of the particles on
the phase spaces. Also, stellar evolution in the two phase spaces
enables us to understand how strengthened tangential components of
gravitational force can cause significant changes of these four variables.
Fig. 1  illustrates a major difference in the distributions of stellar particles
on the $t_{\rm a}-\varphi_{\rm a}$ plane  between disks with and without bars.
The significantly smaller number of grid points occupied by stellar particles
in the bar model in this figure might get readers  to think of analogy
between bar formation and Fermi Degeneracy.
Fig. 2 illustrates (A) the required synchronization of $\varphi_{\rm a}$ 
and $\Omega_{\rm pre}$ in bar formation due to APS and (B) the
strengthened $F_{\rm t}$ as the major physical
mechanism of APS.

\begin{figure*}
\psfig{file=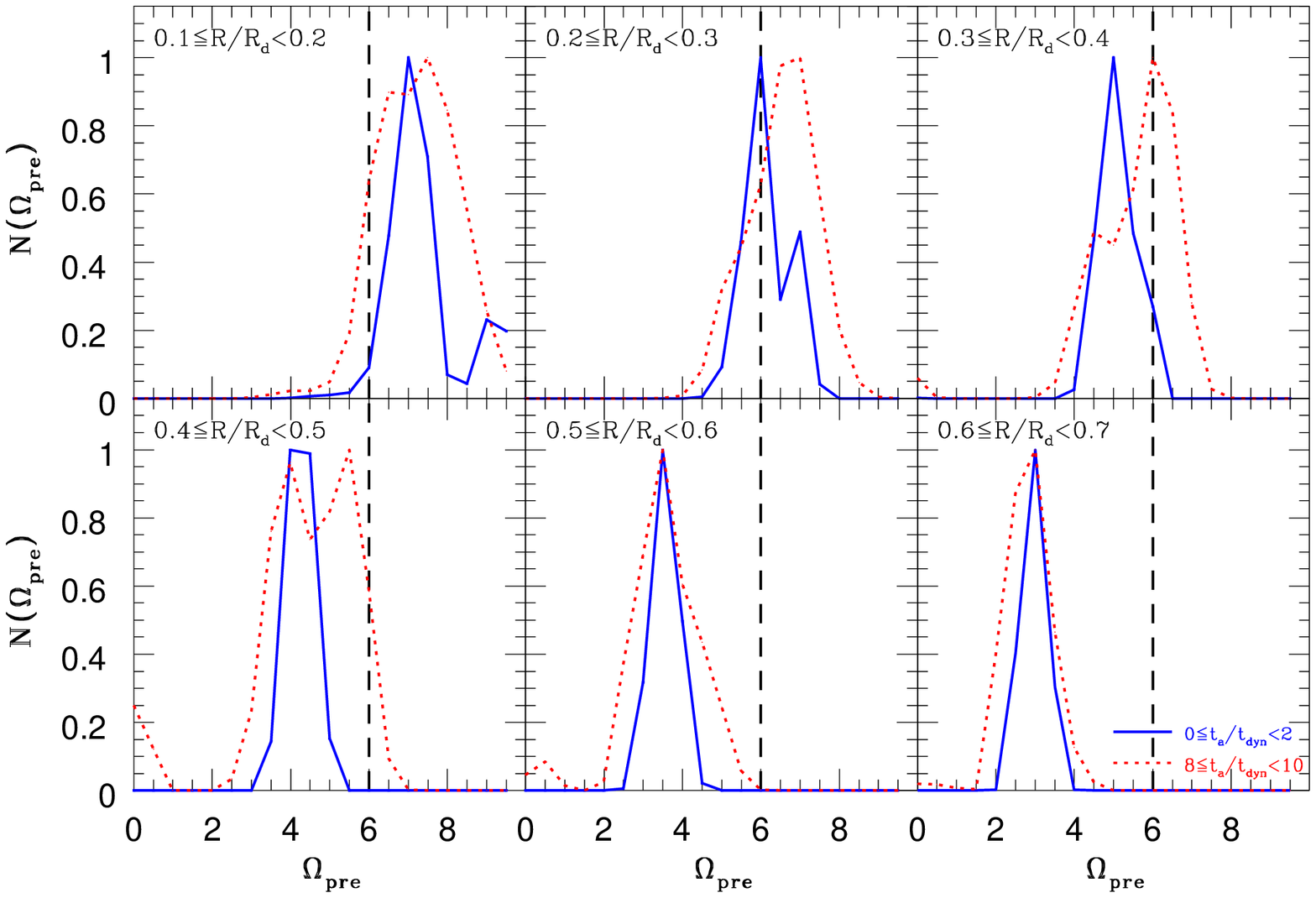,width=18cm}
\caption{
Number distributions of precession rates $N(\Omega_{\rm pre})$ of stellar particles
at different radial ranges
for initial ($0 \le t_{\rm a}/t_{\rm dyn}<2$; blue solid),
and final ($8 \le t_{\rm a}/t_{\rm dyn}<10$; red dotted) time periods
in the fiducial model. $N(\Omega_{\rm pre})$ is normalized by its maximum value
for each of the two time periods. The adopted radial ranges are shown
in the upper left corner of each panel and the thick dashed line represents
the pattern speed of the bar estimated at $R=0.2R_{\rm d}$ at $T=4t_{\rm dyn}$
in this model.
Clearly, $\Omega_{\rm pre}$  moves toward the bar pattern speed
for a significant fraction of stars at $0.3 \le R/R_{\rm d} < 0.5$ during bar formation, 
which is evidence for synchronization of $\Omega_{\rm pre}$.
}
\label{Figure. 8}
\end{figure*}

\begin{figure*}
\psfig{file=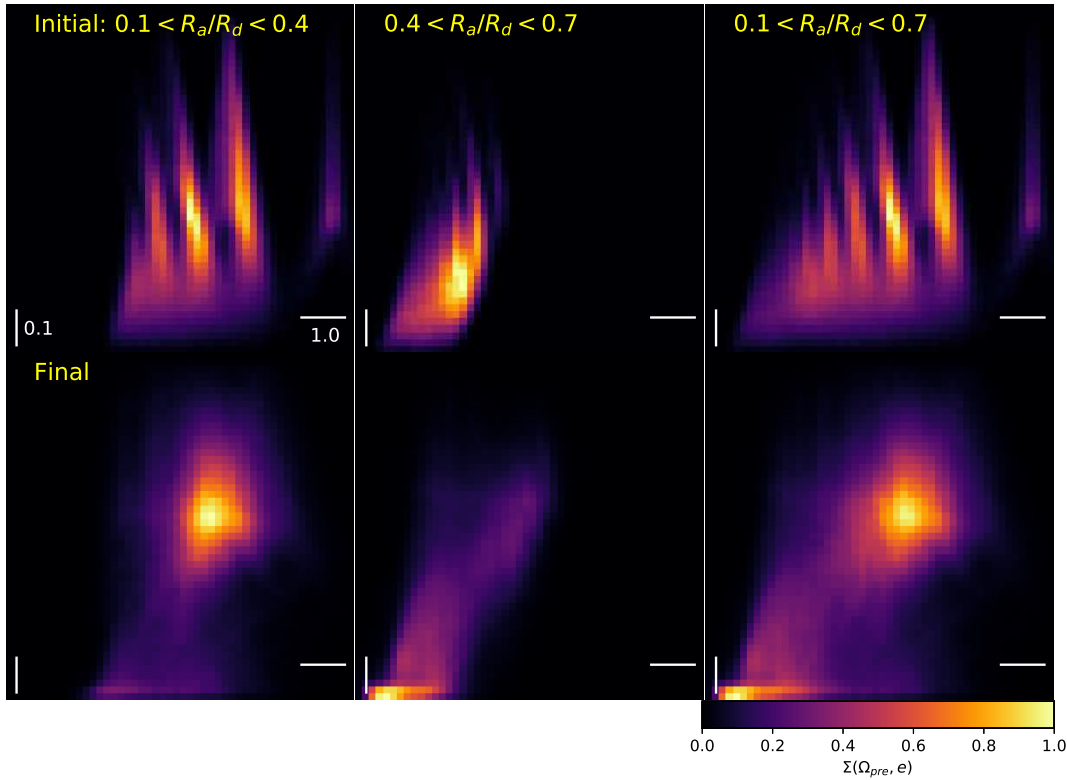,width=18cm}
\caption{
Surface number densities of
stellar particles ($\Sigma(\Omega_{\rm pre}, e)$)
on the $\Omega_{\rm pre}-e$ plane
at different radial bins
in the fiducial model. For clarity,  $e$ ranges from 0 (bottom) to 1 (top)
whereas $\Omega_{\rm pre}$
ranges from 2 (left) to 10 (right) in these 2D maps. The vertical and
horizontal scale bars measure 0.1 in $e$ and 1.0 in $\Omega_{\rm pre}$, respectively.
The upper and lower panels show the initial and final
distributions, respectively, and the left, middle, and right panels shows
the results for particles with $0.1<R_{\rm a}/R_{\rm d}<0.4$,
$0.4<R_{\rm a}/R_{\rm d}<0.7$,
and $0.1<R_{\rm a}/R_{\rm d}<0.7$, respectively.
The surface number density at each grid point ($\Sigma(\Omega_{\rm pre},e)$)
is normalized in each frame so that the detailed distribution in each model can be
more clearly
seen.
}
\label{Figure. 9}
\end{figure*}

\begin{figure}
\psfig{file=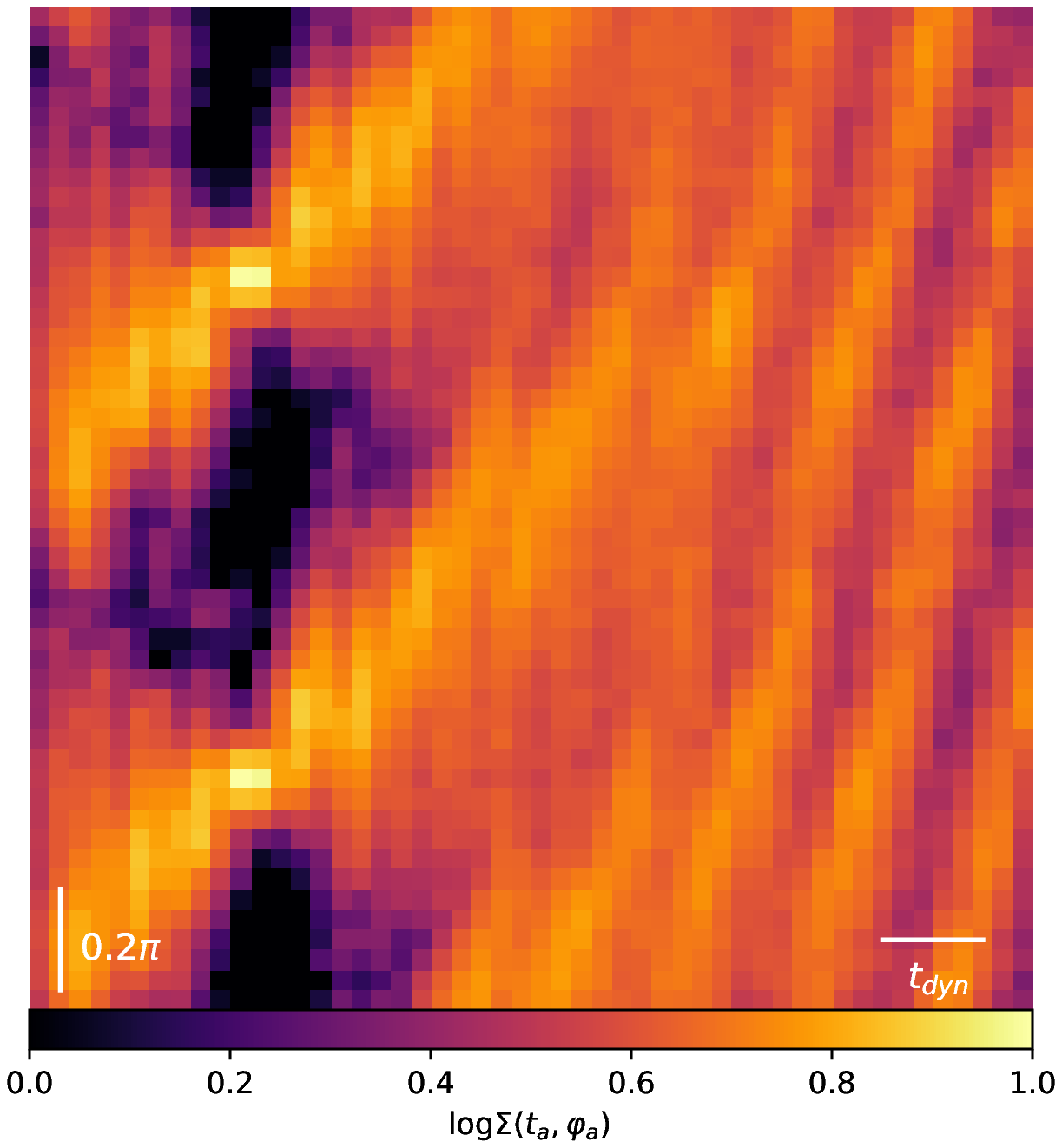,width=8.5cm}
\caption{
The same as Fig. 5 but for the seed bar particles in the fiducial model.
These particles have $t_{\rm a}/t_{\rm dyn}=2.2$
at $0.2<R_{\rm a}/R_{\rm d}<0.3$.
The details of the method to select the particles are given in the main
text.
}
\label{Figure. 10}
\end{figure}

\section{Results}

\subsection{Fiducial model}

\subsubsection{Dynamics of bar formation and growth}

Figs. 3 and 4  show the time evolution of the  two-dimensional (2D) maps of
stellar mass distributions
and  $F_{\rm t}$ 
in the fiducial model with $f_{\rm b}=0$ 
(i.e., bulgeless) and $f_{\rm d}=0.35$. 
The face-on views of the stellar disk
clearly indicate that the disk first develops an oval-like structure
in the inner region ($T=5.7t_{\rm dyn}$) and then
a longer and stronger bar finally ($T=10t_{\rm dyn}$).
In this particular bulgeless model,  strong spiral arms cannot be seen
in the early dynamical evolution of the disk 
($4.4t_{\rm dyn} \le T\le 5.7t_{\rm dyn}$),
which implies that dynamical action of spiral arms  on disk field stars
is irresponsible for the early growth of the oval-shaped 
structure (and thus for the transformation from the oval to the bar).

Fig. 4 demonstrates that the $|F_{\rm t}|$ is rather small ($<0.1$)
 in the early
formation phases of the bar ($T<5.7t_{\rm dyn}$) owing to the weak tangential
force of the disk with a seed bar.
As the stellar bar grows, both the strength of $F_{\rm t}$
and the 2D map of $F_{\rm t}$ steadily evolves. A unique shape 
looking like a four-leave clover (or an isocontour of an electric quadrapole field)
can be clearly seen in the 2D map of $F_{\rm t}$
at $T=10t_{\rm dyn}$ when the bar is fully developed.
Stellar particles with $\varphi$ (azimuthal angle, $-\pi/2<\varphi<\pi/2$)
larger than $\varphi_{\rm bar}$ ($\varphi$ for the major axis of the bar)
can have negative $F_{\rm t}$ whereas those with 
$\varphi$ 
smaller than $\varphi_{\rm bar}$ (below the bar in the right half of this  figure)
can have positive $F_{\rm t}$.
This results means that  stellar particles leading (trailing) the bar can experience
negative (positive) $F_{\rm t}$: this is quite important in bar growth,
as discussed later.

\begin{figure*}
\psfig{file=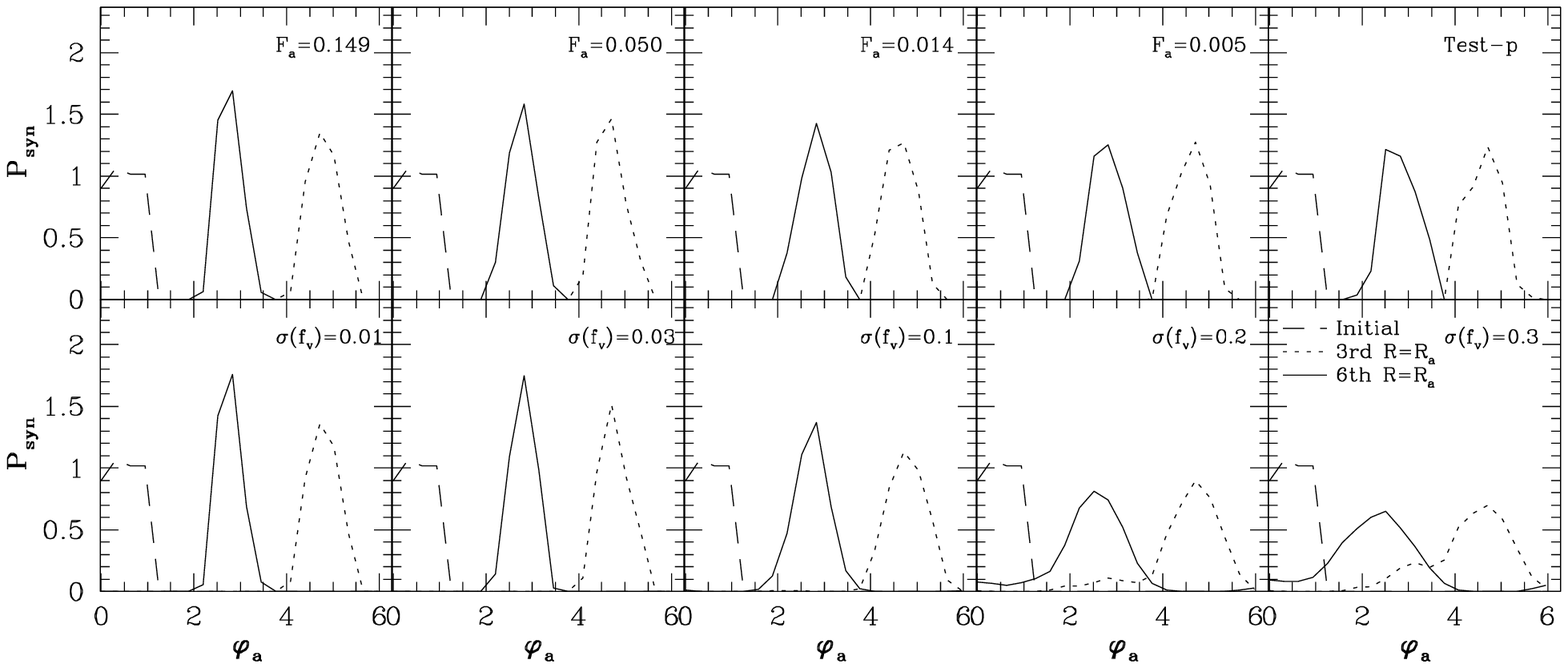,width=18.0cm}
\caption{
Distributions of $P_{\rm syn}$ at initial (dashed), 3rd (dotted)
and 6th (solid) apocenter passages of stars
in the adopted local region with $0.19R_{\rm d} \le R \le 0.21 R_{\rm d}$
and $0 \le \varphi \le 0.33 \pi$ (radian)
for the ten comparative models.
These $P_{\rm syn}$ distributions are shown for the models
with the same $\sigma(f_{\rm v})$ (=0) yet different $F_{\rm a}$,
i.e.,  $F_{\rm a}=0$
0.005, 0.014, 0.050,
and 0.149
in the upper panels,
and for those
with the same $F_{\rm a}$ (=0.149) yet different $\sigma(f_{\rm v})$,
i.e.,  $\sigma(f_{\rm v})=0.01$
0.03, 0.1 0.2,
and 0.3
in the lower panels.
$P_{\rm syn}$ is the ratio of $N(\varphi_{\rm a})$ to its average
and thus describes how much $\varphi_{\rm a}$ of stellar particles  are synchronized
(or aligned) at each $\varphi_{\rm a}$.
$F_{\rm a}$ is the number fraction of the particles with identical $t_{\rm a}$
in the adopted local area: $F_{\rm a}$ is not just a mass fraction of disk
particles.
Therefore, stronger gravitational interaction between the stars with
very similar $t_{\rm a}$ yet different $\varphi_{\rm a}$ is expect to occur
in the models with higher $F_{\rm a}$.
In the model with $F_{\rm a}=0$, the particles are assumed to be
mass-less (i.e, test particles, shown as ``Test-p'') so that gravitational interaction between
the particles cannot occur.
Thus a comparison between these models with different $F_{\rm a}$
enables us to understand the importance of mutual gravitational interaction
of stars in local areas (of a disk galaxy)
in the synchronization of precession (APS). More details on the dependence
of local APS on $F_{\rm a}$ and $\sigma(f_{\rm v})$ are given in the main text.
}
\label{Figure. 11}
\end{figure*}

\begin{figure}
\psfig{file=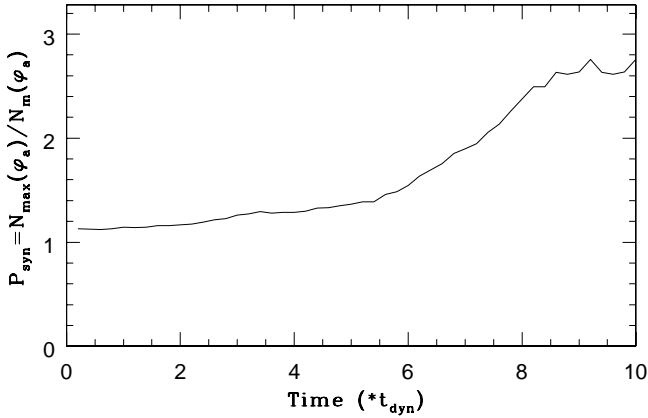,width=8.5cm}
\caption{
Time evolution of $P_{\rm syn}$ at $0.2R_{\rm d}\le R \le 0.3R_{\rm d}$
in the fiducial model.
Here $P_{\rm syn}$ evolution smoothed out over $0.5t_{\rm dyn}$ is plotted,
because the original $P_{\rm syn}$ shows a short-term fluctuation.
}
\label{Figure. 12}
\end{figure}

\begin{figure*}
\psfig{file=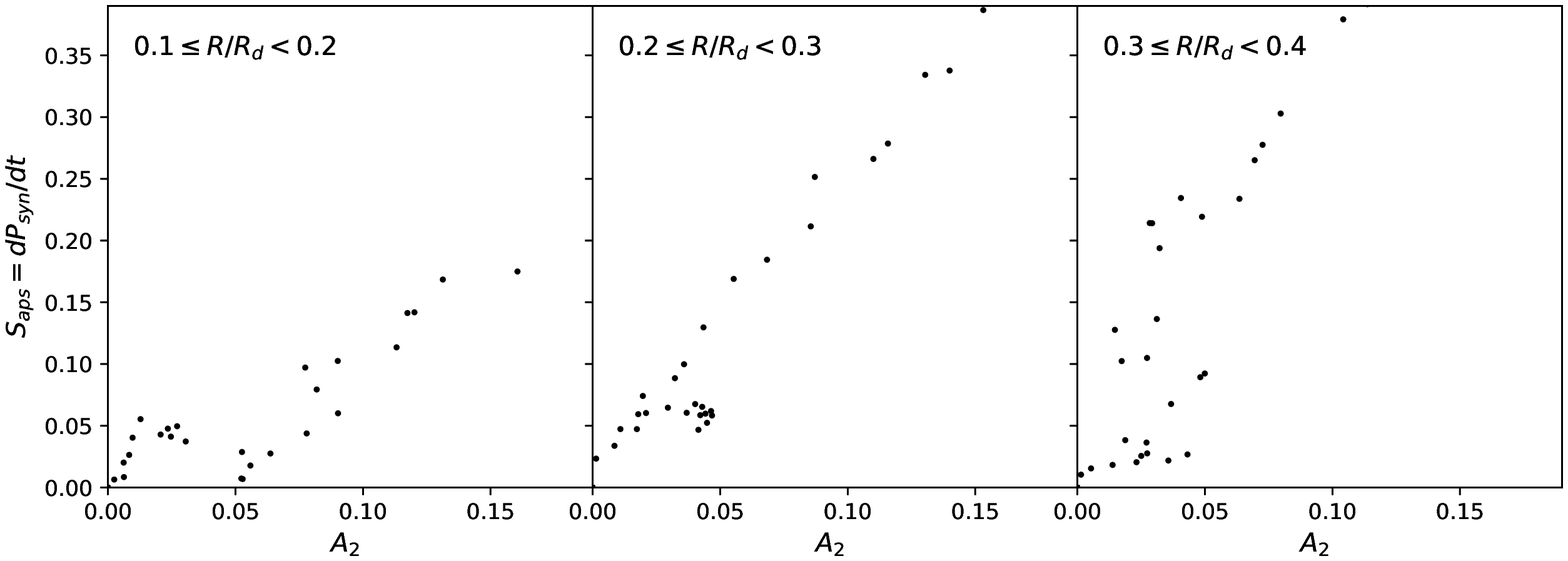,width=18.0cm}
\caption{
A correlation
between the ``$S$'' parameters
($S_{\rm aps}=dP_{\rm syn}/dt$) and $A_2$ (bar strength)
during bar growth for three radial bins,
(i) $0.1 < R/R_{\rm d} < 0.2$ (left), (ii) $0.2 <R/R_{\rm d} < 0.3$ (middle),
and (iii) $0.3<R/R_{\rm d}<0.4$ (right) in the fiducial model.
A positive correlation can be clearly seen for $A_2>0.05$ in this model.
}
\label{Figure. 13}
\end{figure*}

Fig. 5 demonstrates how $\varphi_{\rm a}$ of disk stars 
with $0.1 \le  R_{\rm a}/R_{\rm d} \le 0.7$ can be synchronized 
(i.e., APS during bar formation)  in the fiducial model.
Below, we first describe the APS in $0.2 \le R_{\rm a}/R_{\rm d} <0.3$,
because the transition from unbarred to barred structures on this $t_{\rm a}-\varphi_{\rm a}$ plane
can be more clearly seen for such a radial range.
Initially different stars have different $\varphi_{\rm a}$
and $\Omega_{\rm pre}$ so that
the phase-space distribution $\Sigma(t_{\rm a},\varphi_{\rm s})$
cannot show a striped pattern ($t_{\rm a}/t_{\rm dyn}<2$).
As APS proceeds in the disk, a striped pattern with a higher 
number density of stellar particles
appears in the phase space ($2<t_{\rm dyn}/t_{\rm a}<3$), 
though the pattern
is not so clearly seen.
The multiple stripes in this figure are all steeply inclined, and 
each stripe appears to consist of  $3-4$ small rectangular-shaped regions with
higher densities
in the earlier evolution ($3<t_{\rm a}/t_{\rm dyn}<6$).
Two density peaks in this $\Sigma(t_{\rm a},\varphi_{\rm s})$ map
at a given $t_{\rm a}$ can be found  in the later dynamical evolution,
because the two stripes are separated by $\approx \pi$ in $\varphi_{\rm a}$
and sharply inclined. The two peaks correspond to the two edges of the
growing bar and  
the inclination angle of the striped pattern with respect 
to the $t_{\rm a}$ axis
corresponds to the pattern speed of the stellar bar.
A larger number of stars can have identical $\varphi_{\rm a}$ due to APS
in the disk so that the stripped pattern can become quite remarkable in
the later phase of the bar formation ($T>8t_{\rm dyn}$).
Thus Fig. 4 demonstrates that APS during bar formation can be clearly seen  
in the stellar distribution in the  $t_{\rm a}$-$\varphi$ phase space.

However, such clear APS in the $\varphi_{\rm a}-t_{\rm a}$ map for  $0.2 \le R_{\rm a}/R_{\rm d}<0.3$
cannot be clearly seen in $0.6 \le R_{\rm a}/R_{\rm d}$, which suggests that APS cannot proceed
efficiently in the outer part of the disk.
The distinct two density peaks 
for $0.4 \le R_{\rm a}/R_{\rm d}<0.6$ are weaker compared with those
for $0.2 \le R_{\rm a}/R_{\rm d}<0.4$
in the later dynamical evolution. This result implies that APS induced by bars can proceed
more efficiently in the inner region of the disk.

In the present study, if a stellar disk has two distinct peaks in
the $N(\varphi_{\rm a})$ distribution (normalized by its mean,
$N_{\rm m}(\varphi_{\rm a}$)) at a given $R$, it can be regarded
as having a bar at $R$. 
Accordingly, we investigate when a stellar disk  can 
have such two peaks 
in a simulation  in order to identify the formation epoch of the seed bar.
Fig. 6 shows that (i)  the $N(\varphi_{\rm a})$ distribution at 
$0.2R_{\rm d}\le R \le 0.3R_{\rm d}$ is initially
very flat with no distinct peak
and (ii) it has two clear  peaks at $T=4.4t_{\rm dyn}$.
The two peaks become higher and narrower  as the bar grows due to  APS
in the later epoch of bar formation ($5t_{\rm dyn} \le T \le 10t_{\rm dyn}$).
The $N(\varphi_{\rm a})$ distribution is not so symmetric after the formation
of the  two peaks: this asymmetry can be seen in other radii of the disk.
Such broken symmetry is a signature of a young bar that has just been formed 
and is still growing due to APS.

\begin{figure*}
\psfig{file=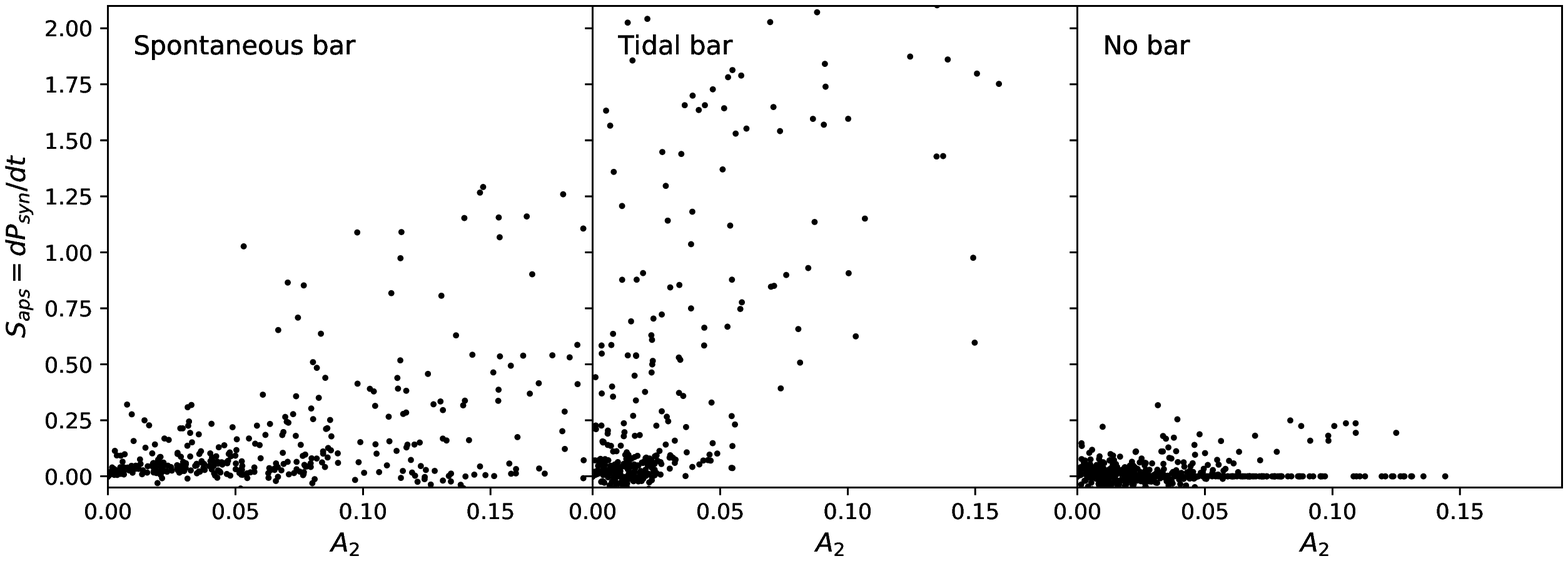,width=18cm}
\caption{
The same as Fig. 13 but for the models with spontaneous bar formation (left),
tidal bar formation (middle), and no bar formation (right).
120 data points are plotted for each of
the three categories for a fair comparison.
The date is from the models with strong spontaneous bars
(IA1, IA9, IA11, and IA15), those with strong tidal bars
(TA1, TA2, TA4, and TA5), and those without bars (IA3, IA5,
IA14, and IA16). The models without bars include lower $f_{\rm d}$ ($<0.2$)
and higher $Q$ (=2), which means that APS can be strongly suppressed in
these models.
The tidal bar model shows higher $S$ parameters even for lower  $A_2$,
which is different from the spontaneous bar model.
}
\label{Figure. 14}
\end{figure*}

Fig. 7 indicates that the epoch of seed bar formation characterized by
two distinct peaks in the normalized
 $N(\varphi_{\rm a})$ distribution is around $T=2.2t_{\rm dyn}$
for $R=[0.2-0.3]R_{\rm d}$ in this model.
The distribution has just one strong and wide peak at $\varphi_{\rm a} \approx 1$ (radian)
at $T=2t_{\rm dyn}$. Although it has multiple peaks at $T=2.2t_{\rm dyn}$,
the highest and the 2nd highest peaks are separated by $\approx 3$ radian, which is
an indication of seed bar formation. The distribution shows two more distinct peaks
at $T=2.4t_{\rm dyn}$ with the peak positions moving right from those at $T=2.4t_{\rm dyn}$
and $3.6t_{\rm dyn}$.
Since these two peaks can be clearly seen in the distribution at $T>2.4t_{\rm dyn}$,
the epoch of seed bar formation can be identified as  $T=2.2t_{\rm dyn}$ in this model.

It should be stressed here that 
$N(\varphi_{\rm a})/N_{\rm m}(\varphi_{\rm a})$  during the seed bar formation
($T=2.2t_{\rm dyn}$)  is only 1.1-1.2,
which is by a factor of $\sim 3$ lower than that in the fully developed bar 
($T=10t_{\rm dyn}$). Therefore, such a structure with two peaks
in  $N(\varphi_{\rm a})$  cannot be clearly  seen as a bar 
in the projected
distribution of stellar particles at $T\sim 2t_{\rm dyn}$: it might not be so reasonable
to call it  ``bar''. Nevertheless, the seed bar can steadily grow stronger
due to APS.
It would be interesting that such  hidden seed bars are  growing in disk galaxies with
apparently no clear barred structures ($A_2<0.05$): it would be our future work
to find a way to detect
such pre-bar phases in observational images of galaxies.

Fig. 8 describes how the number distribution of $\Omega_{\rm pre}$ 
of stellar particles evolves with time at different radii.
$\Omega_{\rm pre}$  needs to become smaller (larger)
to synchronize with $\Omega_{\rm bar}$, 
if $\Omega_{\rm pre}$ is larger (smaller) than  $\Omega_{\rm bar}$.
Since initial $\Omega_{\rm pre}$ is 
systematically larger in the inner region of the disk
for the adopted mass and rotation profiles,
this means that the inner (outer) stellar particles need to decrease
(increase) their $\Omega_{\rm pre}$ through some physical processes
in order to synchronize their $\Omega_{\rm pre}$ with $\Omega_{\rm bar}$.
Fig. 8 indeed confirms that such $\Omega_{\rm pre}$ changes are possible
for some particles at some radii.
For example,
the normalized $N(\Omega_{\rm pre}$) at $\Omega_{\rm  pre}=\Omega_{\rm bar}$
is initially very small ($\approx 0.1$)
at $0.1 \le R/R_{\rm d}<0.2$, it can become as large as 0.6 finally. 
Also the normalized $N(\Omega_{\rm pre}$) 
at $\Omega_{\rm  pre}=\Omega_{\rm bar}$
increases from 0.3 to 1 (i.e., the maximum peak) within $10t_{\rm dyn}$
at $0.3 \le R/R_{\rm d}<0.4$. This significant increase can be seen 
at $0.4 \le R/R_{\rm d}<0.5$ but cannot  be seen
at $0.5 \le R/R_{\rm d}$.
Synchronization of $\Omega_{\rm pre}$  with $\Omega_{\rm bar}$ (i.e., APS)
required for bar growth
can indeed occur during the dynamical evolution of the disk,
though it is limited to a certain radial range.
A physical reason for no/little synchronization for $R>0.5R_{\rm d}$ 
is that the initial $\Omega_{\rm pre}$ is too small to be aligned with
$\Omega_{\rm bar}$ 
even through the stronger tangential force of the growing bar is 
acting on the stars.

Synchronization of $\Omega_{\rm pre}$ can be clearly seen in Fig. 9 too,
which describes  the distribution of stellar particles on
the $\Omega_{\rm pre}-e$ plane for 6 radial bins 
($0.1<R_{\rm a}/R_{\rm d}<0.7$).
The initial distribution with four vertical stream-like features 
at the inner regions 
($0.1<R_{\rm a}/R_{\rm d}<0.4$) can be transformed into
a triangle-like distribution with a single strong yet broad peak
during bar formation.
Furthermore, the mean $e$ can become significantly larger during
bar formation, which means that  the orbits of stellar particles
are significantly elongated during this $\Omega_{\rm pre}$  synchronization.
These results suggest 
that $\Omega_{\rm pre}$ synchronization is closely associated
with orbital elongation.
Intriguingly, the particles with $0.4<R/R_{\rm d}<0.7$ shows lower final
$e$ (less elongated orbits), which suggests that the orbits are circularized
during bar formation: these outer particles can gain angular  momentum during
bar formation.
The final two peaks in this diagram for $0.1<R/R_{\rm d}<0.7$ is characteristic
for spontaneous bars, which reflects the differences of bars' dynamical
influences on disk particles  between inner and outer disk regions.

\subsubsection{Seed bar formation due to local APS}

Although Figs. 6 and 7 have shown that a seed bar can be formed
at $T\le 2.2t_{\rm dyn}$,
it is yet to be clarified how APS is possible in such an early dynamical
phase of the stellar disk: see also Fig. B1,
which indicates that the bar growth rate is quite different between
$T<4t_{\rm dyn}$ ($A_2<0.1$)  and $T>4t_{\rm dyn}$ ($A_2>0.1$):
As shown in Fig. 4,
the tangential force ($F_{\rm t}$) due to a bar-like structure is rather weak
in the early phase of bar formation ($T<4t_{\rm dyn}$).
These results suggest that the bar formation and growth processes can be divided into 
the following two stages: one is ``from noise to seed'' stage ($T<4t_{\rm dyn}$) 
and the other is ``from seed to
fully grown'' (or ``seed to matured'') stage ($T>4t_{\rm dyn}$).
In the present study, we discuss the bar formation/growth  processes
separately for the two stages.

In order to reveal the formation mechanism of
the seed bar, 
we select the stellar articles that constitute the seed bar at $T=2.2t_{\rm dyn}$
(referred to as ``seed bar particles'' for simplicity),
and thereby investigate
the orbital evolution of the particles.
This investigation can reveal the initial locations of the stars and how
the seed bar grows from the initial density enhancement(s) of the stellar disk.
We select the seed bar particles,
if $\varphi_{\rm a}$ of stellar particles is either 
$|\varphi_{\rm a}-\varphi_{\rm a, p1}| \le 0.126$ (radian) or
$|\varphi_{\rm a}-\varphi_{\rm a, p2}| \le 0.126$,
where $\varphi_{\rm a, p1}$ is the peak $\varphi_{\rm a}$ in
the $N(\varphi_{\rm a})$ distribution (see Fig. 7) and
$\varphi_{\rm a, p2}$ is the second peak 
($\varphi_{\rm a, p2}=\varphi_{\rm a, p1}+\pi$ if $\varphi_{\rm a}\le \pi$).
Therefore, these particles are within the seed bar at $T=2.2t_{\rm dyn}$.
It should be noted here that we can select other seed bar particles 
at different times (e.g., $T=2.4$, 2.6, $2.8t_{\rm dyn}$ etc).
We here describe the results for the seed bar particles selected at
$0.2R_{\rm d} \le R \le 0.3R_{\rm d}$  (where
the bar has the highest $A_2$). The time evolution of the spatial
distribution of the particles projected onto the $x$-$y$ plane is 
described in Appendix A.

Fig. 10 shows that the seed bar particles initially 
have wide ranges of initial $t_{\rm a}$
and $\varphi_{\rm a}$, though there are two
inclined stripe-like shapes with higher phase space densities that finally
form two peaks at $T=2.2t_{\rm dyn}$.
This synchronization (or narrowing) of $\varphi_{\rm a}$ should be caused by some local
physical processes, because a global bar does not exist at $T<2.2 t_{\rm dyn}$.
After the formation of the seed bar, the particles continue to show two peaks
at a given time, though the peak becomes weaker possibly due to phase mixing
within the disk. The particles finally show the same pattern as seen in
$\Sigma (t_{\rm a},\varphi_{\rm a})$ for all particles at $0.2\le R/R_{\rm d} \le 0.3$
(see Fig. 7).

It is important here to point out that a significant fraction
of the seed bar particles
at $T=2.2t_{\rm dyn}$
originate from the highest density peak of
$\Sigma (t_{\rm a},\varphi_{\rm a})$  at 
$t_{\rm a} \approx 0.5t_{\rm dyn}$ and $\varphi_{\rm a} \approx 0.52$ (radian).
This means that
such the highest phase space density can grow fastest to form (the part) of the 
seed bar.
For a given $R$,  there is almost no difference
in the surface number densities of stellar particles at different
$\varphi$ ($\delta \Sigma(R, \varphi) \approx 0$).
However, there is a very large difference (a factor of more than 10 
for $50 \times 50$ $(t_{\rm a},\varphi_{\rm a})$ grid points) in 
the initial $\Sigma (t_{\rm a},\varphi_{\rm a})$ at different $t_{\rm a}$
and $\varphi_{\rm a}$. This difference can cause a difference in
the dynamical evolution of stellar populations with different $t_{\rm a}$
and $\varphi_{\rm a}$.
A key question here is as to how the synchronization of $\varphi_{\rm a}$
leading to the seed bar formation is possible in the local regions
with initially high $\Sigma (t_{\rm a},\varphi_{\rm a})$.

Here, we hypothesize that
mutual gravitational interaction between particles
with very similar $t_{\rm a}$ at near $R_{\rm a}$ is responsible for the 
synchronization of $\varphi_{\rm a}$ (referred to as ``local APS'' from now on),
because stronger gravitational interaction between these particles 
lasts longer at $R_{\rm a}$:
in particular, the tangential component of the force ($F_{\rm t}$) becomes stronger.
As discussed later in \S 4,
$\varphi_{\rm a}$ and $\Omega_{\rm pre}$
can be much more significantly influenced  by
$F_{\rm t}$.
Therefore,  
strong mutual gravitational  interaction between paricles 
at $R_{\rm a}$ (at $t_{\rm a}$) can possibly cause
APS between the local particles.

In order to investigate the above hypothesis,
we run comparative models
in which stellar particles in a local region
move under the fixed gravitational potential that is
exactly the same as the initial potential of the fiducial model.
The key parameter in these comparative experiments is $F_{\rm a}$
(defined in the model section), which controls
the initial fraction of stars that have a same $t_{\rm a}$ yet different
$\varphi_{\rm a}$ at a given $R$: these particules do not show strong 
orbital alignment initially.
Although there are many particles with different $t_{\rm a}$ and $\varphi_{\rm a}$ in
a local small area of a stellar disk,  the number of stars with almost identical
$t_{\rm a}$ in the area should be  small.
By changing $F_{\rm a}$, we can possibly reveal the threshold $F_{\rm a}$ above which
stars with initially different $\varphi{\rm a}$
can finally have similar $\varphi_{\rm a}$ due to local APS (caused by mutual gravitational
interaction between local particles).

As shown in Fig. 11,  the model with $F_{\rm a}=0.149$ has a narrow peak
(i.e., higher $P_{\rm syn}$) in the $N(\varphi_{\rm a})$ distribution,
whereas the model with $F_{\rm a}=0$ that does not include
gravitational interaction between particles 
does not show such a narrow peak.
These results accordingly reveal  that mutual gravitational interaction between
particles in a local small area is a key for the alignment of 
$\varphi_{\rm a}$ (i.e., APS). 
The comparative model with $F_{\rm a}=0.005$ does not show a narrower
$N(\varphi_{\rm a})$ distribution (very similar to the test particle model)
whereas the model with $F_{\rm a}=0.014$ shows the narrowing more clearly.
These results 
imply that $F_{\rm a}$ should be 
as large as $\approx 0.014$ for APS to be effective.
Clearly, $P_{\rm syn}$ is higher for larger $F_{\rm a}$, though
$P_{\rm syn}$ can become  at most 1.8 within $4t_{\rm dyn}$.

The models with smaller  $\sigma (f_{\rm v})$ show narrower
final $\varphi_{\rm a}$ distributions, and
$\sigma (f_{\rm v})$ should be as small as $\approx 0.1$ so that
$\varphi_{\rm a}$ synchronization can occur for $F_{\rm a}=0.149$.
This implies that APS works only between particles with similar
$t_{\rm a}$ and  similar $e$ that is determined by $f_{\rm v}$
at a given $R_{\rm a}$.
This also implies that APS cannot proceed so efficiently
in stellar disks with higher $Q$ values in which
a wider range of orbits is possible (owing to more random motions of stars).
Thus, these comparative experiments demonstrate that  APS can proceed
in a local area due to mutual gravitational interaction between the particles
in the area, though $\sigma (f_{\rm v})$ needs to be smaller. 

\begin{figure}
\psfig{file=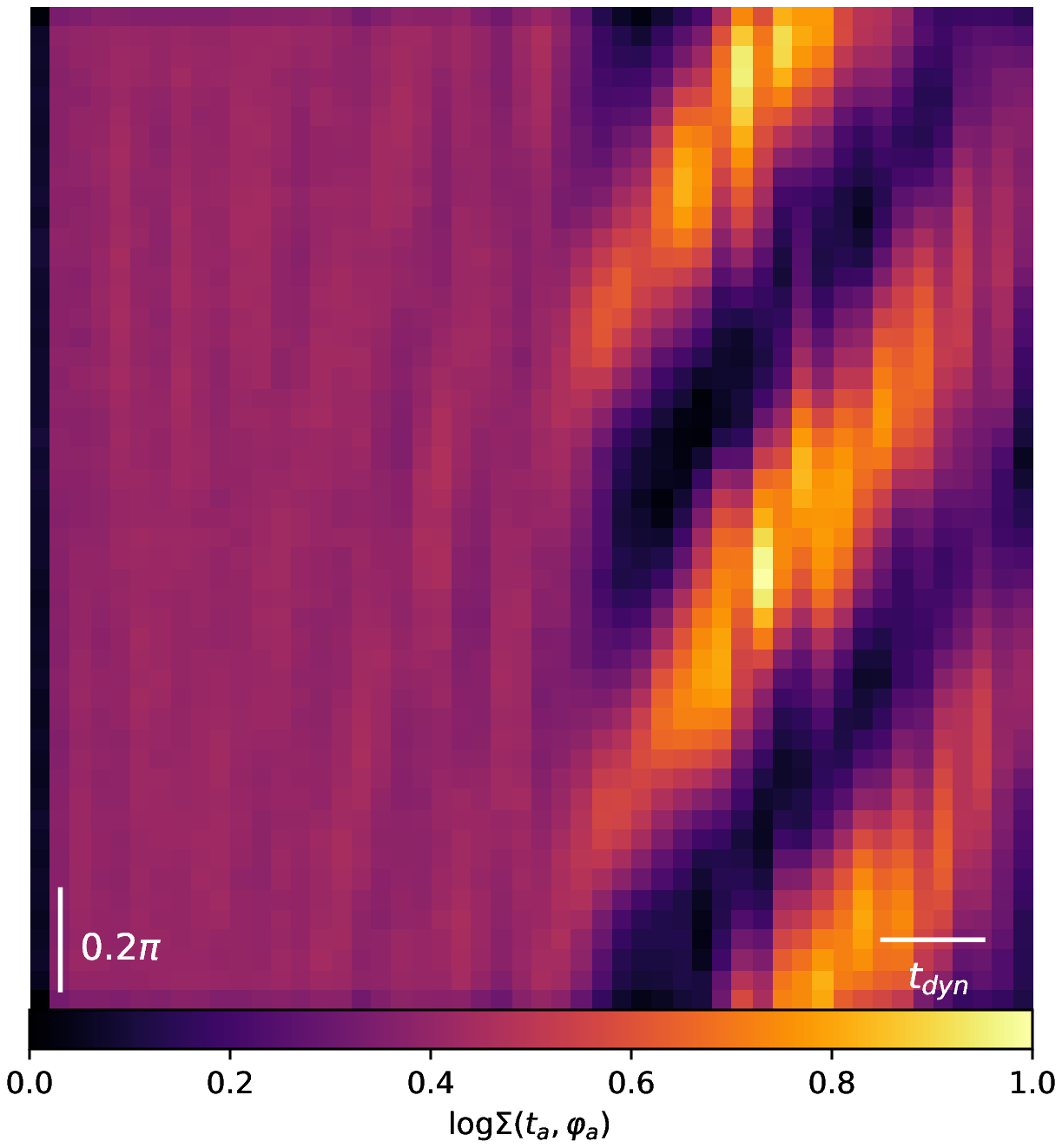,width=8.5cm}
\caption{
The same as Fig. 5 but for the tidal interaction model TA1.
}
\label{Figure. 15}
\end{figure}

\begin{figure*}
\psfig{file=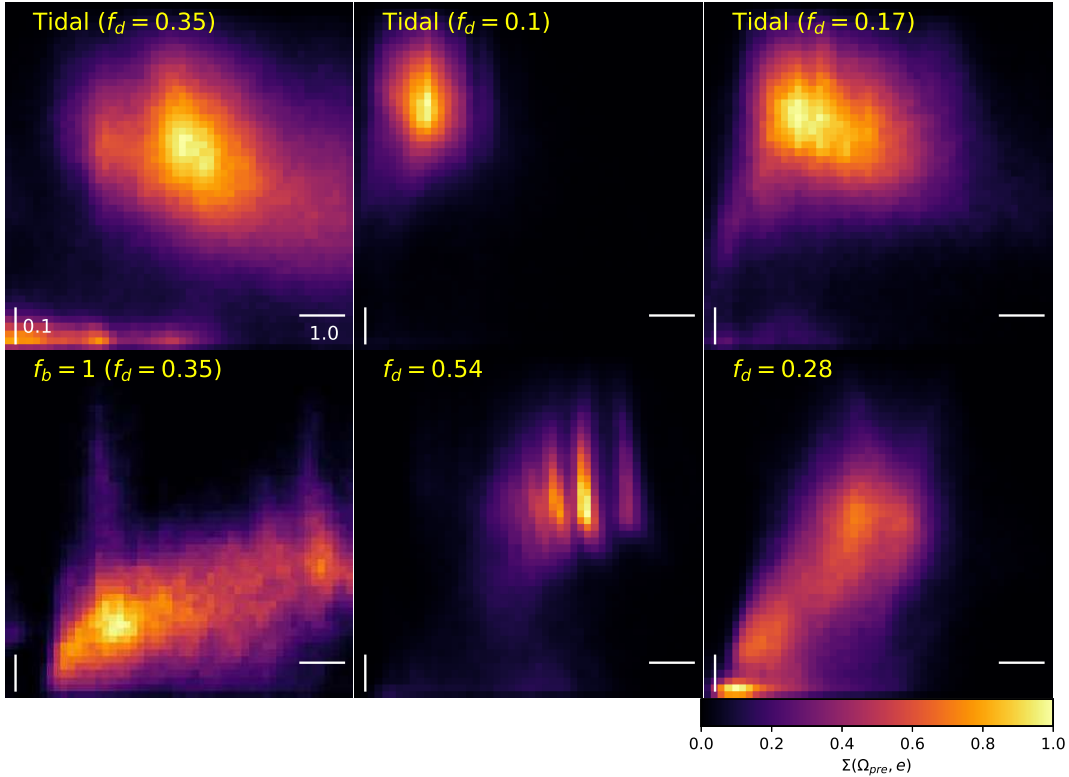,width=18cm}
\caption{
The same as Fig. 11 but for three tidal bars (upper, TA1, TA2, and TA5)
and three spontaneous
ones (lower; IB3, IA8, and  IA7). Strong distinct peaks at higher $e$
in lower disk mass fractions ($f_{\rm d}=0.1$ and 0.17) suggests that tidal bar
formation in these disks are associated with elongation of orbits in a significant
number of stars.
A bar can be formed
in the model with $f_{\rm b}=1$ (left in the lower panel)
only after $\approx 30t_{\rm dyn}$ of the disk
evolution. The bar in this massive bulge model has a high $\Omega_{\rm bar}$,
which corresponds to the faint peak at $\Omega_{\rm pre}\approx 9$
on the $\Omega_{\rm pre}-e$ plane: the strongest peak at lower $e$
and lower
$\Omega_{\rm pre}$ on this plane does not correspond to the mean $\Omega_{\rm pre}$
of stellar  particles
in the bar. The central peak in the model with $f_{\rm d}=0.28$ is weaker
owing to the weaker bar. More details are found in the main text.
}
\label{Figure. 16}
\end{figure*}

\begin{figure}
\psfig{file=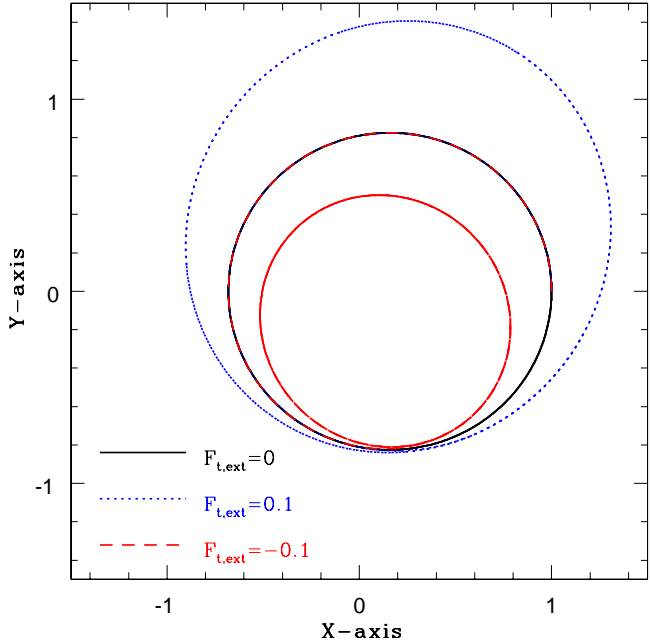,width=8.5cm}
\caption{
Orbital evolution of stellar particles under a point-mass Kepler potential
with $F_{\rm t, ext}=0$ (black solid; no external tangential force),
0.1 (blue dotted), and $-0.1$ (red dashed).
This $F_{\rm t, ext}$ measures the ratio of the tangential component
of external force to the total gravitational force, and positive values
mean that the direction of the external force is aligned with
$v_{\varphi}$ (azimuthal component of velocity).
In a point-mass potential,  apsidal precession never occurs, as shown in
the model with $F_{\rm t,ext}=0$: the major axis of the orbital ellipse is
parallel to the $x$ axis during orbital evolution ($\varphi_{\rm a}=0$).
 However,  the major axis of the ellipse is rotated
in  the model with $F_{\rm t, ext}=\pm 0.1$ (i.e., $\varphi_{\rm a}$ becomes
positive and negative, respectively).
Accordingly this figure illustrates that strengthened $F_{\rm t}$ by some
mechanisms (e.g., dynamical action of bars)
can change $\varphi_{\rm a}$ in real galactic disks.
}

\label{Figure. 17}
\end{figure}

\begin{figure}
\psfig{file=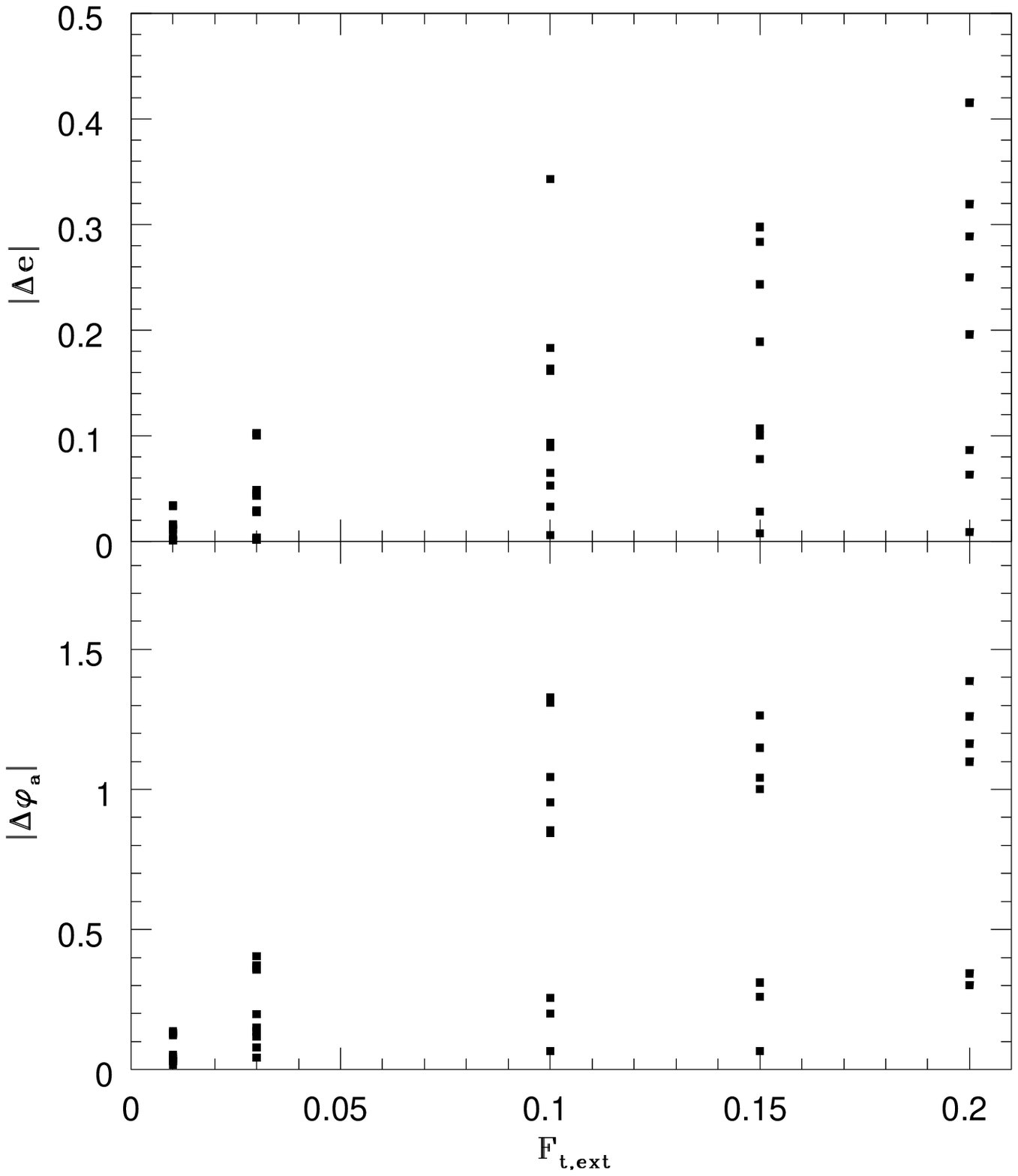,width=8.5cm}
\caption{
Dependence of $|\Delta e|$ and $\Delta \varphi_{\rm a}$ on $F_{\rm t, ext}$ in
the orbital models with a point-mass potential.
Here $|\Delta e|$ and $\Delta \varphi_{\rm a}$ describe the difference
in $e$ and $\varphi_{\rm a}$, respectively,  between orbits for
$F_{\rm t, ext}=0$
(i.e., purely Kepler orbits) and those for non-zero $F_{\rm t, ext}$.
For each $F_{\rm t, ext}$.
Different models adopt  different times at which external tangential force
starts to influence on the orbits.
}
\label{Figure. 18}
\end{figure}

It is the tangential component of gravitational force ($f_{\rm t}$)  between the
local  particles that
causes  the local APS in these models,
because the radial component ($f_{\rm r}$)  
is much weaker than the gravitational force  from the 
potential.
 At $R_{\rm a}$, particles with similar $t_{\rm a}$ can mutually interact
with one another for longer timescales so that the tangential force can change
$\varphi_{\rm a}$. Although this explanation would be quite reasonable,
it is yet to be demonstrated in the present study why
the tangential force of the particles can cause such $\varphi_{\rm a}$ alignment
if $F_{\rm a}$ exceeds a certain threshold value.
Probably, the tangential gravitational force  acting on particles can 
always move $\varphi_{\rm a}$ of the particles
toward the direction of the tangential force at $R \approx R_{\rm a}$.
In this mechanism, the change of $\varphi_{\rm a}$ should be proportional
to the relative strength of $F_{\rm t}$ among particles in the local area:
\begin{equation}
\Delta \varphi_{\rm a} \propto F_{\rm t}.
\end{equation}
Therefore, if $F_{\rm t}$ is positive (e.g., +0.5)
, then the change of $\varphi_{\rm a}$ should
be positive (e.g., 0.2 radian).
This  shift of $\varphi_{\rm a}$ toward the direction of tangential force
could be the main physical reason for local APS.
We discuss this point in \S 4 using idealized models.

\begin{figure}
\psfig{file=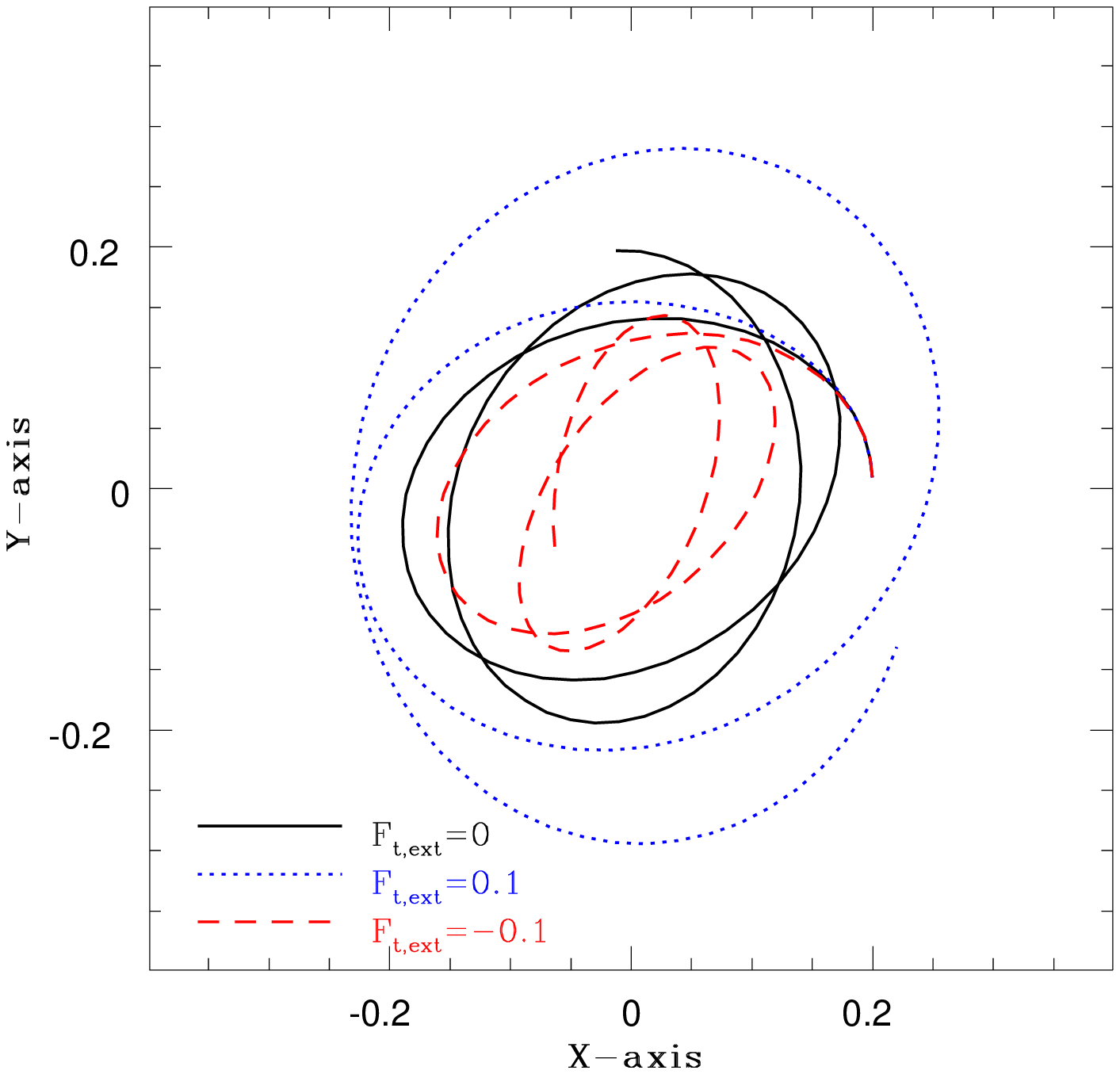,width=8.5cm}
\caption{
The same as Fig. 17 but for the fixed gravitational potential
that is exactly the same as that at $T=0$ in the fiducial Nbody model.
Clearly, $\varphi_{\rm a}$ changes due to external tangential force and thus
$\Delta \varphi_{\rm a}$ that is the $\varphi_{\rm a}$ difference between
two consecutive apocenter passages.
In this model,
$\Delta \varphi_{\rm a}$ for $F_{\rm t}=0$, 0.1, and $-0.1$
are 3.56, 3.63, and 3.51 (radian), respectively. This means
that positive (negative)
$F_{\rm t}$ can increase (decrease) $\Delta \varphi_{\rm a}$.
}
\label{Figure. 19}
\end{figure}

\subsubsection{Bar growth  due to bar-induced APS}

Fig. 12 shows that $P_{\rm syn}$ can very slowly increases during the 
formation/growth phases of the
seed bar ($T<4t_{\rm dyn}$), which is consistent with $A_2$ evolution of the stellar disk 
(in Fig. 5). However, $P_{\rm  syn}$ can rapidly increase after $T\approx 4t_{\rm dyn}$,
which means that the $S$-parameter ($S_{\rm aps}$) can be significantly different
between the two stages of bar formation. 
In order to clarify the physical mechanism
of bar growth after $T=4t_{\rm dyn}$, we investigate the correlation  
between $A_2$ and $S_{\rm aps}$ in this and other models.
In the following discussion, the (two) azimuthal angles of the two edges of a 
stellar bar are referred to as $\varphi_{\rm bar}$ for convenience
(no distinction between the two).

Fig. 13   demonstrates that $S_{\rm aps}$ 
is linearly proportional
to  $A_2$ for $0.05<A_2<0.16$ (after seed bar formation) in the three
selected radial bins,
though the correlation between $S_{\rm aps}$ and $A_2$
is not so tight in this model.
The correlation can be approximated by the following equation:
\begin{equation}
S_{\rm aps}=\alpha A_2+\beta,
\end{equation}
where $\alpha$ and $\beta$ can be different in different radial bins 
in a same galaxy model.
For example, $\alpha \approx 2$ and $\beta \approx 0.03$ for
$ 0.2 \le  R/R_{\rm d} < 0.3$ in the fiducial model.
Similar positive correlations between $S_{\rm aps}$ and $A_2$
can be seen in other models with strong bars,
though the slopes ($\alpha$) of the relations are quite different, as discussed later.
It should be pointed out here that such a positive correlation
cannot be clearly seen in the three radial bins for $A_2<0.05$ 
when the bar has not been developed yet.
The small $S_{\rm aps}$ for $A_2<0.05$ is due to local APS that is 
less efficient than the bar-induced one.

Given that the $S_{\rm aps}$ is the time derivative of $P_{\rm syn}$,
the correlation in Fig. 13  indicates that the  strength of the bar itself
is a key parameter for the synchronization
of $\varphi_{\rm a}$ during the growth stage of the bar.
The radial component of gravitational force of the weak bar with $A_2<0.2$
cannot be significantly stronger than that of the background axisymmetric potential.
Accordingly, the tangential component of the bar's force and its time evolution
should be the main cause of this correlation.
As the bar in a disk becomes stronger, the tangential force of the bar becomes also stronger
to influence the orbits of the disk stars more strongly.
As a result of this, a larger number of stars can be gravitationally
captured by the bar and  also have $\varphi_{\rm a}$ at $T=t_{\rm a}$
aligned to $\varphi$ of the either side of the bar major axis
($\varphi_{\rm bar}$).
Therefore, APS can proceed more efficiently in stronger bars, and more efficient APS can cause
the stronger growth of bars. Thus we can conclude
that  this positive feedback loop is the essence of bar instability
in disk galaxies.

APS due to dynamical action of growing bars themselves is quite different from
local APS that can be effective during seed bar formation. 
It is yet to be clarified why APS can proceed more efficiently (i.e., $S_{\rm aps}$ is higher),
if the tangential force is stronger (i.e., $A_2$ is higher).
In order to solve this problem, we first need to understand why and how the tangential force
of a bar can cause APS.
Although it would be  reasonable that (i) the tangential force of bars can act to reduce
$|\varphi_{\rm a}-\varphi_{\rm bar}|$ and 
(ii) the stronger force can trigger the greater reduction of 
$|\varphi_{\rm a}-\varphi_{\rm bar}|$,
we currently do not have a quantitative answer for this problem.
Although it  is  our future study to investigate
how stars with different $\varphi_{\rm a}$
response to stellar bars with different bar strength to finally have 
$\varphi_{\rm a} \approx \varphi_{\rm bar}$,
we discuss this issue qualitatively in \S 4 using idealized models.

\subsection{Bar formation in other models}

\subsubsection{Spontaneous bar formation}

The basic physical mechanisms and processes of bar formation in other isolated models with
bars
are confirmed to be the same as those revealed in the fiducial model: APS can be clearly
seen in $N(\varphi_{\rm a})$ distribution in the early and late 
phases of the bar formation
(i.e., higher $P_{\rm syn}$ in later phases).
 It is also found that bar formation is not possible
or strongly suppressed  (i.e., very long formation time of $>20 t_{\rm dyn}$)
in the models with lower $f_{\rm d}$.
For example,
the maximum $A_2$ within $30t_{\rm dyn}$ 
in the models IA2 and IA3 with $f_{\rm d}=0.04$ and 0.11
are 0.05 and 0.09, respectively,  which means that bar formation is not possible
in these models with low $f_{\rm d}$ owing to much less efficient APS.
Bars cannot be formed within $10t_{\rm dyn}$ in the model IA4, however, 
they can be slowly developed within $30t_{\rm dyn}$ in the model.
IA6 - IA 10 all show bar formation, and 
the bar formation time scale is shorter for larger $f_{\rm d}$.
No bar formation in smaller $f_{\rm d}$ ($< 0.25$) is confirmed in the models
with more diffuse dark matter halos (IB1, IB2, and IB5).

Although the central bulges can suppress the bar formation in IA12, IA13, and IA14
with larger $f_{\rm b}$ within $10t_{\rm dyn}$,
a bar can be formed in the Milky Way type 
model IA11 ($f_{\rm b}=0.17$).It should be stressed that IB3 with 
a big massive bulge ($f_{\rm b}=1$)
and a higher disk mass fraction ($f_{\rm d}=0.41$) can form a bar within
$30t_{\rm dyn}$ ({\it not} $10t_{\rm dyn}$). 
These results suggest that if $f_{\rm b} \le 1$,
then the combination of massive bulges and more compact
dark matter halos (not just massive bulges alone) are required for the severe
suppression of bar formation: just compact bulges alone cannot completely
suppress bar formation. 
Such slow bar formation in IB3 is quite intriguing, because it indicates
that APS is slowly ongoing in such a model.
IB4 with $f_{\rm b}=2$ corresponding to bulge-dominated S0 galaxies
cannot show bar formation within $30t_{\rm dyn}$, which may explain why
the observed bar fraction in S0s is lower.
IA16 and IA17 with higher $Q$ ($\ge 2$) do not show bar formation, because
APS can be severely suppressed by the higher degrees of random motion of stars in
the models.

IC1 and IC2 with $c=16$ have smaller  $f_{\rm d}$ (by $20\%$) compared to those
with $c=10$ models (IA3 and IA5) so that bar formation can be more severely
suppressed. 
The time scale of bar formation in IC3 with $f_{\rm d}=0.28$ is significantly
longer than the fiducial model, which means that bar formation can be delayed
by more compact dark matter halos.
Given that disk galaxies with lower masses are more likely
to have larger $c$ and smaller $f_{\rm d}$,
these results imply that the bar formation can be severely suppressed
in disk galaxies with lower masses.
Strong suppression of bar formation in disks with lower $f_{\rm d}$, higher $Q$,
and higher $f_{\rm bul}$ found in the present study 
were already reported in previous simulations 
(e.g., AS86; Efstathiou et al. 1982).

Fig. 14 demonstrates that APS is more likely to proceed effectively
for larger $A_2$  in the isolated models with strong bar formation.
The $S_{\rm aps}$ parameter can be different in different models, because
the rapidity of bar formation is quite different: for example, a strong bar
can be more rapidly formed due to lower stellar velocity dispersions
in the model with $Q=1$ (i.e., larger $S_{\rm aps}$). Clearly,
the models with no bar formation do not show larger $S_{\rm aps}$ ($>0.3$) for
a given $A_2$, which confirms that a strong bar cannot grow due to APS
in these models.
Intriguingly, the distributions on this $A_2 - S_{\rm aps}$ map are
quite different between spontaneous and tidal bar models in the sense
that the latter can show very large $S_{\rm aps}$ ($>1$) even
for $A_2<0.05$, when a bar is not developed at all. This is mainly
because the external tidal perturbation in these tidal models 
is the main cause of APS (not due to the growing bar, as seen in spontaneous
bar formation).
Such a sharp increase in $S_{\rm aps}$ is associated with dramatic
orbital elongation and loss of angular momentum of stars during tidal 
interaction, as discussed later in this paper.

\subsubsection{Tidal bar formation}

Galaxy interaction can trigger bar formation even in the disks with lower $f_{\rm d}$
and higher $Q$ for which bars cannot be formed spontaneously (e.g., Noguchi 1987).
This point is confirmed in the present study: for example,
TA2 can form a bar whereas IA3 cannot.
Given that APS is not so efficient for lower $f_{\rm d}$ and higher $Q$,
the mechanism of bar formation in interacting galaxies should be different
from that in isolated galaxies.
Fig. 15 describes the 2D map of
$\Sigma(t_{\rm a},\varphi_{\rm a})$ in the tidal interaction model TA1 in which
the initial disk is exactly the same as that adopted in the fiducial model.
There are three major differences in this map between this model and the fiducial one
(compare Fig. 15 with Fig. 7). First, striped patterns appear suddenly 
around $T \approx 5t_{\rm dyn}$ in the tidal model, which is in a striking contrast with
the fiducial one (i.e., isolated one) in which such patterns slowly build up.
This dramatic change in the 2D map is due to very rapid formation of the bar
triggered by strong tidal force of the companion galaxy.
Second is that the striped patterns in the tidal model
are less inclined with respect to the $t_{\rm a}$
axis (i.e., shallower slopes),
which physically means lower $\Omega_{\rm bar}$.
Such  lower $\Omega_{\rm bar}$ during tidal bar formation
was already reported by a number of previous
simulations (e.g., MN98).
Third is that each strip is wider in the tidal model than in the isolated
one. This is consistent with the morphology of the tidal bar with a  
fatter appearance (e.g., Noguchi 1987).

The very rapid synchronization of $\varphi_{\rm a}$ soon after the pericenter passage
of the companion is due largely to the strong tidal perturbation of the bar
to the disk particles. Such tidal perturbation can forcefully transform the initially
circular disk into a very elongated structure (i.e., bar) and also 
align $\varphi_{\rm a}$ of the particles to a very narrow range of $\varphi_{\rm a}$.
Accordingly,
the rapidity and physical mechanism of APS in this tidal bar formation are quite
different from those of spontaneous bar formation.
Angular momentum redistribution of stars occurs during spontaneous bar formation
(as shown in the present study) whereas a large fraction of disk stars can lose
their angular momentum in a tidal bar (e.g., MN98).
It would be therefore possible to say that the two formation paths are different in terms 
of angular momentum evolution during bar formation.

As shown in Fig. 16,
there is a strong concentration of the particles
in  the distribution of stellar particles on
the $\Omega_{\rm pre}-e$ plane in the tidal bar model TA1,
as is the case with the spontaneous
bar formation.
The details of the  $\Sigma(\Omega_{\rm pre},e)$ map in the tidal bar model, however,
are significantly different from those of the spontaneous one (IA1),
which reflect the difference in APS processes between the two.
The slightly lower $\Omega_{\rm pre}$ at the peak of $\Sigma(\Omega_{\rm pre},e)$ in the tidal bar is
consistent with the shallower slope of the inclined strip-patterns
shown in the $t_{\rm a}-\varphi_{\rm a}$ map
for this model (see Fig. 15).
The systematically higher $e$ in this tidal bar compared to the spontaneous 
bar is consistent with loss of angular momentum (in the tidal bar)
that can cause lower $e$.
It is also confirmed in Fig. 16 that other tidal models show similarly lower
$\Omega_{\rm pre}$ and higher $e$  at the peak of $\Sigma(\Omega_{\rm pre},e)$,
which suggests that $\Omega_{\rm bar}$ 
can be low in tidal bars.
Previous studies found evidence for fast bars in post-interacting
and weakly interacting galaxies (e.g., Debattista et al. 2002;
Cuomo et al. 2019), which is not so consistent with lower $\Omega_{\rm bar}$ 
in Fig. 16.
However, there is no observational study which systematically
investigated the differences in $\Omega_{\rm bar}$ between
spontaneous and tidal bars: even the classification of such two different bars has
not been done for the observed galaxies with bars.
Therefore, it is too early to conclude that the present results
are inconsistent with observations.
It is quite intriguing to compare between these $\Omega_{\rm pre}-e$
maps for bars and those in the models without bars in the present study.
Accordingly, the details of 
the $\Omega_{\rm pre}-e$ maps for non-barred galaxies are described in Appendix B.

Although TA3, TB1, and TB2 with compact bulges can finally form bars,
TB3 (S0 model) with $f_{\rm b}=2$ cannot develop a bar even after strong tidal interaction.
Given that IB4 with $f_{\rm b}=2$ does not show a bar either,
these suggests that S0s with large bulge-to-disk-ratios cannot have bars.
Previous simulations of tidal bar formation revealed the physical properties of bars
and their correlations 
(e.g., between bar strengths and pattern speeds; Berentzen et al. 2004).
It is found in the present study
that stellar kinematics can be different between spontaneous and tidal bars 
(e.g., not so clear ``S-shaped'' line-of-sight velocity profiles in tidal bars).
However we will describes stellar kinematics such as the 2D maps of
velocities and velocity dispersion in many models with bars in our future studies,  because they
are not so relevant to the purpose of this paper.

\begin{figure}
\psfig{file=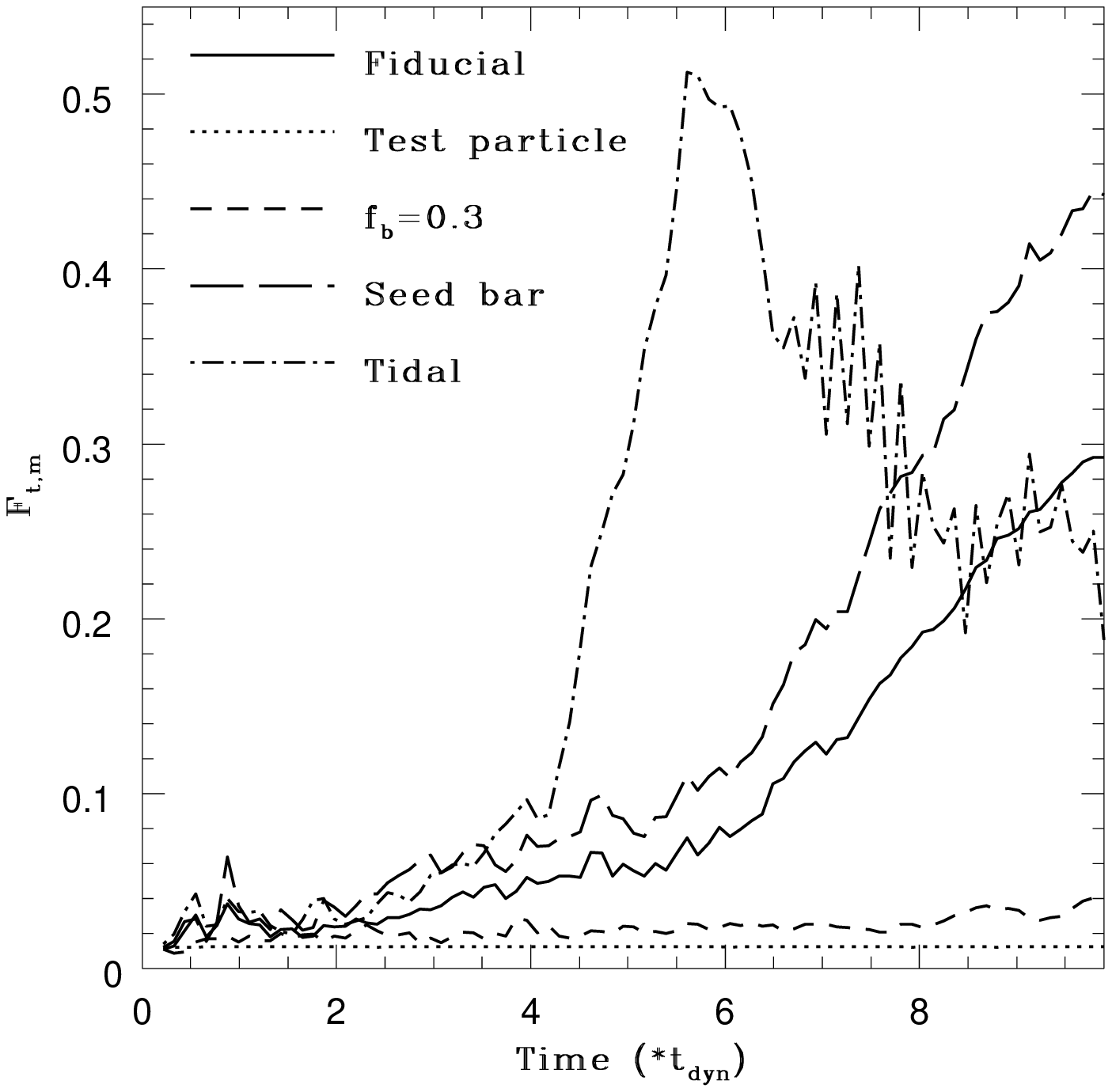,width=8.5cm}
\caption{
Time evolution of $F_{\rm t, m}$ (mean $|F_{\rm t}|$) in five different models:
fiducial model (IA1, solid), comparative model with test particles in which
self-gravity of the particles is switched off (dotted),
early-type spiral model with $f_{\rm b}=0.3$ (IA12, short-dashed),
fiducial model with $F_{\rm t, m}$ being estimated only from seed bar particles
(long-dashed),
and tidal interaction model (TA1, dot-dashed).
}
\label{Figure. 20}
\end{figure}

\section{Discussion}

\subsection{Physical origins of APS}

We have shown that synchronization of
$\varphi_{\rm a}$ and $\Omega_{\rm pre}$  (i.e., APS)
is strongly influenced by (i)  mutual gravitational interaction
of stars in local areas of disks and (ii) by gravitational fields of growing
bars in the present study. However,  the physical mechanism(s) of these 
local and bar-induced APS have not been elucidated well  in this paper. 
In order to understand the physics of APS deeply, we need to 
answer the following three questions: (1) 
why $\varphi_{\rm a}$ can be influenced
by  gravitational interaction of local particles  and growing bars
(``$\varphi_{\rm a}$ shift''),
(2) why $\Omega_{\rm pre}$ can be changed during the dynamical evolution
of stellar disks,
and (3) why $\varphi_{\rm a}$ and $\Omega_{\rm pre}$ change
in such a way that their dispersion, i.e.,
$\sigma(\varphi_{\rm a})$ and $\sigma(\Omega_{\rm pre})$, 
become quite small (i.e., synchronization).

In initial stellar disks, the radial components of gravitational force
are  much stronger than the tangential ones as follows:
\begin{equation}
|F_{\rm r}| \gg |F_{\rm t}|.
\end{equation}
As demonstrated later,
$F_{\rm t}$ evolves with time
more dramatically and also can  be  much more 
strongly influenced by these local particles and growing bars compared
to $F_{\rm r}$  in the present disk galaxy models.
Also $F_{\rm r}$ cannot be significantly
changed by local gravitational interaction  and growing weak bars due to the dominance of 
dark matter in disk galaxies (except in the late phase of bar formation).
We therefore consider that strengthened $F_{\rm t}$ during disk evolution
is the main causes of local and bar-induced APS in the present study.
Accordingly we try to answer the above-mentioned three questions by investigating
how $F_{\rm t}$ can influence $\varphi_{\rm a}$ and $\Omega_{\rm pre}$ using
simple orbital calculations of stellar particles.

\subsubsection{$\varphi_{\rm a}$ shift due to $F_{\rm t}$}

In order to illustrate (i) how this $\varphi_{\rm a}$ shift can occur
and (ii) how the shift  depends on the strengths
of external (tangential/azimuthal) force,
we investigate orbital evolution of stars in a point-mass
(Kepler)  potential in which external
gravitational force is added {\it only} to the azimuthal component
of gravitational force. 
In the Kepler potential without external force,
$\varphi_{\rm a}$ can never change.
We therefore adopt this idealized
model for orbital calculations, because it is straightforward to
quantify the effects of external force on $\varphi_{\rm a}$.
A key parameter in this study is 
$F_{\rm t,ext}$ that is the strength ratio of the external force
to the total gravitational one.
The  tangential force starts to  influence the orbits of stars  
at $T=t_{\rm 1, 1}$ and ends at $T=t_{\rm t, 2}$.
(given in units of $t_{\rm dyn}$).
Although $F_{\rm t,ext}$ is a time-dependent variable in simulations,
we here assume a fixed $F_{\rm t, ext}$ for clarity in this orbital 
calculation. We show the results of the models with
$t_{\rm t,2} -t_{\rm 1, 1}=t_{\rm dyn}$.

In this study, the mass, size, velocity, and time units and the gravitational
constant  are all given as 1 and the initial $R_{\rm a}$ and $f_{\rm v}$ without
external perturbation are set to be 1.0 and 0.7, respectively.
By changing $F_{\rm t, ext}$ for a given set of $t_{\rm t, 1}$ and $t_{\rm t, 2}$,
we can discuss how the tangential force influences
$\varphi_{\rm a}$. 
Initial positions of a particles, $(x,y,z)$
are fixed at $(1,0,0)$ in  all models and the particles 
are initially at their apocenter distances ($R_{\rm a}$).
Given that the first $\varphi_{\rm a}$ is 0 in the model with $F_{\rm t, ext}=0$
(i.e., no external perturbation),
the second $\varphi_{\rm a}$ in the models with non-zero $F_{\rm t, ext}$ is
simply the amount of $\varphi_{\rm a}$ shift ($\Delta \varphi_{\rm a}$).
Although the results depend on $t_{\rm t,1}$ and $t_{\rm t,2}$,
we describe the results only for 
the models with $e=0.19$ (corresponding to the peak $e$ in the $N(e)$
in the fiducial model) and $(t_{\rm t,1},t_{\rm t,2})=(0,1)$,
(1,2), (2,3), (3,4), and (4,5) below.
For these models with $t_{\rm t,2}-t_{\rm t,1}=1$ ($t_{\rm dyn}$),
the mean of $F_{\rm t, ext}$ ($F_{\rm t, m}$) is simply $F_{\rm t, ext}$.
We use these $F_{\rm t, m}$  as a guide to
physical interpretation of APS demonstrated
for the present full Nbody simulations.

Fig. 17 shows that (i) $\varphi_{\rm a}$ shift can be clearly seen in
the models with $F_{\rm t,ext} = 0.1$
and (ii) the direction of the shift can be clockwise or counter-clockwise
depending
on the sign of $F_{\rm t, ext}$.  
As demonstrated in Fig. 18,  there is a positive correlation between
$|F_{\rm t, ext}|$ and $|\Delta \varphi_{\rm a}|$,
thought it is not so strong.
These results strongly suggest that the tangential component of
gravitational force can play a key role in local and bar-induced APS.
Also, negative tangential force can reduce the angular momentum
(and kinetic energy of stars) so that $e$ can be higher (i.e., 
more elongated) in these orbital models,
and $e$ changes ($\Delta e$)  depend on $|F_{\rm t}|$ too.

Although these results are based on the adopted Kepler potential, it is confirmed
in the present study that $\varphi_{\rm a}$, $t_{\rm a}$, $e$, and 
$\Omega_{\rm pre}$ can be influenced by stronger $F_{\rm t}$ in 
realistic galactic potentials too. Fig. 19 shows how stellar orbits can be changed
by stronger positive and negative $F_{\rm t}$ in the fiducial model. In
these orbital calculations, mass-less test particles can move around the 
galactic center being influenced both by
the  potential generated 
by the {\it fixed} Nbody particles and by external $F_{\rm t,ext}$
for $t_{\rm t,1} \le T \le t_{\rm t, 2}$: the model with
$(t_{\rm t,1},t_{\rm t,2})$=(0,1) is shown in Fig. 19.
Clearly, negative $F_{\rm t}$ can (i) elongate the orbits, (ii) shorten the
radial periods of the orbits ($T_{\rm r}$), and (iii) make $R_{\rm a}$ smaller.
It is also found that $\Omega_{\rm pre}$ can be influenced significantly
by stronger $F_{\rm t}$, which is described in detail in Appendix D.
These results for Kepler and more realistic galactic potential models
suggest that 
if the net  $F_{\rm t}$ for a star becomes negative due to dynamical
actions by local density
enhancement or an existing bar, then   the star becomes able to have an elongated
orbit.  Such a star  with a higher $e$ can thus become the major constituent of a bar.

Although these results depend also on $t_{\rm t,1}$ and $t_{\rm, t, 2}$
($\delta t=t_{\rm t, 2}-t_{\rm t, 1}$),
we can discuss how $\Delta \varphi_{\rm a}$,
which is the $\varphi_{\rm a}$ difference between
two consecutive apocenter passages,
depends on $F_{\rm t,ext}$ using the models with different 
$F_{\rm t, ext}$ (0, $\pm 0.1$, $\pm 0.2$, $\pm 0.4$, and $\pm 0.6$)  and 
a fixed $(t_{\rm t,1},t_{\rm t,2})$=(0,0.5).
Although we did not run many models, we could find the following 
relation;
\begin{equation}
\Delta \varphi_{\rm a} \approx 0.24 F_{\rm t}\delta t + 3.47.
\end{equation}
It should be stressed here that
this linear relation is introduced only for the purpose of discussing
the local and bar-induced APS in this paper.
We need to do a thorough parameter survey in order to understand this
$F_{\rm t, ext}-\Delta \varphi_{\rm a}$ relation.

If the outer parts of stellar disks with $\Omega_{\rm pre} < \Omega_{\rm bar}$
can experience net positive $F_{\rm t}$,
stellar bars would not grow due to APS  in the outer parts of disks,
because $\Omega_{\rm pre}$ becomes smaller: the initially 
lower $\Omega_{\rm pre}$ in the outer parts
needs to become larger to be similar  with $\Omega_{\rm bar}$ to join the bars.
Accordingly, we need to understand how APS depends on $\Omega_{\rm pre}-\Omega_{\rm bar}$
in order to understand what determines the lengths of stellar bars.
Thus both APS and  elongation of orbits that are important in bar formation,
and they can be caused by stronger tangential force during disk evolution.

Here we can compare  $|F_{\rm t, ext}|$ 
in these orbital studies with the mean of $|F_{\rm t}|$ 
(simply referred to as $F_{\rm t, m}$)
from the present Nbody simulations in order to discuss whether
$F_{\rm t,m}$ in the simulated galaxies can be as strong as
$F_{\rm t, ext}$ required for $\varphi_{\rm a}$ shift 
(i.e., $F_{\rm t,m}$ as large as $0.02-0.03$).
Fig. 20 shows that $F_{\rm t,m}$  can be as large as 0.02
during the early formation of the seed bar ($T<4t_{\rm dyn}$) in the fiducial model. 
Given that the test particle model shows that $F_{\rm t,m}$ is almost constant and low
($<0.01$), the increasing $F_{\rm t,m}$ in this model is due to 
gravitational interaction (i.e., self-gravity) between stellar particles.
The stellar particles in the seed bar shows a larger $F_{\rm t,m}$ compared
with the mean of all particles in the fiducial model, which is consistent with
the faster growth of the seed bar.
in the fiducial model. It is also clear in Fig. 20 that
$F_{\rm t, m}$ can be significantly larger  in the later epoch of bar formation,
due to the stronger tangential force from the stronger bar
(see Fig. 5 for the time evolution of $A_2$).

As described in Fig. 20,  $F_{\rm t, m}$ in the tidal bar model
becomes significantly larger than that 
in the fiducial model, which demonstrates that much larger $F_{\rm t,m}$ during
tidal interaction is the main cause of APS in the tidal bar formation.
This rather large $F_{\rm t, m}$ can cause loss of angular momentum and
significant elongation of stellar orbits, which is described  in Appendix D in detail.
The model with $f_{\rm b}=0.3$
have smaller $F_{\rm t, m}$, 
which is consistent with no  bar formation within $10t_{\rm dyn}$ in this model.
Although  we do  not show the time evolution of 
$F_{\rm t,m}$ for all models with bars,
these results clearly demonstrate that stronger $F_{\rm t}$ is
a main cause for $\varphi_{\rm a}$ shift thus for  local and bar-induced APS.
Appendix E  also explains  why $\varphi_{\rm a}$ shift 
in bar-induced APS  proceeds
in such a way that $|\varphi_{\rm a}-\varphi_{\rm bar}|$
(where $\varphi_{\rm bar}$ is the azimuthal angle of the bar's major axis)
can  become smaller in the present models of disk galaxies.

\subsubsection{$\Omega_{\rm bar}$ change due to $F_{\rm t}$}

It is confirmed in the present study that $\Omega_{\rm pre}$ can be 
significantly changed due largely to stronger
$F_{\rm t,ext}$,  as long as $|F_{\rm t,ext}|\delta t$
is larger than a threshold value of this  (see Appendix C for the details ).
As long as $|F_{\rm t}|$ is not very large ($>0.5$),
positive $F_{\rm t}$ can decrease both $\Omega_{\rm pre}$ and $e$ due to increase
in angular momentum. On the other hand,
negative $F_{\rm t}$ can increase both $\Omega_{\rm pre}$ and $e$. 
Even if $F_{\rm t, ext}$ is as small as $0.02$,  
$\Omega_{\rm pre}$ can be significantly changed for $\delta t=5t_{\rm dyn}$.
These results clearly show that strengthened $F_{\rm t}$ by local particles
and growing bars are able to change $\Omega_{\rm pre}$.

\subsubsection{Local synchronization} 

Now that we have shown that $\varphi_{\rm a}$ can change
due to stronger $F_{\rm t}$,
we need to explain why dispersions in $\varphi_{\rm a}$ ($\sigma(\varphi_{\rm a})$)
can be smaller in 
local particles with similar $t_{\rm a}$, as shown already in Fig. 11.
We here again adopt the model with $F_{\rm a}=0.149$ shown in Fig. 11 in order
to provide a qualitative explanation for the smaller $\sigma(\varphi_{\rm a})$.
We assume that the local area consists of two groups of stellar particles, i.e.,
one with  initial $\varphi_{\rm a}$ ranging from 0 to 0.52 and the other 
with  initial $\varphi_{\rm a}$ ranging from 0.52 to 1.05. We here
consider mutual gravitational interaction between the two groups  
for convenience and clarity (not between
individual particles of the two groups) and define the
first and second groups' initial $\varphi_{\rm a}$ as $\varphi_{\rm a,1}(1)$ and
$\varphi_{\rm a,2}(1)$, respectively
($\varphi_{\rm a,2}(1)-\varphi_{\rm a,1}(1)=0.52$): the number in ``()'' indicates
the $n$-th apocenter passage and the first apocenter passage (i.e., $n=1$)
corresponds to $T=0$ 
(start of the simulation).
The two groups increase their $\varphi_{\rm a}$ by 
$\delta \varphi_{\rm a,1}$ and 
$\delta \varphi_{\rm a,2}$, respectively, 
in the next (2nd) apocenter passage
as follows;
\begin{equation}
\varphi_{\rm a,1}(2)=\varphi_{\rm a, 1}(1)+\delta\varphi_{\rm a,1},
\end{equation}
and 
\begin{equation}
\varphi_{\rm a,2}(2)=\varphi_{\rm a, 2}(1)+\delta\varphi_{\rm a,2}.
\end{equation}
If there is no $F_{\rm t}$ between the two groups, then
$\delta \varphi_{\rm a,1}$ is equal to
$\delta \varphi_{\rm a,2}$: 
\begin{equation}
\delta\varphi_{\rm a,2}=\delta\varphi_{\rm a,1}=\delta \varphi_{\rm a, 0},
\end{equation}
where $\delta\varphi_{\rm a, 0}$ is constant.
Accordingly, 
$\varphi_{\rm a,2}(2)-\varphi_{\rm a,1}(2)$ can be constant ($=0.52$) in the 
second apocenter passage.
However, 
$\delta \varphi_{\rm a,1}$ is not equal to
$\delta \varphi_{\rm a,2}$
if $F_{\rm t}$ is not 0:
\begin{equation}
\varphi_{\rm a,2}(2)-\varphi_{\rm a, 1}(2)=0.52 + \delta\varphi_{\rm a,2}
- \delta\varphi_{\rm a,1}.
\end{equation}
As shown in Fig. 18, positive and negative $F_{\rm t}$ can lead to larger and
smaller $\varphi_{\rm a}$, respectively, in the second apocenter
passage (i.e., larger and smaller amount of $\varphi_{\rm a}$ increment).
Since group 1 and 2 experience positive and negative $F_{\rm t}$ by definition,
$\delta\varphi_{\rm a,2}
- \delta\varphi_{\rm a,1}$ can be negative, which means that
$\varphi_{\rm a,2}(2)-\varphi_{\rm a, 1}(2)$ can be smaller than
the initial value of 0.52 (at the first apocenter passage).
Thus $\sigma(\varphi_{\rm a})$ should be able to be smaller owing to
mutual gravitational interaction between local particles with
similar $t_{\rm a}$.
In the early formation of seed bars, $\Omega_{\rm pre}$ is yet to be synchronized.
Therefore, synchronization in $\Omega_{\rm pre}$ is due largely to dynamical
action of bars on stellar particles.

\section{Conclusions}

Using idealized collisionless Nbody simulations of disk galaxies,
we have investigated the orbital properties of the individual stellar particles
and their time evolution in order to understand
the physical mechanisms of bar formation.
Based on these orbital properties,
the number distributions of stellar
particles  with $\varphi_{\rm a}$ ($N(\varphi_{\rm a})$) and
the phase space densities of stellar particles in the $t_{\rm a}-\varphi$
(i.e., $\Sigma(t_{\rm a},\varphi_{\rm a})$) 
and the $t_{\rm a}-\varphi$ planes
(i.e., $\Sigma(\Omega_{\rm pre},e)$) 
have been also investigated.
Although bars can be formed from a number of galaxy-scale physical processes
(global instabilities, tidal interaction, mergers, and dynamical action of triaxial
dark matter halos etc),
we have exclusively focused on two formation paths of bars, i.e.,
(i) global bar instability (``spontaneous bars'')
and (ii)  tidal interaction (``tidal bars'').

Our numerical simulations have investigated,
for the first time,  how and why
the apsidal precession synchronization (``APS'') of stars
can occur during  bar formation in disk galaxies. 
This APS has been quantified by newly introduced two parameters, i.e., (i)  $P_{\rm syn}$,
which is defined as the ratio of maximum to average $N(\varphi_{\rm a})$,
and (ii) $S_{\rm aps}$ ($=dP_{\rm syn}/dt$) in the present study.
Disk galaxies with stronger bars tend to have higher $P_{\rm syn}$,
and $S_{\rm aps}$ can be higher in bars that are more rapidly forming.
These two parameters are investigated for different radii ($R$) in each bar model
and cross-correlated with the physical properties of bars such as
the relative amplitudes of  $m=2$
Fourier mode (corresponding to the bar strength).
The main results are as follows:

(1) APS is the main mechanism of bar formation in isolated and tidally interacting
disk galaxies, though the physical processes that cause APS are quite different between
these spontaneous and tidal bar formation.
APS can be clearly seen in the time evolution of the
$\Sigma(t_{\rm a},\varphi_{\rm a})$ map
as characteristic stripped patters in these two formation paths.
During and after bar formation within a stellar disk,
the striped patterns can have two distinct $\varphi_{\rm a}$  peaks with
the peak locations being roughly separated by $\pi$ (radian) for a given $t_{\rm a}$,
which is a clear reflection of very similar $\varphi_{\rm a}$
(i.e., synchronized precession)  at the two edges of
the bar.

(2) Spontaneous bar formation in isolated stellar disks
consists of (i) seed bar formation due to APS in
local regions with higher phase space densities ($A_2<0.1$) 
and (ii) bar growth due to APS caused
by existing (growing) stellar bars ($A_2>0.1$).
Stars  with initially different $\varphi_{\rm a}$ yet similar $t_{\rm a}$
in a local region of a stellar disk can have similar $\varphi_{\rm a}$
due to mutual gravitational interaction of the stars within a few $t_{\rm dyn}$.
This relatively rapid APS is
a key physical process for the formation of seed bars,
though $P_{\rm syn}$ can be only moderately high ($\approx 1.8$).
This seed bar formation due to local APS is strongly suppressed or much
delayed in disks with lower $f_{\rm d}$ ($<0.2$), because 
$S_{\rm aps}$ is very low  in such disks.

(3) Stellar disks with growing bars show
a positive correlation between $S_{\rm aps}$ and $A_2$,
which demonstrates that APS is caused by dynamical action of existing (weak)
bars on stellar orbits.
APS in a growing bar physically means the alignment of $\varphi_{\rm a}$ of stars  to 
 the azimuthal angles of the bar's two edges ($\varphi_{\rm bar}$).
As a stellar bar grows (higher $A_2$), the power of the bar
to align $\varphi_{\rm a}$ of stars
to $\varphi_{\rm bar}$ becomes greater due to the stronger tangential gravitational
force (which can change $\varphi_{\rm a}$ significantly): bar growth can enhance further
growth of bars through APS.
This positive feedback loop can be 
called ``gravitational self-proliferation'' in the present study.
This self-proliferation can continue until bars show their maximum possible strengths.

(4) APS in a tidally interacting pair of disk galaxies
can proceed more rapidly during the pericenter passages of its companion
galaxy. Accordingly, the time evolution of 
the $\Sigma(t_{\rm a},\varphi_{\rm a})$ map is quite different from that of
isolated disk models, which confirms that bar formation mechanisms
are different between the two formation paths.
Stellar disks with tidal bars do not show a strong correlation
between $A_2$ and $S_{\rm aps}$, because bar formation processes do not
depend strongly on the structures of tidally perturbed galaxies.
Both $P_{\rm syn}$ and $S_{\rm aps}$ are higher in tidal interaction models than
in isolated ones.
The strong time-changing tidal force of galaxy interaction
is the main cause of such efficient and rapid  APS seen in interaction models.

(5) Thus APS in the formation of spontaneous and tidal bars has been clearly shown,
for the first time, in the present numerical simulations of disk galaxies.
The speed and efficiency of APS have been demonstrated to be different between
spontaneous and tidal bar formation, which reflects the differences in physical
processes that cause APS in the two formation paths.
APS can depend on the physical parameters of disk galaxies, such as $f_{\rm d}$,
$f_{\rm b}$, and $Q$ parameters. Therefore,
one can discuss why bars can be formed in certain ranges of these parameters
(or particular combinations of these) in the context of APS. \\
\section{DATA AVAILABILITY}
The data used in this paper (outputs from computer simulations) 
will be shared on reasonable request
to the corresponding author.
\section{Acknowledgment}
I (Kenji Bekki; KB) am   grateful to the referee  for  constructive and
useful comments that improved this paper.

\appendix

\section{Description of models}

\subsection{Disk galaxies}

A disk  galaxy is assumed to consist of dark matter and stars only (i.e.,
purely collisionless system), because we only investigate the roles
gravitational dynamics of stars in bar formation in the present study.
The total masses of dark matter halo, stellar disk, 
bulge are denoted as $M_{\rm h}$, $M_{\rm d}$, 
and $M_{\rm b}$, respectively.
In order to describe the initial mass density profile of the dark matter halo,
we adopt the density distribution of the so-called ``NFW''
halo (Navarro, Frenk \& White 1996) suggested from CDM simulations:
\begin{equation}
{\rho}(r)=\frac{\rho_{0}}{(r/r_{\rm s})(1+r/r_{\rm s})^2},
\end{equation}
where  $r$, $\rho_{0}$, and $r_{\rm s}$ are
the spherical radius,  the characteristic  density of a dark halo,  and the
scale
length of the halo, respectively.
The $c$-parameter ($c=r_{\rm vir}/r_{\rm s}$, where $r_{\rm vir}$ is the virial
radius of a dark matter halo) and $r_{\rm vir}$ are chosen appropriately
for a given dark halo mass ($M_{\rm dm}$). We mainly investigate
the models with $c=10$ and 16 in the present study.

The stellar bulge of a disk galaxy is assumed to (i) have
a spherical shape  with a size of $R_{\rm b}$
and a scale-length of $a_{\rm b}$ and
(ii) be represented by the Hernquist
density profile with $R_{\rm b}=5a_{\rm  b}$.
The adopted scaling relation between masses (or surface mass densities) and 
sizes for bulges is consistent with the observed one (e.g., Kormendy 1977).
For example, the models with $f_{\rm d}=0.35$ (for $M_{\rm d}=1$)
have $R_{\rm b}=0.2$, 0.35, and 0.49 
for $f_{\rm b}=0.17$, 0.5, and 1.0, respectively.
The spherical bulge is assumed to have isotropic velocity dispersions dependent
on the distance from the bulge's center 
and the radial profile is given according to the Jeans equation
for a spherical system.
The stellar disk with an initial  mass of $M_{\rm d}$ and an initial
size of $R_{\rm d}$ 
is assumed to have
the standard exponential profile and 
the radial ($R$) and vertical ($Z$) density profiles  are
assumed to be proportional to $\exp (-R/a_{\rm s}) $ with scale
length $a_{\rm s} = 0.2R_{\rm d}$  and to ${\rm sech}^2 (Z/Z_{0})$ with scale
length $Z_{0} = 0.04R_{\rm d}$, respectively.
Rotational velocity caused by the gravitational field of disk,
bulge, and dark halo components, and the initial radial and azimuthal
velocity dispersions are assigned to the disc component according to
the epicyclic theory with a given Toomre's ``$Q$''parameter ranging from
1.0 to 3.0 in the present study.
The vertical velocity dispersion at a given radius is set to be 0.5
times as large as the radial velocity dispersion at that point.

We investigate various disk galaxies models with different bulge-to-disk-mass-ratios
($f_{\rm b}=M_{\rm b}/M_{\rm d}$) and disk mass fraction 
disk mass fraction,
and $Q$ 
and $c$ parameters. 
A particularly important parameter in the present study is the disk mass
fraction within $R_{\rm d}$, which is defined as follows:
\begin{equation}
f_{\rm d}=
\frac{ M_{\rm d}(R<R_{\rm d}) }{ M_{\rm t}(R<R_{\rm d}) },
\end{equation}
where $M_{\rm d}(R<R_{\rm d})$ and $M_{\rm t}(R<R_{\rm d})$
represent the total mass of a stellar disk in a disk galaxy
within $R_{\rm d}$ and that of the disk galaxy
including dark matter,
respectively.
We mainly investigate a bulge-less disk model with $f_{\rm d}=0.35$,
$f_{\rm bul}=0$, and $Q=1.5$ in the present study, and the model 
is referred to as the fiducial model. The number of particles
and the spatial resolution of the stellar particles  in
the fiducial model are $1500000$, $0.03R_{\rm d}$, respectively.
Since we focus exclusively on global dynamics ($> 1$kpc-scale)
 of bar formation in disk galaxies,
this adopted number is enough for the required spatial resolution in
this investigation.
The number of particles used for bulges in other models are simply
proportional to $f_{\rm b}N_{\rm d}$, where $N_{\rm d}$ is the number 
of particles used for stellar disks.

In the present study,
all masses and lengths are measured in units of $M_{\rm d}$ and $R_{\rm d}$, 
respectively, unless specified. 
Velocity and time are measured in units of $V_{\rm d} =\sqrt{GM_{\rm d}/R_{\rm d}}$ 
and $t_{\rm dyn} =R_{\rm d}/V_{\rm d}$, respectively, 
where G is the gravitational constant that is  assumed to be 1.0. 
These means that if we adopt $M_{\rm d}  = 6.0 \times 10^{10} {\rm M}_{\odot}$
and $R_{\rm d} = 17.5$  kpc in the fiducial model (like  the Galaxy),
then $v_{\rm d} = 1.21 \times 10^2$ km s$^{-1}$  
and $t_{\rm dyn} = 1.41 \times  10^8$ yr, respectively. 
The pattern speed of a bar and precession and
angular speeds  of stars, which are defined later,
are all given in these dimensionless units (e.g., $V_{\rm d}/R_{\rm d}$ for
the angular speeds, which is 6.91 km s$^{-1}$ kpc$^{-1}$). 
The mass resolution is $10^{-6}M_{\rm d}$, which corresponds
to $6\times 10^4 {\rm M}_{\odot}$ in the above specific disk galaxy mass.
The size resolution for stellar particles (i.e., gravitational
softening length) is $6.7\times 10^{-3} R_{\rm d}$ (117pc for 
$R_{\rm d}=17.5$ kpc), which is enough to discuss global dynamic of
galaxies.

\subsection{Tidal interaction}

One of the two galaxies (`primary galaxy') in a pair of interacting galaxies
is represented by the above-mentioned disk galaxy model
whereas the interacting companion galaxy is represented
by a point-mass particle. 
Although the mass-ratio of the companion to the primary ($m_2$) can be a free parameter,
we investigate the only tidal  models with $m_2=1$, because we focus exclusively
on the formation mechanism of tidal bars and its comparison with that of 
spontaneous bars: the details of the physical properties of tidal bars
were already discussed in previous simulations (e.g., Miwa \& Noguchi 1998; MW98).
We investigate
only ``prograde interaction'' models  in which
the spin axis of the primary in a pair of interacting galaxies
is parallel to the orbital spin axis (thus tidal perturbation to the primary is quite strong).
The initial distance of the two galaxies,  the pericenter distance,
and the orbital eccentricity
are set to be $16R_{\rm d}$, $2R_{\rm d}$, and 1.1
respectively. The disk of the primary is included by 30 degrees with respect to the 
orbital plane of the interacting pair.

\subsection{Simulation code}

We will use our original  code adopted in our previous
studies (Bekki 2013, 2015) in order to perform numerical
simulations of bar formation.
Since the details of the simulation code used in the present
study  are already given in Bekki (2013), we only briefly
describe the code in the present study.
The code  adopts a direct-summation $N$-body algorithm for
gravitational interaction between dark matter and  stars.
The calculation speed of a $N$-body system
is proportional to $N^2$ in the code, however, the 
speed is significantly increased by using GPU.
We adopt the  multiple gravitational softening lengths, i.e.,
the gravitational softening length ($\epsilon$)
can be different
for each component in a galaxy:
$\epsilon$ for  dark matter  ($\epsilon_{\rm dm}$),
disk stars ($\epsilon_{\rm s}$), and bulge stars ($\epsilon_{\rm b}$)
are determined separately by the initial mean separation of
each component.

Furthermore, when two different components interact gravitationally in a galaxy, 
the mean softening length for the two components is 
applied for the gravitational calculation. 
For example, $\epsilon=(\epsilon_{\rm dm} + \epsilon_{\rm s})/2$
is used for gravitational interaction between dark matter and disk stars.
The code also adopts
(i) the models for chemical evolution and galaxy-wide star formation,
and (ii) the smoothed-particle hydrodynamics (SPH) method for following the time
evolution of gas dynamics in galaxies.
However, the modules of the code  related to gas dynamical processes, star formation,
and chemical enrichment
are ``switched off'' in the present study,
because we only analyze gravitational dynamics of disk galaxies.
The roles of gas dynamics and star formation in bar formation should be discussed
in our next papers, though there are many previous papers discussing
how bar formation is influenced by dissipative gas processes
and star formation (e.g., Shlosman \& Noguchi 1993; 
Bekki  1997; Bournaud \& Combes 2002).

\subsection{Comparative models with fixed potentials}

In order to understand the physical mechanisms of bar formation more clearly,
we run the following  comparative models with fixed gravitational potentials in addition to the 
above-mentioned full Nbody simulations.
We consider that APS needs to proceed locally through dynamical processes
in stellar disks when the disks have 
no global stellar  bars.
It would be possible that stars with initially slightly
different $\varphi_{\rm a}$  in local regions with higher mass densities
(or phase densities)  can finally have
similar $\varphi_{\rm a}$ (and $\Omega_{\rm pre}$) due to their mutual gravitational interaction within
local regions.
This local APS can end up with the formation of seed bars from noisy
initial distribution of stars in disks.
In order to demonstrate that 
APS can really proceed in local regions with higher mass densities of stars,
we run ``fixed potential models'' in which (i) the initial mass distributions
of stars and dark matter do not evolve with time at all and (ii)
``local particles'' are newly added to a narrow area
of the above-mentioned full Nbody disk in the fiducial model.
In these comparative experiments, only the local particles
can move under the fixed potential.
and their orbits 
are investigated to derive $\varphi_{\rm a}$ and $t_{\rm a}$ 
of the particles.
In the present study, 
these local particles do not necessarily mean
mass-less particles that do not mutually interact with other 
particles. Instead, local particles with  masses
are assumed to be  gravitationally influenced both by
other local particles and by
the fixed gravitational potentials of the disks.
We also run models in which local particles do not have masses
(i.e., test particles) in order to understand the roles of 
mutual gravitational interaction of the particles in local APS.

These local particles are distributed in a small portion of the  exponential
disk of a galaxy.
The local particles are distributed in a local area with
$0.16R_{\rm d} \le R \le 0.24R_{\rm d}$ and $0 \le \varphi \le 0.2 \pi$
(i.e., in an arc-like region of the disk).
The initial distribution are  drawn from 
the initial distributions
of the Nbody disk model.
They are assumed to be  at their apocenter distances initially (i.e., $T=0$)
and have their initial velocities ($v_{\varphi}$) of $f_{\rm v} v_{\rm c}$
where $v_{\rm c}$ is the 
circular velocity at their initial $R$
and $f_{\rm v}$ is defined as follows:
\begin{equation}
f_{\rm v}=\frac{ v_{\varphi} }{ v_{\rm c} }.
\end{equation}
We show the new results only for the models with $f_{\rm v}=0.7$
for all stellar particles,  mainly because
the results are more than enough to demonstrate the roles of self-gravity
among local particles in 
APS (i.e., synchronization of $\varphi_{\rm a}$).
We also investigate more realistic  models in which different stars have different
$f_{\rm v}$ at $R=R_{\rm a}$ in order to understand how local APS depends on
the dispersion in orbital properties of particles. 
The models with dispersions in $f_{\rm v}$
(denoted by $\sigma(f_{\rm v}$)) being 0.01, 0.03, 0,1, 0.2, and 0.3 are mainly
investigated.

We consider that the ratio of the total mass (number) of the
local particles in the local area 
to that of the same local area including all components
(dark matter, disk, bulge)  in the  fixed Nbody disk 
is also  a key parameter denoted as
$F_{\rm a}$. 
We therefore investigate
the time evolution of $\varphi_{\rm a}$ distribution functions
for models with different $F_{\rm a}$  to confirm that 
the distributions have sharper peaks in the later phase of orbital
evolution of local particles.
In order to derive a reasonable range of $F_{\rm a}$ in the fiducial model,
we here calculate $F_{\rm a}$ for a given $R$ and $\varphi$ at each time step ($t$)
as follows. First, we count the number of particles with $t_{\rm a}$ ranging from
$t-0.1t_{\rm dyn}$ to $t+0.1t_{\rm dyn}$ at each $R$ and  $\varphi$ bin and 
divide the number by the total number of particles located at $R$ and $\varphi$
(at the time step $t$). Accordingly this fraction corresponds to the 
number fraction of stars that have passed their apocenter within $\pm 0.1t_{\rm dyn}$
at each time step ($t$).

We confirm that $F_{\rm a}$ is quite different at different $R$, $\varphi$,
and $t_{\rm a}$ in the fiducial model and accordingly consider that
$F_{\rm a}$ is a key parameter that can control the formation of seed bars
in stellar disks. The average  value $F_{\rm a}$ at $R=0.2R_{\rm d}$
($F_{\rm a, mean}$) 
is 0.28 in the fiducial model:
\begin{equation}
F_{\rm a, mean} = 0.28.
\end{equation}
About 30\% of stars with different $e$ and $\Omega_{\rm pre}$ 
can pass through their apocenters at $|t-t_{\rm a}|\le  0.1t_{\rm dyn}$.
We thus try to demonstrate that mutual gravitational interaction between particles
with similar $t_{\rm a}$ at near apocenters can be crucial
in the formation of seed bars by investigating the models with
different $F_{\rm a}$ in these comparative experiments.

\section{Quantification of  bars}

\begin{figure*}
\psfig{file=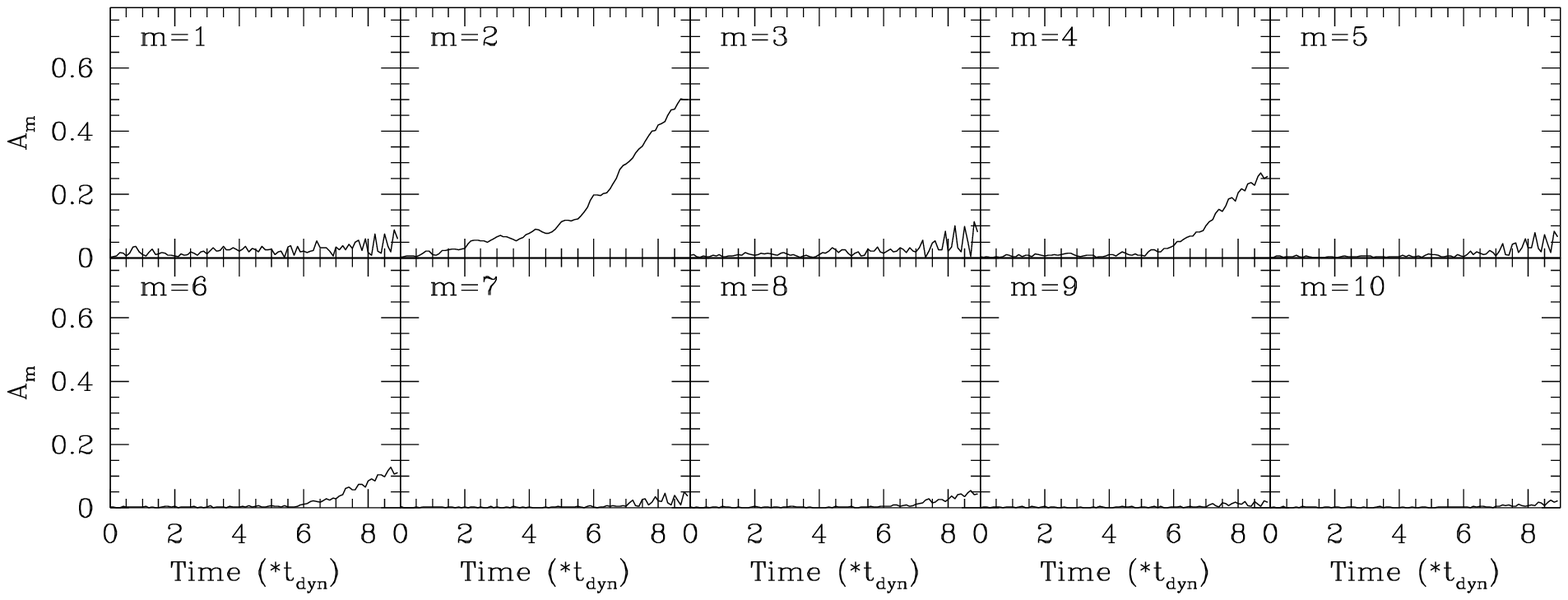,width=18cm}
\caption{
Time evolution of the relative strengths of ${\rm m}=i$ ($i=1$, 2, .... 10)
Fourier components ($A_i=a_i/a_0$) in the fiducial model. $A_2$ corresponding
to the strength of the bar evolved with time differently between $T<4t_{\rm dyn}$
and $T>4t_{\rm dyn}$.
}
\label{Figure. 21}
\end{figure*}

\subsection{Bar strength and pattern speed}

Although there are a number of ways to quantify the strength of
a stellar bar (e.g., Lee et al. 2020),
we adopt the $m=2$ relative Fourier amplitude
($A_2$ as the bar strength as done in other previous simulations
(e.g., Athanassoula et al. 2002).
We investigate the relative Fourier amplitudes at each radius ($R$),
as follows:
\begin{equation}
A_{\rm m}(R)=\frac{ \sqrt{ a_{\rm m}(R)^2 + b_{\rm m}(R)^2 } }{ a_0 },
\end{equation}

where $a_{\rm m}(R)$,  $b_{\rm m}(R)$,
and $a_0$  are the Fourier coefficients estimated
at $R$ from the projected surface density
$\Sigma(R,\phi)$ in the polar coordinate. These coefficient are given
as follows:

\begin{equation}
a_{\rm m}(R)=\frac{1}{\pi}\int_{0}^{2\pi} \Sigma(R,\phi)\cos(m\phi)d\phi
\end{equation}

\begin{equation}
b_{\rm m}(R)=\frac{1}{\pi}\int_{0}^{2\pi} \Sigma(R,\phi)\sin(m\phi)d\phi
\end{equation}

\begin{equation}
a_{\rm 0}=\frac{1}{\pi}\int_{0}^{2\pi} \Sigma(R,\phi)d\phi .
\end{equation}
In order to estimate
the pattern speed of a bar at each radius  ($\Omega_{\rm bar}$(R))
in a simulation at each time step, we estimate the phase of the
$m=2$ component at each time step ($t)$;

\begin{equation}
\phi_{2}(R)=\tan^{-1} \frac{b_2(R)}{a_2(R)}.
\end{equation}
The bar pattern speed at $j$-th time step
is therefore derived as follows:
\begin{equation}
\Omega_{\rm bar}(R)=\frac{ \phi_{2,j}(R) - \phi_{2,j-1}(R) }
{dt},
\end{equation}
where $\phi_{2,j}(R)$ and $\phi_{2,j-1}(R)$ are $\phi_2(R)$ estimated
at $j$-th and ($j-1$)-th time steps, respectively,
and $dt$ is the time step width.

\subsection{Definition of seed bar formation}

A disk galaxy would develop a very weak bar initially,
and the ``seed bar'' grows
though some physical processes in the disk. As described later, the strength of
this seed bar is crucial in the later bar growth processes in the present study.
We therefore need to define the epoch of seed bar formation in order to demonstrate
how the bar is formed.
However,
it would be fairly ad hoc for the present study to use the $A_2$ parameter
(e.g., $A_2=0.2$), which is often used for the strengths of stellar bars
both in observations and simulations,
to define the bar formation epoch.
We consider that when a disk has a seed bar, then (i) there should be two distinct
peaks in the $N(\varphi)$ distribution and
(ii) the two peaks are separated roughly by $\pi$ radian.  Accordingly,
we try to find the bar formation epochs in the simulated disk galaxies
based on these two requirements.

\subsection{Evolution of $A_2$ and $\Omega_{\rm bar}$}

As shown in Fig. B1, $A_2$ of the stellar bar
(i.e., bar strength) rapidly increases
after $T\approx 4t_{\rm dyn}$ in the fiducial model,
however, a bar-like shape cannot be clearly seen
at $T=4.4$ and $5.5 t_{\rm dyn}$. This means that it is difficult to
find morphological evidence for  ``pre-bar'' phases (or ``seed bar'' phases) in 
the observational images of disk galaxies by human eye inspection.
If disk  galaxies with apparent no bars
and $A_{2}<0.1$ observed at different redshifts ($z$)
can be selected as ``pre-bar'' galaxies,
then one can discuss the redshift evolution of the number fraction of forming bars. 
The bar growth rate, $dA_2/dt$ (i.e., slopes in the $A_2$ evolution),
is significantly different
between  $T<4t_{\rm dyn}$ (shallower) and $T>4t_{\rm dyn}$ (steeper),
which indicates the two stages of bar formation.
Fig. B1 
 also clearly demonstrates that other even modes, $m=4$, 6, and 8 can start to grow
significantly later than $m=2$ bar mode.
This result suggests that $A_4/A_2$ (also $A_6/A_2$ and $A_8/A_2$) can be used
to measure the dynamical ages of stellar bars: younger bars that have formed recently
can have smaller $A_4/A_2$, for example. The physical reason for this time evolution
of mode ratios is yet to be clarified, however.

Fig. B2 shows rapid changes of $\Omega_{\rm bar}$
during the early dynamical evolution  ($T<3t_{\rm dyn}$) in this model,
because a seed bar is not so fully developed that the $m=2$ phase ($\phi_2$) 
cannot be accurately defined.
Clearly, $\Omega_{\rm bar}$ estimated
at $R=0.2R_{\rm d}$ ($a_{\rm s}$) and $0.4R_{\rm d}$ ($2a_{\rm s}$)
becomes almost identical only after $T\approx 4t_{\rm dyn}$
when the seed bar starts to grow rapidly (see Fig. B1).
In this bulgeless model, $\Omega_{\rm bar}$ is slightly
lower than $\Omega$ at $R=0.2R_{\rm d}$, which is consistent with
$\Omega_{\rm bar}$ derived in  many previous simulations (e.g., Sellwood 1981).
Given that $\Omega_{\rm bar}$ is much larger than $\Omega - 0.5 \kappa$,
it is not be  possible that the bar gravitationally captures
the resonant orbits of stellar orbits to grow stronger:
the mechanisms proposed by L79 does not work in this model.
Instead, other mechanisms need to be responsible for the bar formation
with $\Omega_{\rm bar}=0.9\Omega$ at $R=0.2R_{\rm d}$, as discussed later in
this paper.

\subsubsection{A correlation between bar pattern speeds and peak precession rates}

Fig. B3 shows a positive and almost linear  correlation between $\Omega_{\rm bar}$ and 
$\Omega_{\rm pre}$ at the peaks of $N(\Omega_{\rm pre})$ distributions
($\Omega_{\rm pre, p}$) in the models in which strong bars
are developed. Although the correlation is not so strong with 
a significant dispersion in $\Omega_{\rm bar}$ for a given $\Omega_{\rm pre, p}$, 
this result indicates that
$\Omega_{\rm bar}$ can be determined by initial $N(\Omega_{\rm pre})$ from which
$\Omega_{\rm p}$ can be estimated directly.
As shown in Fig. 16,
detailed morphological properties of particles distribution on 
the $\Omega_{\rm pre}-e$ plane is quite diverse, which reflects the differences both
in the initial $\Omega_{\rm pre}$ distributions and the bar formation processes
(e.g.,  tidal or spontaneous bar formation).

\begin{figure}
\psfig{file=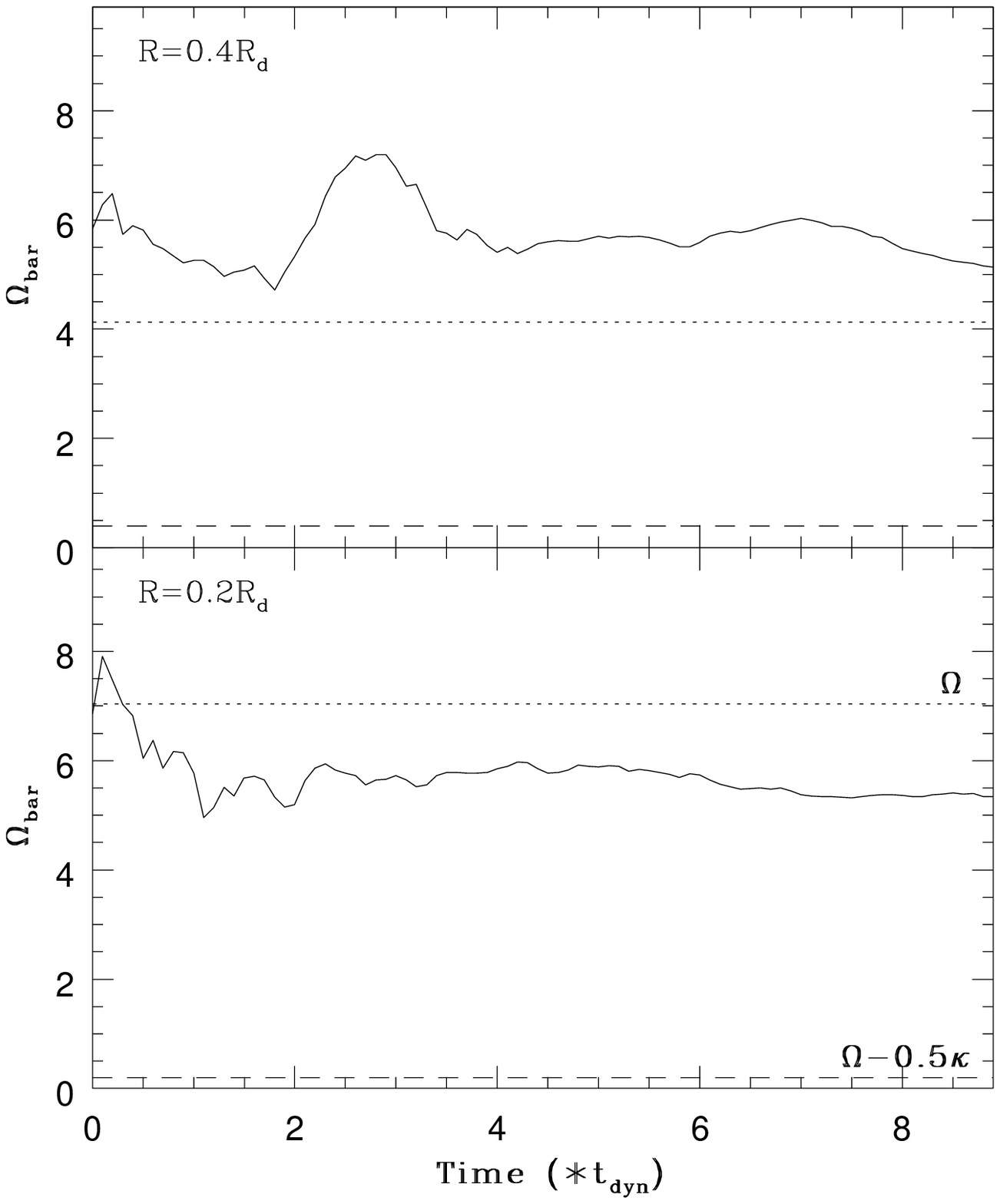,width=8.5cm}
\caption{
Time evolution of bar pattern speeds ($\Omega_{\rm bar}$) estimated
at $R=0.2R_{\rm d}=a_{\rm s}$ (lower) and $R=0.4R_{\rm d}$ (upper)
in the fiducial model. These $\Omega_{\rm bar}$ are given in simulations units.
$\Omega$ and $\Omega-0.5 \kappa$ estimated from the epicyclic theory for
the adopted galactic potential at $T=0$ are shown by dotted and dashed lines,
respectively, in each panel.
The derived pattern speed of $\Omega_{\rm bar} \approx 6$ corresponds
to 41.7 km $^{-1}$ kpc$^{-1}$ for the adopted units for a Milky Way type
disk galaxy. This $\Omega_{\rm bar}$ is slightly lower than the observed
one of the Galaxy (e.g., Weiner \& Sellwood 1999).
}
\label{Figure. 22}
\end{figure}

\begin{figure}
\psfig{file=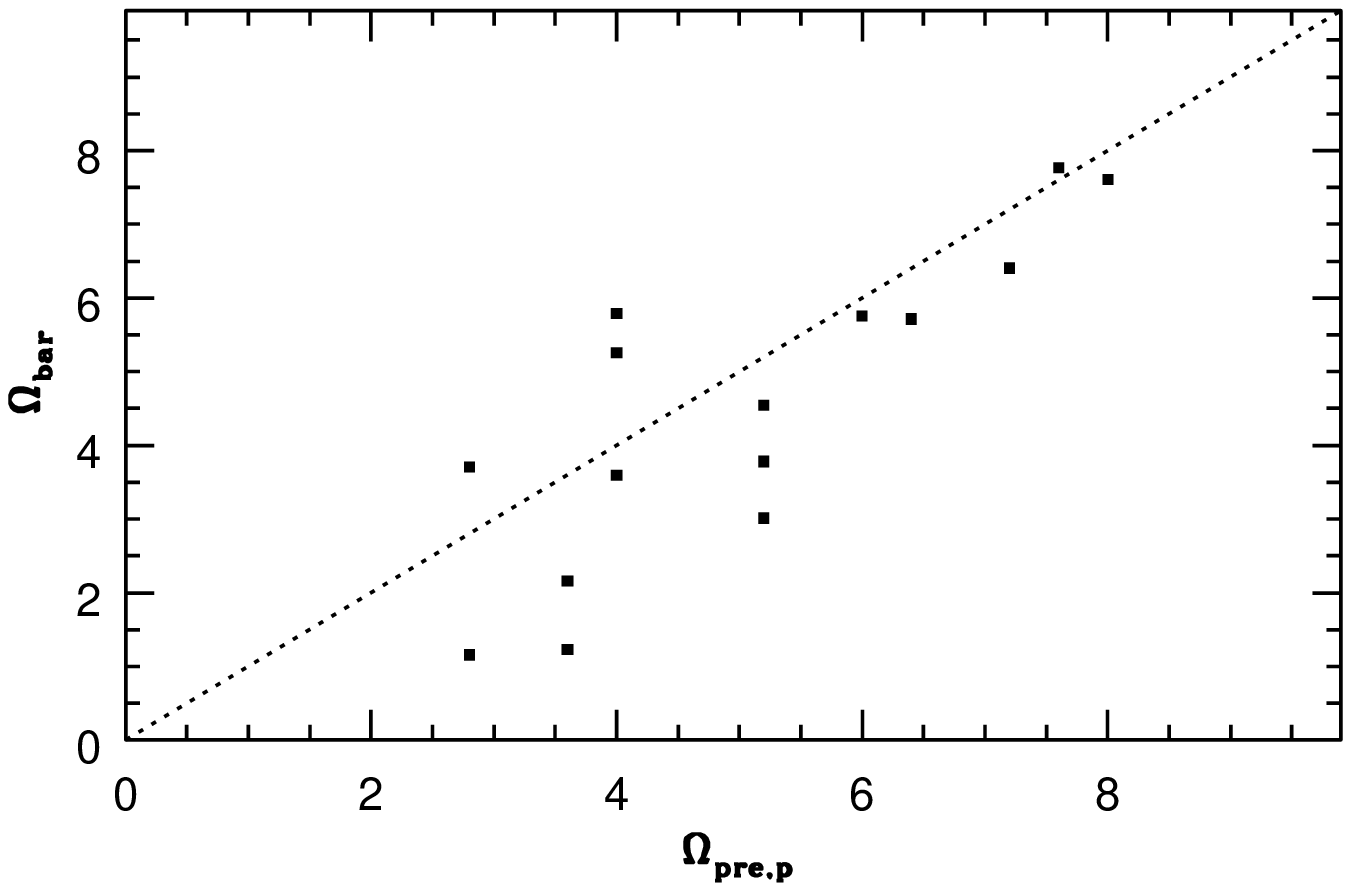,width=8.5cm}
\caption{
A correlation between $\Omega_{\rm bar}$ and $\Omega_{\rm pre}$ averaged
for stellar particles within $R=2a_{\rm s}$ in 9 isolated models with different
model parameters (e.g.,  $f_{\rm d}$
and $f_{\rm b}$) in which strong stellar bars can be formed.
}
\label{Figure. 23}
\end{figure}

\section{$\Omega_{\rm pre}-e$ maps in other models}

\begin{figure*}
\psfig{file=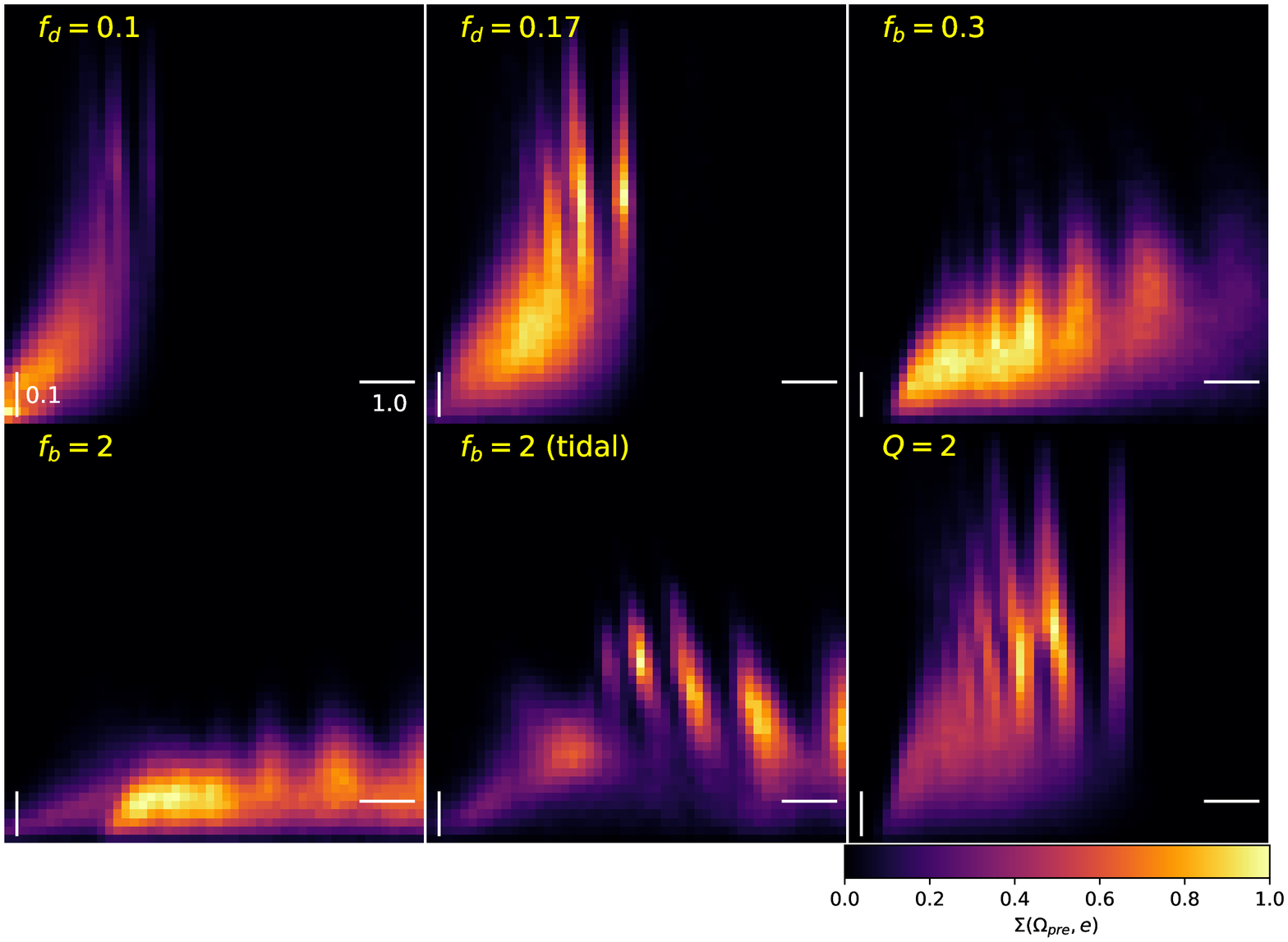,width=16.0cm}
\caption{
The same as Fig. 9 but for six models in which stellar bars cannot be formed
within $10 t_{\rm dyn}$ (IA3, IA5, IA12, IB4, TB3, and IA16).
 The key parameters in these models are shown in the upper
left corner of each frame. The models with lower disk mass fractions ($f_{\rm d}=0.1$
and 0.17) cannot form bars owing to the smaller degrees of self-gravitation in the
stellar components. The models with higher $f_{\rm b}$ cannot form bars either
mainly because the combination of gravitational force from their dark matter
halos and central bulges can prevent local APS from working efficiently for
bar formation. The higher degree of random motion in disk field stars in the model
with $Q=2$ can also suppress the bar formation due to much less efficient APS.
}
\label{Figure. 24}
\end{figure*}

It is clearly shown in the main text that APS is responsible for
the formation of a strong single peak in the 2D
$\Omega_{\rm pre}-e$ map for a simulated disk galaxy: this single peak
can be regarded as a signature of bar formation due to APS.
Since the time evolution of $\Omega_{\rm pre}-e$ maps in the simulated
disk galaxies with bars is well presented and discussed in the main text,
here we focus mainly on the results for disk models  without bars.
Fig. C1 describes the  $\Omega_{\rm pre}-e$ maps for six models in which
stellar bars cannot be formed within $20 t_{\rm dyn}$.
Clearly, all of these models do not show single strong peak in the map
as the models with bars show. Instead, they all show multiple stripe-like
patterns in the map, which is a clear indication that APS cannot proceed
efficiently in the stellar disks.  The models with lower disk mass 
fractions ($f_{\rm d}=0.1$ and 0.17) cannot form bars mainly because the weaker
self-gravity of the disks cannot allow local APS to work to form seed bars.
The central compact bulge in the model with $f_{\rm b}=0.3$ can also
severely suppress APS mechanisms so that a bar cannot be formed within
$20 t_{\rm dyn}$. It should be stressed here that no bar formation in
disks with lower $f_{\rm d}$ and higher $f_{\rm bul}$ was
already pointed out by previous studies (e.g., see a review by Sellwood 2015).
The present study clearly visualize this no formation of bar in the 
$\Omega_{\rm pre}-e$ maps.

It is intriguing that both isolated and tidal interaction models have no bars
if $f_b$ is as large as 2. As a result of this, a single peak
characteristic for bar formation due to APS cannot be seen
in the $\Omega_{\rm pre}-e$ maps of these two models (see Fig. C1).
Given that these models are for bulge-dominated S0s, these results indicates that
bar formation due to spontaneous bar instability and tidal interaction is unlikely
in S0s thus provides a physical reason why the bar fraction in S0s is small both
in observational studies and in cosmological simulations (e.g., Cavanagh et al. 2022 for 
recent discussion on this topic).
It can be seen clearly, however, that the orbital eccentricities of stars ($e$) 
in the tidal model become significantly higher: local APS do not work to form
seed bars though.
The model with higher $Q$ (=2) cannot form a bar, which is consistent with
lack of a single peak (i.e., no APS)  in the 
$\Omega_{\rm pre}-e$ map of the model in Fig. B1.
Thus, the results of these six models with no bars clearly demonstrate that 
APS is one of essential ingredients in bar formation.

\section{A physical origin of $\Omega_{\rm pre}$ change}

\begin{figure*}
\psfig{file=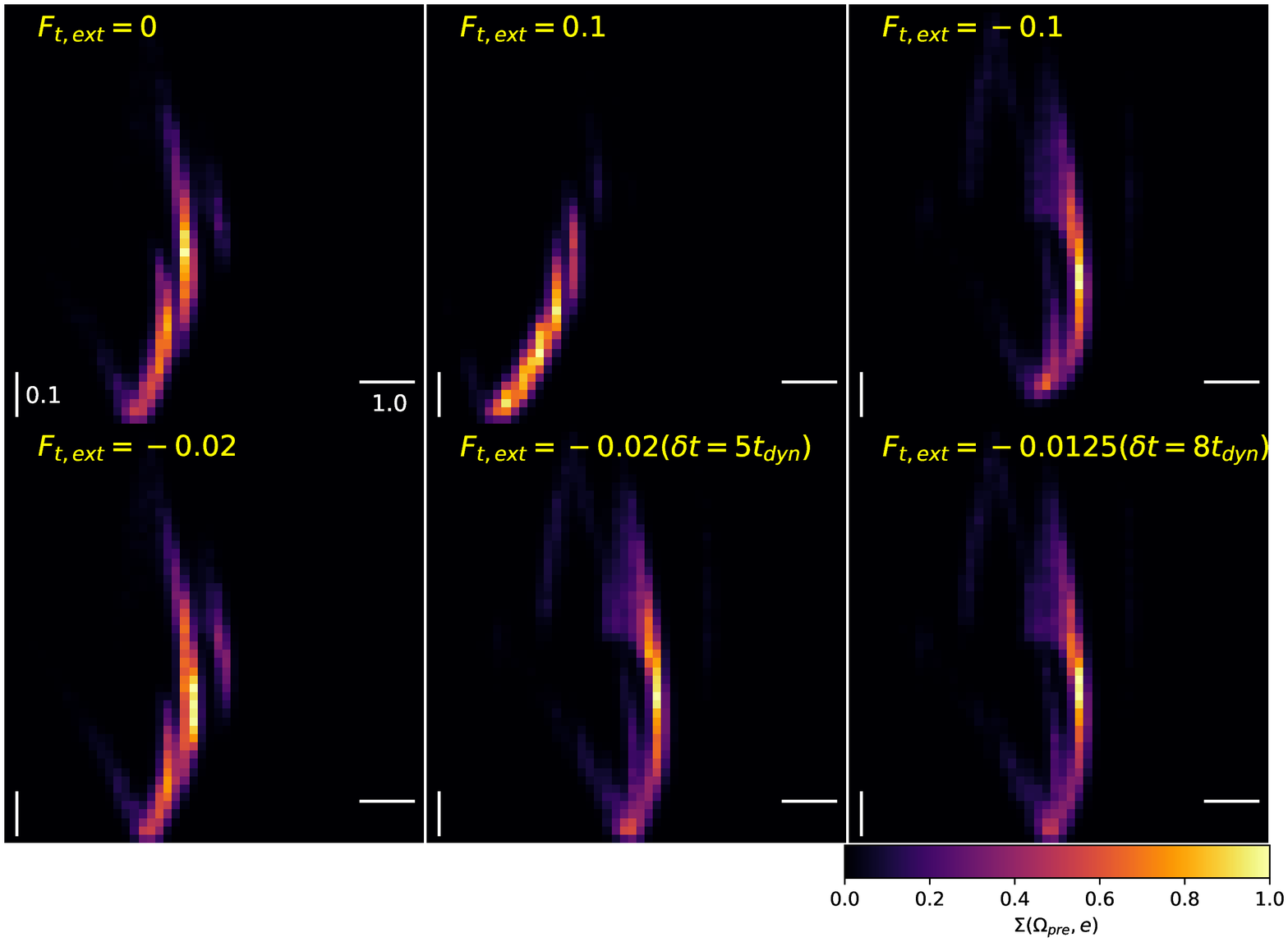,width=16.0cm}
\caption{
The same as Fig. 9 but for six comparative models with different $F_{\rm t, ext}$ and
$\delta t$ (i.e., duration of external force).
Stellar particles in these models initially  have $0.18 \le R/R_{\rm d} \le 0.22$
and $f_{\rm v}=0.7$ and selected from the fiducial model
(i.e., not the entire disk).
The same $\delta t$ ($=t_{\rm dyn}$) is adopted  for the three models
with different $F_{\rm t, ext}$
(upper three frames) whereas two values of $\delta t$ are adopted for the models
with $F_{\rm t, ext}=-0.02$ (lower left and middle frames).
The model with very weak external force ($F_{\rm t, ext}=-0.0125$) yet
very long $\delta t$ ($=8t_{\rm dyn}$) is also shown in the lower right frame.
}
\label{Figure. 25}
\end{figure*}

Although
We have
already shown how $\varphi_{\rm a}$ and $e$ can be influenced
by external gravitational force ($F_{\rm t, ext}$),
we have not
shown how $\Omega_{\rm pre}$ can be influenced by $F_{\rm t, tid}$
in the main text.
In the models described in the main text, this $\delta t$ is fixed to 
be $t_{\rm dyn}$ whereas $F_{\rm t, ext}$ is changed in different models.
We here consider that a change of $\Omega_{\rm pre}$ depends on
$F_{\rm t, ext}$ multiplied by $\delta t$ as follows;
\begin{equation}
\Delta \Omega_{\rm pre} \propto F_{\rm t, ext}\times \delta t.
\end{equation}
Therefore, we investigate models with different $F_{\rm t, ext}$ and
$\delta t$ using the fiducial model below
Fig. D1 describes how the distribution of test stellar
particles  with $0.4 \le R/R_{\rm d} \le 0.5$ in
the $\Omega_{\rm pre}-e$ map 
Fig. D1 describes how the distribution of test stellar
particles  with $0.4 \le R/R_{\rm d} \le 0.5$ in
the $\Omega_{\rm pre}-e$ map depend on $F_{\rm t, ext}$ and $\delta t$.
In these  six models, test stellar particles can move under
the fixed gravitational potential generated by
Nbody particles of the fiducial model: they do not mutually interact through
gravity. Therefore, the effects of mutual gravitational interaction of the particles
on their orbital evolution
are artificially ``switched off'' so that the effects of external force can be 
on $\Omega_{\rm pre}$ and $e$ can be clearly inferred in these models.
It is assumed that $v_{\rm f}$ and its dispersion
are 0.7 and 0.3, respectively, for these stellar particles.

Fig. D1 describes the results of the three models
with $F_{\rm t, ext}=0$ (no external gravitational force),
0.1 (positive tangential force), and $-0.1$ (negative one) in which
$\delta t$ is $t_{\rm dyn}$.
This figure clearly shows that
positive $F_{\rm t, ext}$ (=0.1)  can shift stellar particles toward 
lower $e$ and lower $\Omega_{\rm pre}$ 
on the $\Omega_{\rm pre}-e$ map  due largely to the increased
total angular momentum of the particles. On the other hand,
the particles are moved toward higher $\Omega_{\rm pre}$ without 
any significant change of $e$ in the model with
negative $F_{\rm t, ext}$ (-0.1). 
If $\Omega_{\rm bar}$ of a barred disk galaxy is larger than
$\Omega_{\rm pre}$ of its stars, then the stars need to increase $\Omega_{\rm pre}$
to match with $\Omega_{\rm bar}$ (i.e., to join the bar).
The results in Fig. D1 suggest that the stars can join the bar if they experience
negative $F_{\rm t, ext}$ from the existing bar: stars with 
$\Omega_{\rm pre}> \Omega_{\rm bar}$, on the other hand,
need to experience positive $F_{\rm t,ext}$
from the bar to align with the bar.

Fig. D1 also describes how $\delta t$ influence the evolution of stellar orbits
on the $\Omega_{\rm pre}-e$ map in the models with $F_{\rm t,ext}=-0.02$.
The distribution of stellar particles on the map in the model
with $F_{\rm t, ext}=-0.02$ and $\delta t=t_{\rm dyn}$ is almost identical
to that of the model with $F_{\rm t, ext}=0$.
However, the density peak in the map is moved toward higher $\Omega_{\rm pre}$
in the model with
$F_{\rm t, ext}=-0.02$ and $\delta t=5t_{\rm dyn}$ in which
$F_{\rm t, ext}\delta t$ is the same as that in the model 
with $F_{\rm t, ext}=-0.1$ and $\delta t=t_{\rm dyn}$.
This result implies that $|F_{\rm t, ext}|$ needs to be larger
for a certain period of time in order for external gravitational force
to change $\Omega_{\rm pre}$ effectively.
This value of $F_{\rm t, ext}$ can be achieved for mutual gravitational interaction
of local high-density regions in stellar disks, as demonstrated in the main text.
Fig. D1 also shows that if pretty weak $F_{\rm t, ext}$ can influence stellar
orbits for a longer time scale ($8t_{\rm dyn}$), then their
$\Omega_{\rm pre}$ can be changed significantly by such weak force.
Thus it can be concluded that strengthened tangential/azimuthal component
of gravitational force by local density enhancements and existing bars
is the main cause of $\Omega_{\rm pre}$ changes that are necessary
to APS.

\section{Bar-induced $\varphi_{\rm s}$ synchronization} 

It needs to be clarified why $\varphi_{\rm a}$ shift 
in bar-induced APS  proceeds
in such a way that $|\varphi_{\rm a}-\varphi_{\rm bar}|$
can  become smaller.
This is a complicated problem, because this alignment process
of $\varphi_{\rm a}$ to $\varphi_{\rm bar}$ depends on
orbital parameters of stars and bar pattern speeds
(e.g.,  such as $\Omega-\Omega_{\rm bar}$). 
We here quantify $|\varphi_{\rm a}-\varphi_{\rm bar}|$ at two consecutive
($j$-th and $(j+1)$-th) apocenter passages  of stars at $t_{\rm a, \it j}$ and $t_{\rm a, \it j+1}$ in a growing
bar
by adopting the following  assumptions for clarity:
(1) the bar major axis is aligned with the $x$-axis at $t_{\rm a, \it j}$ 
(i.e., $\varphi_{\rm bar, \it j}=0$),
(2) $0<\varphi_{\rm a, \it j}<\pi/2$,
(3) $\Omega_{\rm pre}$ and $\Omega_{\rm bar}$ are constant and similar 
initially ($\Omega_{\rm pre} \approx \Omega_{\rm bar}$),
and (4) stars are within or near the bar region.
In this qualitative discussion, the difference in $\varphi$ between
a star and the bar at $j$-th apocenter passage  
($|\varphi_{\rm a, \it j}-\varphi_{\rm bar, \it j}|$)
is simply $\varphi_{\rm a, \it j}$.
At $(j+1)$-th apocenter passage, it is described as follows:
\begin{equation}
|\varphi_{\rm a, \it j+1}-\varphi_{\rm bar, \it j+1}|= \varphi_{\rm a, \it j}
+\Delta \varphi_{\rm a}-T_{\rm r}\Omega_{\rm bar},
\end{equation}
where $\Delta \varphi_{\rm a}$ is the $\varphi_{\rm a}$ difference between
two consecutive apocenter passages and $T_{\rm r}$ is the radial 
period of a star.
If there is no (bar's) $F_{\rm t}$
\begin{equation}
\Delta \varphi_{\rm a}= T_{\rm r} \times \Omega_{\rm pre}.
\end{equation}
Accordingly,
\begin{equation}
|\varphi_{\rm a, \it j+1}-\varphi_{\rm bar, \it j+1}|= \varphi_{\rm a, \it j}
+T_{\rm r}\Omega_{\rm pre}-T_{\rm r}\Omega_{\rm bar} \approx \varphi_{\rm a, \it, j},
\end{equation}
because of $\Omega_{\rm pre} \approx \Omega_{\rm bar}$. This means that
$|\varphi_{\rm a}-\varphi_{\rm bar}|$ does not change at all. 
However, as shown in Fig. 23, 
this $\Delta \varphi_{\rm a}$ can change depending on the sign and the magnitude
of $F_{\rm t}$. 
For $\varphi_{\rm a, \it j}>0$ (i.e., star leading the bar), 
$F_{\rm t}$ is negative (see Fig. 4) so that 
$\Delta \varphi_{\rm a}$ can be smaller than $T_{\rm r}  \Omega_{\rm pre}$
by a small amount of $\Delta \varphi$ ($>0$).  Accordingly,
$\Delta \varphi_{\rm a}$ ) for $F_{\rm t}=0$ is  as follows:
\begin{equation}
\Delta \varphi_{\rm a}= T_{\rm r} \times \Omega_{\rm pre}-\Delta \varphi .
\end{equation}
Therefore, it follows from these equations that 
\begin{equation}
|\varphi_{\rm a, \it j+1}-\varphi_{\rm bar, \it j+1}|
<|\varphi_{\rm a, \it j}-\varphi_{\rm bar, \it j}|
\end{equation}

If $-\pi/2<\varphi_{\rm a, \it j}<0$ (the bar leading stars),
then the net $F_{\rm t}$ due to the bar's gravitational force can be positive.
Therefore $\Delta \varphi_{\rm a}$ can be larger
than $T_{\rm r}  \Omega_{\rm pre}$, which means that 
$|\varphi_{\rm a, \it j+1}-\varphi_{\rm bar, \it j+1}|
<|\varphi_{\rm a, \it j}-\varphi_{\rm bar, \it j}|$.
Clearly, these discussion is valid only 
for $\Omega_{\rm pre} \approx \Omega_{\rm bar}$, 
and we did not consider that  {\it net} $F_{\rm t}$ effect on the long-term  orbits of stars  depends on how long the stars
lead or trail the bar. It is possible that
the ``donkey effect'' (Lynden-Bell \& Kalnajs 1972)
can influence the time duration when a star leads/trails the bar
(due to positive or negative $F_{\rm t}$ by the bar).
More quantitative discussion needs further investigation of 
$F_{\rm t}$ evolution as a function of $\varphi_{\rm a}$, $\Omega_{\rm pre}$,
and $R_{\rm a}$ of stars
and $\Omega_{\rm bar}$.

It also needs to be clarified why $|\Omega_{\rm pre}-\Omega_{\rm bar}|$ 
become small due to dynamical action of bars  on stars.
As discussed in Appendix C,  $\Omega_{\rm pre}$ can change significantly
due to stronger $F_{\rm t}$ ($>0.02$), if stars can experience such tangential
force (due to bar in this discussion) for certain duration of time.
Accordingly,
if $\Omega_{\rm pre} > \Omega_{\rm bar}$,
the net $F_{\rm t}$ needs to be positive so that 
$\Omega_{\rm pre}$ can become smaller (to be more similar to $\Omega_{\rm bar}$).
On the other hand,
If $\Omega_{\rm pre} < \Omega_{\rm bar}$,
the net $F_{\rm t}$ needs to be negative so that 
$\Omega_{\rm pre}$ can becomes larger.
As shown in Fig. 8, a  fraction of
stellar particles with $0.1 \le R/R_{\rm d} < 0.2$
decrease their $\Omega_{\rm pre}$ to be more similar to $\Omega_{\rm bar}$.
Also, a significant fraction of  stellar particles with $0.3 \le R/R_{\rm d} < 0.4$
increase their $\Omega_{\rm pre}$ to be more similar to $\Omega_{\rm bar}$ in Fig. 8.
Thus, a very qualitative explanation for this evolution of $\Omega_{\rm pre}$
depending on $\Omega_{\rm pre}$  is that  net $F_{\rm t}$ of stars with
different $R$ can depend largely on
their $\Omega_{\rm pre}$ (that depends on $R$).
It is our future study to provide
a more quantitative explanation for the bar-induced APS
for the formation of bars  with different bar properties 
(lengths, pattern speeds, and shapes etc).

\end{document}